\documentclass[aps,onecolumn]{revtex4}

\usepackage[dvips]{graphicx}
\def\theequation{\arabic{section}.\arabic{equation}}
\newcommand{\Slash}[1]{\ooalign{\hfil/\hfil\crcr$#1$}}
\begin{document}

\title{ {\tt {\small \begin{flushright}
MAN/HEP/2012/22, December 2012
\end{flushright} } }
\vspace{5mm}
Charged Lepton Flavour Violation in Supersymmetric Low-Scale Seesaw Models}

\author{Amon Ilakovac$^{\,a}$, Apostolos Pilaftsis$^{\,b}$ and Luka
  Popov$^{\,a}$\vspace{2mm}} 

\affiliation{
${}^a$University of Zagreb, Department of Physics,
  Bijeni\v cka cesta 32, P.O. Box 331, Zagreb,
  Croatia\vspace{1mm}\\ 
${}^b$Consortium for Fundamental Physics,
  School of Physics and Astronomy, University of Manchester,
  Manchester M13 9PL, United Kingdom
}

\begin{abstract}
\noindent
We study  charged lepton flavour violation in  low-scale seesaw models
of minimal supergravity, which realize large neutrino Yukawa couplings
thanks  to  approximate   lepton-number  symmetries.   There  are  two
dominant  sources of  lepton flavour  violation in  such  models.  The
first  source originates  from the  usual  soft supersymmetry-breaking
sector,  whilst the second  one is  entirely supersymmetric  and comes
from the supersymmetric neutrino  Yukawa sector.  Within the framework
of minimal  supergravity, we consider  both sources of  lepton flavour
violation, soft and supersymmetric, and calculate a number of possible
lepton-flavour-violating transitions,  such as the  photonic decays of
muons and  taus, $\mu \to e\gamma$,  $\tau \to e\gamma$  and $\tau \to
\mu\gamma$, their neutrinoless three-body decays, $\mu \to eee$, $\tau
\to~eee$,  $\tau \to  \mu\mu\mu$,  $\tau  \to  e e\mu$  and  $\tau\to
e\mu\mu$, and  the coherent $\mu  \to e$ conversion in  nuclei.  After
taking into account the exclusion bounds placed by present experiments
of lepton flavour violation,  we derive combined theoretical limits on
the universal  heavy Majorana mass scale~$m_N$  and the light-to-heavy
neutrino  mixings.   Supersymmetric   low-scale  seesaw  models  offer
distinct    correlated   predictions    for   lepton-flavour-violating
signatures,  which  might  be  discovered  in  current  and  projected
experiments, such as MEG, COMET/PRISM, Mu2e, super-BELLE and LHCb.

\medskip

\end{abstract}

\maketitle 

\medskip

\setcounter{equation}{0}
\section{Introduction}\label{intro}

Neutrino                                                    oscillation
experiments~\cite{sol-nuosc,atm-nuosc,acc-nuosc,reac-nuosc}        have
provided undisputed evidence of  Lepton Flavour Violation (LFV) in the
neutrino sector,  pointing towards  physics beyond the  Standard Model
(SM).   Recent  reactor  neutrino oscillation  experiments~\cite{nu13}
have     shown    that    the     angle    $\theta_{13}$     of    the
Pontecorvo--Maki--Nakagawa--Sakata~(PMNS) mixing matrix~\cite{PMNS} is
nonzero, thus hinting at  a non-trivial neutrino-flavour structure and
possibly  at   CP  violation.   Nevertheless,  in   spite  of  intense
experimental searches~\cite{LFV-tau,SINDRUM88,Titanium,Gold,MEG11}, no
evidence of LFV has been found yet in the charged lepton sector of the
SM, implying conservation of  the individual lepton numbers associated
with the  electron $e$, the muon  $\mu$ and the  tau~$\tau$.  All past
and  current experiments  were only  able  to report  upper limits  on
observables of Charged Lepton Flavour Violation~(CLFV).

Recently, the MEG collaboration~\cite{MEG11} has announced an improved
upper limit on the branching ratio of the CLFV decay $\mu\to e\gamma$,
with $B(\mu\to  e\gamma) < 2.4\times 10^{-12}$ at  the 90\% confidence
level~(CL).  As  also shown in  Table~\ref{TableI}, future experiments
searching  for the  CLFV  processes, $\mu\to  e\gamma$, $\mu\to  eee$,
coherent    $\mu    \to   e$    conversion    in   nuclei,    $\tau\to
e\gamma/\mu\gamma$,   $\tau\to   3\   \mbox{leptons}$   and   $\tau\to
\mbox{lepton$+$light  meson}$, are  expected to  reach branching-ratio
sensitivities    to    the    level   of    $10^{-13}$~\cite{MEGPhD12}
($10^{-14}$~\cite{Hew12}),  $10^{-16}$ \cite{mu3eBerger11} ($10^{-17}$
\cite{Hew12}),                             $10^{-17}$~\cite{Mu2e,COMET}
($10^{-18}$~\cite{PRISM,Hew12}),          $10^{-9}$~\cite{Bon07,Hay09},
$10^{-10}$~\cite{Hay09}   and  $10^{-10}$~\cite{Hay09},  respectively.
The values in parentheses indicate the sensitivities that are expected
to  be achieved by  the new  generation CLFV  experiments in  the next
decade.   Most interestingly,  the projected  sensitivity  for $\mu\to
eee$ and $\mu  \to e$ conversion in nuclei is  expected to increase by
five and  six orders of  magnitude, respectively.  Given that  CLFV is
forbidden  in  the  SM,  its  observation  would  constitute  a  clear
signature for New Physics, thus rendering this field of investigations
even more exciting.

\begin{table}[ht]
 \begin{tabular}{rlclcl}
\hline
 No.~ & Observable &~~~~~~& Upper Limit &~~~~~&Future Sensitivity \\
\hline
 1.~  & $B (\mu \to e \gamma)$ & &$2.4 \times 10^{-12}$ \cite{MEG11}
      & &$1-2\times 10^{-13}$ \cite{MEGPhD12}, $10^{-14}$ \cite{Hew12}\\
 2.~  & $B (\mu \to eee)$ & &$10^{-12}$ \cite{SINDRUM88}
      & &$10^{-16}$ \cite{mu3eBerger11}, $10^{-17}$   \cite{Hew12} \\
 3.~  & $R_{\mu e}^{\rm Ti}$ & &$4.3\times 10^{-12}$ \cite{Titanium}, 
      & &$3-7\times 10^{-17}$ \cite{Mu2e,COMET}, $10^{-18}$ \cite{PRISM,Hew12}\\
 4.~  & $R_{\mu e}^{\rm Au}$ & &$7 \times 10^{-13}$ \cite{Gold}
      & &$3-7\times 10^{-17}$ \cite{Mu2e,COMET}, $10^{-18}$ \cite{PRISM,Hew12}\\
 5.~  & $B(\tau \to e \gamma)$ & & $3.3 \times 10^{-8}$ \cite{LFV-tau}
      & & $1-2\times 10^{-9}$ \cite{Bon07,Hay09}\\
 6.~  & $B(\tau \to \mu \gamma)$ & &$4.4 \times 10^{-8}$ \cite{LFV-tau}
      & & $2\times 10^{-9}$ \cite{Bon07,Hay09} \\
 7.~  & $B(\tau \to eee)$ & & $2.7 \times 10^{-8}$ \cite{LFV-tau}
      & & $2 \times 10^{-10}$ \cite{Bon07,Hay09} \\
 8.~  & $B(\tau \to e\mu\mu)$ & & $2.7 \times 10^{-8}$ \cite{LFV-tau}
      & & $10^{-10}$ \cite{Hay09} \\
 9.~  & $B(\tau \to \mu\mu\mu)$ & & $2.1 \times 10^{-8}$ \cite{LFV-tau}
      & & $2 \times 10^{-10}$ \cite{Bon07,Hay09} \\
10.~  & $B(\tau \to \mu ee)$ & & $1.8 \times 10^{-8}$ \cite{LFV-tau}
      & & $10^{-10}$ \cite{Hay09}\\
\hline
\end{tabular}
\caption{Current upper limits and future sensitivities of CLFV observables 
under study.}\label{TableI}
\end{table}

Although forbidden  in the SM, CLFV  is a generic feature  for most of
its extensions.   One such  well-motivated extension is  the so-called
Minimal  Supersymmetric  Standard  Model  (MSSM)~\cite{Review},  where
supersymmetry  (SUSY) is  softly broken  at  the 1--10  TeV scale  for
phenomenological  reasons. The  MSSM  provides a  quantum-mechanically
stable  solution to the  gauge hierarchy  problem and  predicts rather
accurate  unification of  the SM  gauge couplings  close to  the Grand
Unified Theory  (GUT) scale.  If  R-parity is conserved,  the lightest
supersymmetric particle (LSP)  is stable and, if neutral,  such as the
neutralino, it  could play  the role  of the Dark  Matter (DM)  in the
Universe.  Finally, the MSSM  typically predicts a SM-like Higgs boson
lighter than 135~GeV, in agreement  with the recent observations for a
125--GeV  Higgs  boson,  made  by  the ATLAS  \cite{HdisATL}  and  CMS
\cite{HdisCMS} collaborations.

In the MSSM with R-parity conservation, the lepton number is preserved
and   all  left-handed   light  neutrinos   $\nu_{e,\mu,\tau}$  remain
massless, exactly  as in  the SM.  To  account for the  observed light
neutrino  masses and  mixings, while  maintaining R-parity,  the field
content of the MSSM needs  to be extended. An interesting extension to
the MSSM  is provided  by the so-called  seesaw mechanism.   There are
three  realizations  of  the  seesaw  mechanism:  the  seesaw  type  I
\cite{seesaw}, the seesaw type  II \cite{seesawII} and the seesaw type
III  \cite{seesawIII}.  The three  scenarios differ  by the  nature of
their seesaw messengers that are  needed to explain the small neutrino
masses. In this study, we will adopt a low-scale variant of the seesaw
type-I   realization,  whose  messengers   are  three   singlet  heavy
neutrinos~$N_{1,2,3}$.

In the usual seesaw type-I mechanism, the heavy singlet neutrinos must
assume  masses of  order~$\sim 10^{12-14}$~GeV,  for electroweak-scale
Dirac  neutrino masses,  in order  to account  for the  observed light
neutrino mass  spectrum. The mixing between light  and heavy neutrinos
is of the order  $\xi_{\nu N}\sim \sqrt{m_\nu/m_N} \sim 10^{-12}$, for
light-neutrino masses $m_\nu\sim  10^{-1}$~eV. As a consequence, heavy
neutrinos decouple  from low-energy processes  of CLFV in the  SM with
right-handed  neutrinos,  giving  rise  to  extremely  suppressed  and
unobservable rates.  In the MSSM with right-handed neutrinos, however,
the singlet  heavy neutrinos do  not fully decouple.  They  impact the
low-energy  sector, through  renormalization  group~(RG) effects  that
induce  sizeable LFV  in  the slepton  sector  for a  SUSY mass  scale
$M_{\rm SUSY}\sim 1-10$~TeV.

A potentially interesting alternative to the ordinary seesaw mechanism
may   arise   from   the   presence   of   approximate   lepton-number
symmetries~\cite{WW,MV,BGL,AZPC} in  the theory. The  smallness of the
light neutrino  masses is a consequence of  these approximate leptonic
symmetries  which are  radiatively  stable~\cite{AZPC,APRLtau}, whilst
the heavy neutrino  mass scale could be as low  as 100~GeV.  Unlike in
the    usual   seesaw    scenario,    the   light-to-heavy    neutrino
mixings~$\xi_{\nu  N}$  are  not  correlated  to  the  light  neutrino
masses~$m_\nu$.    Instead,  $\xi_{\nu   N}$   are  free   parameters,
constrained by  experimental limits on deviations of  the $W^\pm$- and
$Z$-boson   couplings   to   leptons   with  respect   to   their   SM
values~\cite{LFVp,BK}.   Approximate lepton-number  symmetries  do not
restrict the size  of LFV, and so potentially  large phenomena of CLFV
may  be  predicted.   This  feature  is  quite  generic  both  in  the
SM~\cite{IPNPB}  and in  the  MSSM~\cite{IPPRD,IPNPBP} augmented  with
low-scale right-handed  neutrinos.  It is  this new source for  LFV in
the MSSM that we wish to study  in detail here, in addition to the one
resulting   from  the   frequently  considered   soft   SUSY  breaking
sector~\cite{BM,HMTY,softLFV,EHLR}.

In this article, we denote for brevity the SM and the MSSM extended by
low-scale   right-handed  neutrinos   and   approximate  lepton-number
symmetries by  $\nu_R$SM and  $\nu_R$MSSM, respectively. Our  study in
this  paper  focuses  on  the $\nu_R$MSSM  with  constrained  boundary
conditions  at  the   gauge-coupling  unification  scale,  within  the
framework  of  minimal  supergravity  (mSUGRA). However,  the  results
presented  here  are applicable  to  more  general soft  SUSY-breaking
scenarios.

The  $\nu_R$MSSM has  some interesting  features with  respect  to the
MSSM.  In particular,  the heavy singlet sneutrinos may  emerge as new
viable candidates of Cold  Dark Matter (CDM)~\cite{CDM}.  In addition,
the mechanism  of low-scale resonant leptogenesis~\cite{ResLG,APRLtau}
could provide a possible explanation for the observed Baryon Asymmetry
in  the   Universe~(BAU),  as  the  parameter   space  for  successful
electroweak   baryogenesis   gets   squeezed   by  the   current   LHC
data~\cite{EWBAU}.

Given  the   multitude  of  quantum   states  mediating  LFV   in  the
$\nu_R$MSSM,  the predicted  values for  observables of  CLFV  in this
model turn out to be generically larger than the corresponding ones in
the       $\nu_R$SM,      except       possibly       for      $B(l\to
l'\gamma)$~\cite{IPPRD,IPNPBP},  where $l,\: l'  = e,\:  \mu,\: \tau$.
The  origin  of  suppression  for  the  latter  branching  ratios  may
partially be attributed  to the SUSY no-go theorem  due to Ferrara and
Remiddi~\cite{FR74},  which  states that  the  magnetic dipole  moment
operator necessarily violates  SUSY and it has therefore  to vanish in
the supersymmetric limit of the theory.

The goal of this paper is to improve upon earlier calculations of CLFV
in   supersymmetric   low-scale  seesaw   models   which  are   either
approximate, or incomplete.  Specifically, in~\cite{IPPRD} a low-scale
aligned  neutrino and sneutrino  spectrum was  assumed to  the leading
order      in      the      heavy      neutrino      mass,      whilst
in~\cite{Hirsch:2012ax,Abada:2012cq}  only a  subset  of loop  effects
mediated  by the  photon~$\gamma$, the  $Z$ boson,  the  neutral Higgs
scalars  and box  graphs  was considered.   Instead,  we present  here
detailed  analytical  expressions  for  CLFV observables,  induced  by
$\gamma$- and $Z$-boson formfactors, as well as the {\em complete} set
of box contributions  involving heavy neutrinos, sleptons, sneutrinos,
charginos, neutralinos  and charged Higgs bosons.  We  also derive the
{\em complete}  set of kinematic  formfactors that contributes  to the
three-body LFV decays of the muon and tau, such as $\mu\to eee$, $\tau
\to~eee$,  $\tau  \to  \mu\mu\mu$,  $\tau  \to e  e\mu$  and  $\tau\to
e\mu\mu$.  In our numerical  analysis, we consider benchmark scenarios
of the $\nu_R$MSSM, which are  in agreement with the existing LHC data
from     the     recent    discovery     of     a    SM-like     Higgs
boson~\cite{HdisATL,HdisCMS}  and the  non-observation of  squarks and
gluinos~\cite{m-gtqt}.

The  paper  is organized  as  follows.   In Section~\ref{lowscale}  we
describe the relevant leptonic sector of the $\nu_R$MSSM and introduce
two baseline scenarios  based on approximate lepton-number symmetries.
Section~\ref{CLFV}  contains details  of  the calculation  of the  LFV
amplitudes and  branching ratios of CLFV  decays of the  tau and muon,
where  the  complete  set  of  chiral  structures  of  the  amplitudes
contributing  to  the CLFV  processes  $l  \to  l'\gamma$, $l  \to  l'
l_1\bar{l}_2$  and  $\mu\to  e$  conversion  in  nuclei  are  derived.
Section~\ref{numerics}   presents    numerical   estimates   for   the
aforementioned  processes  of CLFV,  within  the  two baseline  mSUGRA
scenarios introduced  in Section~\ref{lowscale}. Moreover,  we discuss
correlated predictions of the CLFV observables with $B(l \to l' \gamma
)$ and other  kinematic parameters. Section~\ref{concl} summarizes the
results of  our analysis and presents our  conclusions.  All technical
details    have    been    relegated    to   the    appendices.     In
Appendix~\ref{vertices}, we give  all relevant interaction vertices in
the $\nu_R$MSSM.  Appendix~\ref{sec:lf} defines the one-loop functions
that  appear in  our analytic  calculations.  In  terms of  these loop
functions, Appendix~\ref{sec:olff} describes  the analytic results for
all  the  one-loop  formfactors  that  occur in  the  CLFV  transition
amplitudes.

\setcounter{equation}{0}
\section{Low--Scale Seesaw Models and Lepton Flavour Violation}\label{lowscale}

In this section  we describe the leptonic sector  of the MSSM extended
by low-scale right-handed neutrinos, which we call the $\nu_R$MSSM for
brevity. In  addition, we introduce  the neutrino Yukawa  structure of
two baseline  scenarios based on  approximate lepton-number symmetries
and universal Majorana masses at the  GUT scale. We will use these two
scenarios to present generic  predictions of CLFV within the framework
of mSUGRA.

The leptonic superpotential part of the $\nu_R$MSSM reads:
\begin{equation}
  \label{Wpot}
W_{\rm lepton}\ =\  \widehat{E}^C {\bf h}_e \widehat{H}_d
\widehat{L}\: +\: \widehat{N}^C {\bf h}_\nu \widehat{L} \widehat{H}_u\:
+\: \frac{1}{2}\,\widehat{N}^C {\bf m}_M \widehat{N}^C\; ,
\end{equation}
where    $\widehat{H}_{u,d}$,    $\widehat{L}$,   $\widehat{E}$    and
$\widehat{N}^C$  denote the two  Higgs-doublet superfields,  the three
left-  and  right-handed  charged-lepton  superfields  and  the  three
right-handed   neutrino   superfields,   respectively.    The   Yukawa
couplings~${\bf  h}_{e,\nu}$ and  the  Majorana mass  parameters~${\bf
  m}_M$  form $3\times 3$  complex matrices.  Here, the  Majorana mass
matrix ${\bf  m}_M$ is taken to  be SO(3)-symmetric at  the GUT scale,
i.e.~${\bf m}_M = m_N\, {\bf 1}_3$.

In      this     study,      we     consider      low-scale     seesaw
models~\cite{WW,MV,BGL,AZPC},  where   the  smallness  of   the  light
neutrino masses  is protected by  natural, quantum-mechanically stable
cancellations   due   to   the   presence  of   approximate   leptonic
symmetries~\cite{AZPC}, whilst the Majorana mass scale $m_N$ can be as
low as 100~GeV.  In these models, the neutrino induced LFV transitions
from a charged lepton $l=\mu,\,\tau$ to another charged lepton $l'\neq
l$ are functions of the ratios~\cite{KPS,IPNPB,DV}
\begin{equation}
  \label{Omega}
{\bf  \Omega}_{l'l}\ =\  \frac{v^2_u}{2  m^2_N}\ ({\bf  h}^\dagger_\nu
{\bf h}_\nu)_{l'l}\ =\ \sum_{i=1}^3 B_{l'N_i} B_{lN_i}
\end{equation}
and are not constrained by  the usual seesaw factor $m_\nu/m_N$, where
$v_u/\sqrt{2}  \equiv \langle  H_u\rangle$ is  the  vacuum expectation
value (VEV) of the Higgs doublet $H_u$, with $\tan\beta \equiv \langle
H_u\rangle/\langle H_d\rangle$. Moreover, the mixing matrix $B_{lN_i}$
that occurs in the interaction  of the $W^\pm$ bosons with the charged
leptons $l = e,\ \mu,\ \tau$ and the three heavy neutrinos $N_{1,2,3}$
is defined  in Appendix~\ref{vertices}.  Note that  the LFV parameters
$\Omega_{l'l}$ do not directly depend  on the RG evolution of the soft
SUSY-breaking parameters, except through  the VEV $v_u$ at the minimum
of the Higgs potential.

In  the weak  basis $\{  (\nu_{e,\mu,\tau L})^C,  \nu_{1,2,3R}\}$, the
neutrino mass matrix  in the $\nu_R$MSSM takes on  the standard seesaw
type-I form~\cite{seesaw}:
\begin{eqnarray}
  \label{Mnu}
{\bf M}_\nu &=& \left(\begin{array}{cc} 0 & {\bf m}_D \\ {\bf m}_D^T &
  {\bf m}_M^* \end{array}\right) \ , 
\end{eqnarray}
where    ${\bf   m}_D    =   \sqrt{2}M_W\sin\beta\,    g^{-1}_w   {\bf
  h}_\nu^\dagger$ and ${\bf m}_M$  are the Dirac- and Majorana-neutrino
mass matrices, respectively.  In~this paper, we consider two baseline
scenarios of neutrino Yukawa couplings.  The first one realizes a U(1)
leptonic symmetry~\cite{APRLtau} and is given by
\begin{eqnarray}
  \label{YU1}
{\bf h}_\nu &=&
 \left(\begin{array}{lll}
 0 & 0 & 0 \\
 a e^{-\frac{i\pi}{4}} & b e^{-\frac{i\pi}{4}} & c e^{-\frac{i\pi}{4}} \\
 a e^{ \frac{i\pi}{4}} & b e^{ \frac{i\pi}{4}} & c e^{ \frac{i\pi}{4}}
 \end{array}\right)\; .
\end{eqnarray}
In the  second scenario, the  structure of the neutrino  Yukawa matrix
${\bf h}_\nu$  is motivated by  the discrete symmetry group  $A_4$ and
has the following form~\cite{KS}:
\begin{eqnarray}
  \label{YA4}
{\bf h}_\nu &=&
 \left(\begin{array}{lll}
 a & b & c \\
 a e^{-\frac{2\pi i}{3}} & b e^{-\frac{2\pi i}{3}} & c e^{-\frac{2\pi i}{3}} \\
 a e^{ \frac{2\pi i}{3}} & b e^{ \frac{2\pi i}{3}} & c e^{ \frac{2\pi i}{3}}
\end{array}\right)\;.
\end{eqnarray}
In~(\ref{YU1}) and~(\ref{YA4}), the Yukawa parameters $a$, $b$ and $c$
are  assumed  to  be  real.    If  the  above  leptonic  (discrete  or
continuous)  symmetries  are  not  broken,  the  light  neutrinos  are
massless.  The neutrino masses are obtained by adding small terms that
break the  symmetry of  the Yukawa matrix.  It is essential  to remark
here  that the  predictions for  CLFV are  independent of  the flavour
structure  of these  small  symmetry-breaking terms  in our  low-scale
seesaw  models  under  study.   These  terms are  needed  to  fit  the
low-energy neutrino  data.  For  this reason, we  do not  discuss here
particular symmetry breaking patterns of the above two baseline Yukawa
scenarios given in~(\ref{YU1}) and~(\ref{YA4}).

The second source of LFV  in the models under consideration originates
from    scalar-neutrino   (sneutrino)    interactions,    namely   the
supersymmetric   partners   of   the  left-handed   and   right-handed
neutrinos. Specifically,  the sneutrino mass Lagrangian  in flavour and
mass bases is given by
\begin{eqnarray}
{\cal L}^{\tilde{\nu}}
 &=&
 (\tilde{\nu}_L^\dagger,\tilde{\nu}_R^{C\,\dagger},\tilde{\nu}_L^T,\tilde{\nu}_R^{C\,T})   
 \,{\bf M}^2_{\tilde{\nu}}\,
 \left(
 \begin{array}{c}
  \tilde{\nu}_L\\ \tilde{\nu}_R^{C}\\ \tilde{\nu}_L^*\\ \tilde{\nu}_R^{C*}
 \end{array}
 \right)
 \ =\
  \tilde{N}^\dagger {\cal U}^{\tilde{\nu}\dagger}
  {\bf M}^2_{\tilde{\nu}}\, {\cal U}^{\tilde{\nu}} \tilde{N}
 \ =\
 \tilde{N}^\dagger \widehat{{\bf M}}^2_{\tilde{\nu}} \tilde{N},
\end{eqnarray}
where  ${\bf M}^2_{\tilde{\nu}}$  is  a $12\times  12$ Hermitian  mass
matrix in the flavour basis and $\widehat{{\bf M}}^2_{\tilde{\nu}}$ is
the  corresponding  diagonal mass  matrix  in  the  mass basis.   More
explicitly,   in    the   flavour   basis   $\{\tilde{\nu}_{e,\mu,\tau
  L},\tilde{\nu}_{1,2,3R}^C,\tilde{\nu}_{e,\mu,\tau
  L}^*,\tilde{\nu}_{1,2,3R}^{C*}\}$,  the sneutrino mass  matrix ${\bf
  M}^2_{\tilde{\nu}}$ may be cast into the form:
\begin{eqnarray}
   \label{M2snu1}
{\bf  M}^2_{\tilde{\nu}} 
 &=&
 \left(
 \begin{array}{cccc}
 {\bf  H}_1 & {\bf  N} & {\bf  0} & {\bf  M} \\
 {\bf  N}^\dagger & {\bf  H}_2^T & {\bf  M}^T & {\bf  0} \\
 {\bf  0} & {\bf  M}^* & {\bf  H}_1^T & {\bf  N}^* \\
 {\bf  M}^\dagger & {\bf  0} & {\bf  N}^T & {\bf  H}_2
 \end{array}
 \right)\; ,
\end{eqnarray}
where the block entries are the following $3\times 3$ matrices
\begin{eqnarray}
  \label{M2snu2}
{\bf  H}_1 &=& {\bf  m}^2_{\tilde{L}} + {\bf  m}_D {\bf  m}_D^\dagger
+ \frac{1}{2} M_Z^2 \cos 2\beta 
\nonumber\\ 
{\bf  H}_2 &=& {\bf  m}^2_{\tilde{\nu}} + {\bf  m}_D^\dagger {\bf
  m}_D + {\bf  m}_M {\bf  m}_M^\dagger 
\nonumber\\
{\bf  M} &=& {\bf  m}_D ({\bf  A}_\nu - \mu \cot\beta)
\nonumber\\
{\bf  N} &=& {\bf  m}_D {\bf  m}_M\; .
\end{eqnarray} 
Here  ${\bf  m}^2_{\tilde{L}}$,  ${\bf m}^2_{\tilde{\nu}}$  and  ${\bf
  A}_\nu$ are $3\times 3$  soft SUSY-breaking matrices associated with
the  left-handed  slepton doublets,  the  right-handed sneutrinos  and
their trilinear couplings, respectively.  In the supersymmetric limit,
all the soft SUSY-breaking matrices  are equal to zero, $\tan\beta =1$
and  $\mu=0$.   As a  consequence,  the  sneutrino  mass matrix  ${\bf
  M}^2_{\tilde{\nu}}$ can  be expressed in terms of  the neutrino mass
matrix ${\bf M}_\nu$ in~(\ref{Mnu}) as follows:
\begin{eqnarray}
{\bf M}^2_{\tilde{\nu}}
 &\stackrel{\rm SUSY}{\longrightarrow} &
 \left(
 \begin{array}{cc} 
 {\bf M}_\nu {\bf M}_\nu^\dagger & {\bf 0}_{6\times 6} \\ 
 {\bf 0}_{6\times 6} & {\bf M}_\nu^\dagger {\bf M}_\nu
 \end{array}
 \right),
\label{MnutMnu}
\end{eqnarray}
leading  to  the  expected  equality between  neutrino  and  sneutrino
mixings.  Sneutrino LFV  mixings do depend on the  RG evolution of the
$\nu_R$MSSM  parameters, but unlike  the LFV  mixings induced  by soft
SUSY-breaking terms,  the sneutrino LFV  mixings do not vanish  at the
GUT scale.

The  sneutrino LFV  mixings are  obtained as  combinations  of unitary
matrices  that diagonalize  the sneutrino,  slepton and  chargino mass
matrices.  It is interesting to  notice that in the diagonalization of
the sneutrino mass matrix ${\bf M}^2_{\tilde{\nu}}$ in (\ref{M2snu1}),
the      sneutrino      fields      $\tilde{\nu}_{e,\mu,\tau      L}$,
$\tilde{\nu}_{1,2,3R}^C$     and      their     complex     conjugates
$\tilde{\nu}_{e,\mu,\tau   L}^*$,  $\tilde{\nu}_{1,2,3   R}^{C*}$  are
treated   independently.    As   a   result,   the   expressions   for
$\tilde{\nu}_{e,\mu,\tau L}$ and $\tilde{\nu}_{1,2,3R}^C$, in terms of
the real-valued mass eigen\-states~$\widetilde{N}_{1,2,\dots,12}$, are
not  manifestly complex  conjugates to  $\tilde{\nu}_{e,\mu,\tau L}^*$
and   $\tilde{\nu}_{1,2,3R}^{C*}$,   thus   leading  to   a   two-fold
interpretation of the flavour basis fields,
\begin{eqnarray}
\tilde{\nu}^*_{i}  &=& (\tilde{\nu}_i)^* 
 \  =\ {\cal U}^{\tilde{\nu}*}_{i A} \widetilde{N}_A\;,\nonumber\\ 
\tilde{\nu}^*_{i} &=& 
 {\cal U}^{\tilde{\nu}}_{i + 6\, A} \widetilde{N}_A\ ,
\end{eqnarray}
where  $\tilde{\nu}_{1,2,3}  \equiv  \tilde{\nu}_{e,\mu,\tau  L}$  and
$\tilde{\nu}_{4,5,6}      \equiv     \tilde{\nu}^C_{1,2,3R}$,     with
$i=1,2,\dots,6$   and   $A=1,2,\dots,12$.    For   this   reason,   in
Appendix~\ref{vertices}  we  include  all  equivalent forms  that  the
Lagrangians, such as ${\cal L}_{\overline{e}\tilde{\chi}^- \tilde{N}}$
and ${\cal L}_{\tilde{N}\tilde{N}Z}$, can be written down.

One  technical comment is  now in  order.  Unlike~\cite{IPPRD,IPNPBP},
the diagonalization  of the $12\times 12$ sneutrino  mass matrix ${\bf
  M}^2_{\tilde{\nu}}$  and  the  resulting  interaction  vertices  are
evaluated  numerically,   without  approximations.   To   perform  the
diagonalization of  ${\bf M}^2_{\tilde{\nu}}$ numerically,  we use the
method developed in~\cite{P-PRD08} for the neutrino mass matrix.  This
method  becomes  very  efficient,  if the  diagonal  submatrices  have
eigenvalues    larger   than    the   eigenvalues    of    the   other
submatrices. Therefore,  we assume that the heavy  neutrino mass scale
$m_N$ is of the  order of, or larger than the scale  of the other mass
parameters in the $\nu_R$MSSM.

Finally, a  third source of LFV  in the $\nu_R$MSSM  results from soft
SUSY-breaking LFV terms. These LFV terms are induced by RG running and
are absent at the GUT scale in mSUGRA.  Their size strongly depends on
the interval of  the RG evolution from the GUT  scale to the universal
heavy neutrino mass scale $m_N$.

All  the  three  different   mechanisms  of  LFV,  mediated  by  heavy
neutrinos,  heavy  sneutrinos  and  soft SUSY-breaking  terms,  depend
explicitly on the  neutrino Yukawa matrix ${\bf h}_\nu$  and vanish in
the limit ${\bf h}_\nu \to 0$.

%******************************************************************
%%%Figure 1
%******************************************************************
\begin{figure}[t]
 \centering
  \includegraphics[clip,
  width=0.7\textwidth,height=0.5\textheight,
     angle=0]{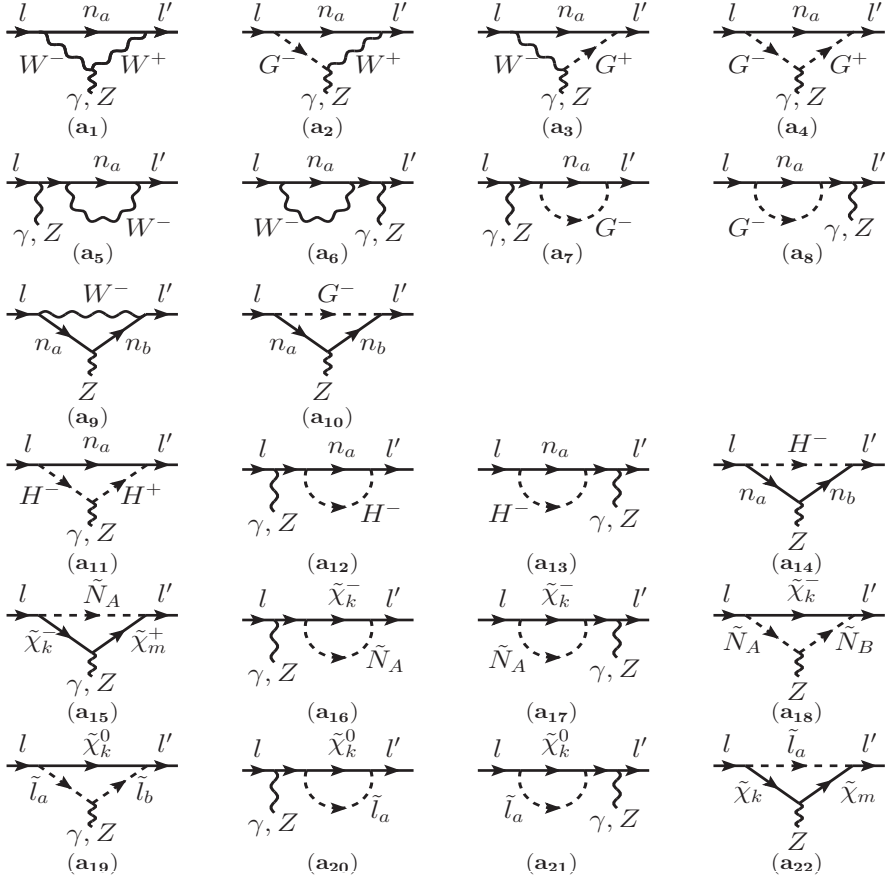} 
\caption{Feynman  graphs contributing to $l\to l'\gamma$ and $Z\to l^Cl'$ 
($l\to Z l'$) amplitudes} \label{f1}
\end{figure}

\vfill\eject

\setcounter{equation}{0}
\section{Charged Lepton Flavour Violation}\label{CLFV}

In  this section, we  present key  details of  our calculations  for a
number   of  CLFV   observables  in   the  $\nu_R$MSSM.    In  detail,
Section~\ref{sec:Zgll} gives analytical  results for the amplitudes of
the  CLFV  decays:  $l\to  l'\gamma$  and $Z\to  l  l'^C$,  and  their
branching     ratios.      Correspondingly,     Sections~\ref{sec:lll}
and~\ref{sec:mueff} give  analytical expressions for  the neutrinoless
three-body decays $l\to l'l_1l_2^C$  pertinent to muon and tau decays,
and  for  coherent  $\mu  \to  e$ conversion  in  nuclei.   All  these
analytical results  are expressed in  terms of one-loop  functions and
composite  formfactors  that  are defined  in  Appendices~\ref{sec:lf}
and~\ref{sec:olff}.

\subsection{The Decays {\boldmath $l\to l'\gamma$} and 
{\boldmath $Z\to l l'^C$}}\label{sec:Zgll}

At the one-loop level, the effective couplings $\gamma l'l$ and $Zl'l$
are  generated by  the  Feynman graphs  shown  in Fig.~\ref{f1}.   The
general  form  of  the  transition amplitudes  associated  with  these
effective couplings is given by
\begin{eqnarray}
  \label{Tll'g}
{\cal T}^{\gamma l'l}_\mu 
 \!&=&\!
 \frac{e\, \alpha_w}{8\pi M^2_W}\: \bar{l}'
 \Big[ (F^L_\gamma)_{l'l}\, (q^2\gamma_\mu-\Slash{q}q_\mu) P_L
     + (F^R_\gamma)_{l'l}\, (q^2\gamma_\mu-\Slash{q}q_\mu) P_R
\nonumber\\&&
     +  (G^L_\gamma)_{l'l}\, i \sigma_{\mu\nu}q^\nu P_L
     +  (G^R_\gamma)_{l'l}\, i \sigma_{\mu\nu}q^\nu P_R \Big]\: l,
\\
  \label{Tll'Z}
{\cal T}^{Z l'l}_\mu 
 \!&=&\!  
 \frac{g_w\, \alpha_w}{8\pi \cos\theta_w}\: \bar{l}'
 \Big[ (F_Z^L)_{l'l}\, \gamma_\mu P_L 
     + (F_Z^R)_{l'l}\, \gamma_\mu P_R \Big]\: l,
\end{eqnarray}
where  $P_{L(R)}   =  \frac{1}{2}\,[1-\!(+)\,\gamma_5]$,  $\alpha_w  =
g^2_w/(4\pi)$, $e$  is the  electromagnetic coupling constant,  $M_W =
g_w \sqrt{v^2_u  +v^2_d}/2$ is the  $W$-boson mass, $\theta_w$  is the
weak mixing angle and $q =  p_{l'} - p_l$ is the photon momentum.  The
form      factors      $(F^L_\gamma)_{l'l}$,      $(F^R_\gamma)_{l'l}$
$(G^L_\gamma)_{l'l}$,    $(G^R_\gamma)_{l'l}$,   $(F_Z^L)_{l'l}$   and
$(F_Z^R)_{l'l}$    receive   contributions   from    heavy   neutrinos
$N_{1,2,3}$, heavy  sneutrinos $\widetilde{N}_{1,2,3}$ and  RG induced
soft SUSY-breaking terms.  The  analytical expressions for these three
individual contributions  are given in  Appendix \ref{sec:olff}.  Note
that  according  to  our  normalization, the  composite  formfactors
$(G^L_\gamma)_{l'l}$ and $(G^R_\gamma)_{l'l}$  have dimensions of mass,
whilst all other formfactors are dimensionless.

It  is  essential  to  remark  here  that  the  transition  amplitudes
(\ref{Tll'g})  and (\ref{Tll'Z})  are  also constituent  parts of  the
leptonic  amplitudes  $l\to  l'l_1l_2^C$ and  semileptonic  amplitudes
$l\to  l'q_1 \bar{q}_2$,  which will  be discussed  in more  detail in
Sections~\ref{sec:lll}  and~\ref{sec:mueff}.   To  calculate the  CLFV
decay  $l\to l'\gamma$,  we only  need to  consider the  dipole moment
operators  associated with  the  formfactors $(G^L_\gamma)_{l'l}$  and
$(G^R_\gamma)_{l'l}$  in  (\ref{Tll'g}). Taking  this  last fact  into
account, the branching ratios for $l\to  l'\gamma$ and $Z \to l l'^C +
l^C l'$ are given by
\begin{eqnarray}
  \label{Bllpg}
B(l\to l'\gamma) 
 \!&=&\!
 \frac{\alpha_w^3 s_w^2}{256\pi^2} \frac{m_l^3}{M_W^4\Gamma_l} 
 \Big( |(G^L_\gamma)_{l'l}|^2 + |(G^R_\gamma)_{l'l}|^2 \Big)\; ,\\
  \label{Blplc}
B(Z\to l'l^C+l'^Cl)
 \!&=&\!
 \frac{\alpha_w^3 M_W}{768 \pi^2 c_w^3\Gamma_Z} \Big(|(F_Z^L)_{l'l}|^2 +
 |(F_Z^R)_{l'l}|^2\Big)\; .
\end{eqnarray}
Observe  that the  above expressions  are  valid to  leading order  in
external charged lepton masses and external momenta, which constitutes
an excellent  approximation for our  purposes.  Thus, in~(\ref{Blplc})
we have assumed that the $Z$-boson mass $M_Z$ is much smaller than the
SUSY and heavy neutrino mass  scales, $M_{\rm SUSY}$ and $m_N$, and we
have kept the leading term in an expansion of small momenta and masses
for  the external  particles. In  the  decoupling regime  of all  soft
SUSY-breaking and charged Higgs-boson masses, the low-energy sector of
the $\nu_R$MSSM becomes the $\nu_R$SM.  In this $\nu_R$SM limit of the
theory, the analytical expressions  for $B(l\to l'\gamma)$ and $B(Z\to
l'l^C+l'^Cl)$ take on the forms presented in~\cite{CL} and~\cite{KPS},
respectively.

\begin{figure}[t]
 \centering
  \includegraphics[clip,
  width=0.7\textwidth,height=0.45\textheight,
     angle=0]{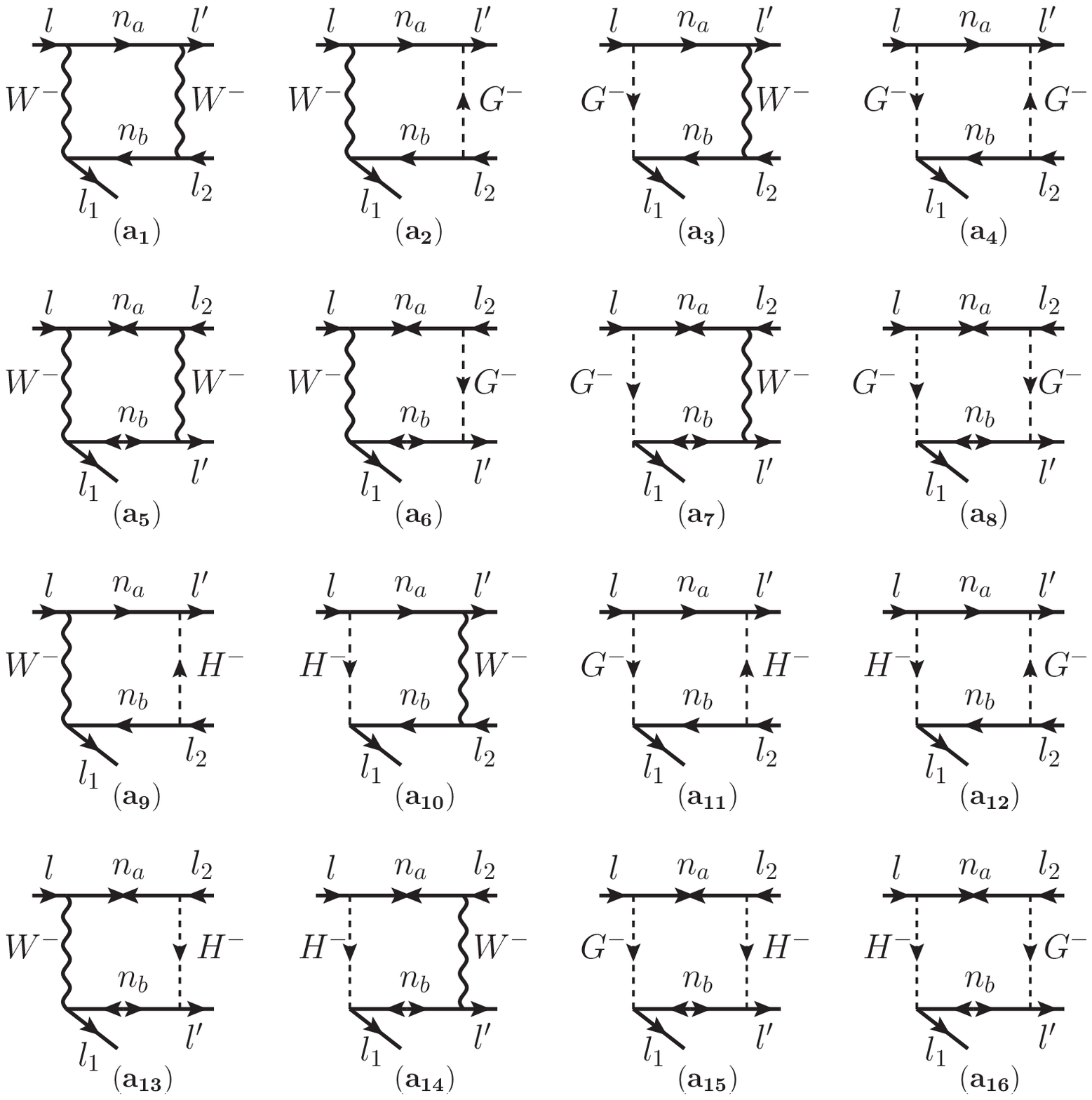}
  \includegraphics[clip,
  width=0.7\textwidth,height=0.27\textheight,
     angle=0]{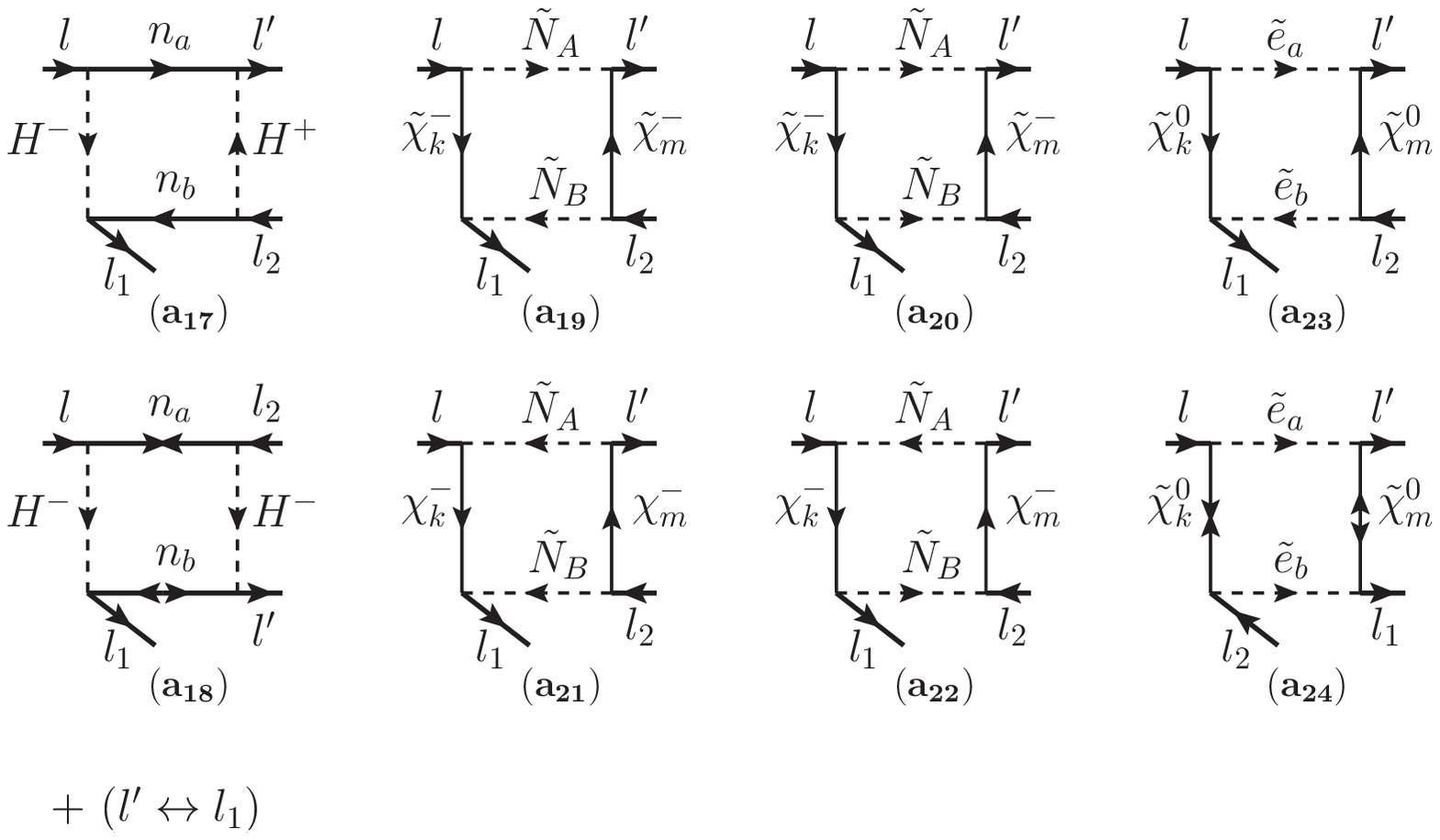}
\caption{Feynman  graphs contributing to the box $l\to l'l_1l_2^C$ 
amplitudes.} \label{f2}
\end{figure}

\subsection{Three-Body Leptonic Decays {\boldmath $l\to
    l'l_1l_2^C$}}\label{sec:lll}  

We now study  the three-body CLFV decays $l\to  l'l_1l_2^C$, where $l$
can be  the muon or tau,  and $l',\, l_1,\, l_2$  denote other charged
leptons to which $l$ is allowed to decay kinematically.

The transition amplitude  for $l\to l'l_1l_2^C$ receives contributions
from  $\gamma$- and $Z$-boson-mediated  graphs shown  in Fig.~\ref{f1}
and from  box graphs displayed  in Fig.~\ref{f2}.  The  amplitudes for
these three contributions are
\begin{eqnarray}
  \label{Tgl3l}
{\cal T}_{\gamma}^{ll'l_1l_2} 
 \!&=&\!
   \label{Tllpg}
 \frac{\alpha_w^2 s_w^2}{2 M_W^2}
 \Big\{ \delta_{l_1l_2}\: \bar{l}'\:
  \Big[ (F_\gamma^L)_{l'l}\, \gamma_\mu P_L + (F_\gamma^R)_{l'l}\,
    \gamma_\mu P_R \nonumber\\
&&
 +  \frac{(\Slash{p}-\Slash{p}')}{(p-p')^2} 
         \Big( (G_\gamma^L)_{l'l}\, \gamma_\mu P_L +
         (G_\gamma^R)_{l'l}\, \gamma_\mu P_R \Big) \Big]\, l\:   
      \: \bar{l}_1\gamma^\mu l_2^C 
  \,-\, [\, l'\leftrightarrow l_1 ] \Big\}\; ,\\
  \label{TZl3l}
{\cal T}_Z^{ll'l_1l_2}
   \label{TllpZ}
 \!&=&\!
 \frac{\alpha_w^2}{2 M_W^2}
 \Big[ \delta_{l_1l_2}\: \bar{l}'\Big( (F_Z^L)_{l'l}\,\gamma_\mu P_L 
+ (F_Z^R)_{l'l}\, \gamma_\mu P_R\Big) l 
    \ \bar{l}_1 \Big(g_L^l\, \gamma^\mu P_L + g_R^l\, \gamma^\mu P_R\Big) l_2^C
- ( l'\leftrightarrow l_1 ) \Big]\; ,\\
  \label{TBl3l}
{\cal T}_{\rm box}^{ll'l_1l_2}
  \label{Tbl_1}
 \!&=&\!
 - \frac{\alpha_w^2}{4 M_W^2}
 \Big(
    B_{\ell V}^{LL}\, \bar{l}'\gamma_\mu P_L l\ \bar{l}_1\gamma^\mu P_L l_2^C
  + B_{\ell V}^{RR}\, \bar{l}'\gamma_\mu P_R l\ \bar{l}_1\gamma^\mu P_R l_2^C
\nonumber\\&&
  + B_{\ell V}^{LR}\, \bar{l}'\gamma_\mu P_L l\ \bar{l}_1\gamma^\mu P_R l_2^C
  + B_{\ell V}^{RL}\, \bar{l}'\gamma_\mu P_R l\ \bar{l}_1\gamma^\mu P_L l_2^C\,
\nonumber\\&&
  + B_{\ell S}^{LL}\, \bar{l}' P_L l\ \bar{l}_1 P_L l_2^C
  + B_{\ell S}^{RR}\, \bar{l}' P_R l\ \bar{l}_1 P_R l_2^C
%\nonumber\\&&
  + B_{\ell S}^{LR}\, \bar{l}' P_L l\ \bar{l}_1 P_R l_2^C
  + B_{\ell S}^{RL}\, \bar{l}' P_R l\ \bar{l}_1 P_L l_2^C
\nonumber\\&&
  + B_{\ell T}^{LL}\, \bar{l}' \sigma_{\mu\nu} P_L l\ \bar{l}_1
  \sigma^{\mu\nu} P_L l_2^C
  + B_{\ell T}^{RR}\, \bar{l}' \sigma_{\mu\nu} P_R l\ \bar{l}_1
  \sigma^{\mu\nu} P_R l_2^C  \Big) \\
\!&\equiv&\!
  \label{Tbl_2}
- \frac{\alpha_w^2}{4 M_W^2} \sum_{X,Y=L,R}\ \sum_{A=V,S,T}
    B_{\ell A}^{XY}\, \bar{l}'\Gamma^X_{A} l\ \bar{l}_1\Gamma^Y_{A}
    l_2^C\; ,
\end{eqnarray} 
where  $g_L^l =  -1/2+s_w^2$ and  $g_R^l=s_w^2$  are $Z$-boson--lepton
couplings  and  $s_w=\sin\theta_w$.   The  composite  box  formfactors
$B_{\ell  A}^{XY}$,   $A=V,S,T$,  $X,Y=L,R$  are   given  in  Appendix
\ref{sec:olff}.  The labels $V$, $S$ and $T$ denote the formfactors of
the vector, scalar and tensor  combinations of the currents, while $L$
and  $R$  distinguish between  left  and  right  chiralities of  these
currents.  We  note that the  box formfactors contain both  direct and
Fierz-transformed contributions.   Equation~(\ref{Tbl_2}) represents a
shorthand   expression   that   takes   account  of   all   individual
contributions  to  the   amplitude  ${\cal  T}_{\rm  box}^{ll'l_1l_2}$
induced  by   box  graphs.   Explicitly,   the  matrices  $\Gamma^X_A$
appearing in~(\ref{Tbl_2}) read:
\begin{equation}
\Big(\,\Gamma^L_V,\,\Gamma^R_V,\,\Gamma^L_S,\,\Gamma^R_S,\,\Gamma^L_T,\,
\Gamma^R_T\,\Big)
\ =\ 
\Big(\, \gamma_\mu P_L,\,\gamma_\mu P_R,\, P_L,\, P_R,\,
\sigma_{\mu\nu}P_L,\,  \sigma_{\mu\nu}P_R\, \Big)\ .\quad 
\end{equation}
Notice that  the tensor formfactors  $B_{\ell T}^{LR}$ and $B_{\ell T}^{RL}$
vanish  in the sum  (\ref{Tbl_2}), i.e.~$B_{\ell T}^{LR}=B_{\ell T}^{RL}=0$,
as   a  consequence   of  the   identity   $\sigma^{\mu\nu}\gamma_5  =
-\frac{i}{2}\varepsilon^{\mu\nu\rho\tau}  \sigma_{\rho\tau}$.   A very
similar  chiral structure  holds also  true for  the  semileptonic box
amplitudes defined  in the next  section.  We should remark  here that
\cite{HMTY} and  \cite{AH05} do not include in  their calculations the
chiral  structures  $P_L\times  P_R$  and $P_R\times  P_L$  and  their
corresponding formfactors $B_{\ell S}^{LR}$ and $B_{\ell S}^{RL}$.

In a  three-generation model, the transition amplitude  for the decays
$l\to l'l_1l_2^C$  may fall in one  of the following  three classes or
categories~\cite{IPNPB}: (i)~$l'\neq l_1=l_2$, (ii)~$l'= l_1=l_2$, and
(iii)~$l'= l_1\neq l_2$.   In the first two classes~(i)  and (ii), the
total lepton number is conserved,  whilst in the third class~(iii) the
total lepton  number is  violated by two  units. Here, we  ignore this
lepton-number  violating class~(iii) of  charged lepton  decays, since
the predictions turn out to  be unobservably small in the $\nu_R$MSSM.
Moreover, we suppress  the universal indices $l'l$ that  appear in the
photon  and  $Z$-boson  formfactors, i.e.~$F_\gamma^L$,  $F_\gamma^R$,
$F_Z^L$ and  $F_Z^R$. Given the above simplification  and the notation
of the box formfactors  in~(\ref{Tbl_2}), the branching ratios for the
class (i) and (ii) of CLFV three-body decays are given by
\begin{eqnarray}
  \label{Bl3l_1}
B(l\to l'l_1l_1^C) 
 \!&=&\! 
 \frac{m_l^5\alpha_w^4}{24576\pi^3 M_W^4\Gamma_l}\: 
 \Bigg\{\, \bigg[\, 
  \Big|2s_w^2 (F_\gamma^L + F_Z^L) - F_Z^L - B_{\ell V}^{LL}\Big|^2
  + \Big|2s_w^2 (F_\gamma^R + F_Z^R) - B_{\ell V}^{RR}\Big|^2
\nonumber\\
&&+\ \Big|2s_w^2 (F_\gamma^L + F_Z^L) - B_{\ell V}^{LR} \Big|^2
  + \Big|2s_w^2 (F_\gamma^R + F_Z^R) - F_Z^R - B_{\ell V}^{RL} \Big|^2\,
  \bigg] 
\nonumber\\
 \!&&\! 
 +\ \frac{1}{4} \Big( |B_{\ell S}^{LL} |^2 + |B_{\ell S}^{RR} |^2 +
  |B_{\ell S}^{LR} |^2 + |B_{\ell S}^{RL} |^2 \Big) 
 + 12 \Big( |B_{\ell T}^{LL}|^2  +  |B_{\ell T}^{RR}|^2 \Big)
\nonumber\\
 \!&&\!
 +\ \frac{32s_w^4}{m_l} \Big[  
 \mbox{Re} \Big( (F_\gamma^R + F_Z^R ) G_\gamma^{L*}\Big)
 + \mbox{Re} \Big( (F_\gamma^L + F_Z^L ) G_\gamma^{R*}\Big)
 \Big] 
\nonumber\\
 \!&&\!
 -\ \frac{8s_w^2}{m_l} \Big[ 
 \mbox{Re}\Big( (F_Z^R + B_{\ell V}^{RR}  + B_{\ell V}^{RL} ) G_\gamma^{L*}\Big)
  + \mbox{Re}\Big( (F_Z^L + B_{\ell V}^{LL} + B_{\ell V}^{LR} )
  G_\gamma^{R*}\Big)\Big] 
\nonumber\\
 \!&&\!
 +\ \frac{32 s_w^4}{m_l^2} \Big(|G_\gamma^L|^2 + |G_\gamma^R|^2\Big)\,
 \bigg(\ln\frac{m^2_l}{m^2_{l'}} - 3\bigg) 
\Bigg\}\; ,\\[3mm]
  \label{Bl3l_2}
B(l\to l'l'l'^C)
 \!&=&\! 
 \frac{m_l^5\alpha_w^4}{24576\pi^3 M_W^4\Gamma_l}\: 
 \Bigg\{ 2 \bigg[\, 
  \Big|2s_w^2 (F_\gamma^L + F_Z^L) - F_Z^L - \frac{1}{2}B_{\ell V}^{LL} \Big|^2
  + \Big|2s_w^2 (F_\gamma^R + F_Z^R) - \frac{1}{2}B_{\ell V}^{RR} \Big|^2\,\bigg] 
\nonumber\\
 \!&&\!
 +\ \Big|2s_w^2 (F_\gamma^L + F_Z^L) - B_{\ell V}^{LR} \Big|^2
  + \Big|2s_w^2 (F_\gamma^R + F_Z^R) - (F_Z^R + B_{\ell V}^{RL} )\Big|^2
  + \frac{1}{8} \Big( |B_{\ell S}^{LL}|^2 + |B_{\ell S}^{RR}|^2 \Big)
\nonumber\\
 \!&&\!
 +\ 6 \Big( |B_{\ell T}^{LL}|^2 +  |B_{\ell T}^{RR}|^2 \Big)
 + \frac{48s_w^4}{m_l} \Big[  
 \mbox{Re} \Big( ( F_\gamma^R + F_Z^R ) G_\gamma^{L*}\Big)
 + \mbox{Re} \Big( ( F_\gamma^L + F_Z^L ) G_\gamma^{R*}\Big)
 \Big] 
\nonumber\\
 \!&&\!
 -\ \frac{8s_w^2}{m_l} \Big[ 
 \mbox{Re} \Big( \big(F_Z^R + B_{\ell V}^{RR} 
     + B_{\ell V}^{RL} \big) G_\gamma^{L*}\Big)
 + \mbox{Re} \Big( \big(2 F_Z^L + B_{\ell V}^{LL} 
     + B_{\ell V}^{LR} \big) G_\gamma^{R*}\Big)\Big]
\nonumber\\
 \!&&\!
 +\ \frac{32 s_w^4}{m_l^2} \Big(|G_\gamma^L|^2 + |G_\gamma^R|^2\Big)\,
 \bigg(\ln\frac{m^2_l}{m^2_{l'}} - \frac{11}{4}\bigg) 
 \Bigg\}\; ,
\end{eqnarray}
where $m_l$ and  $m_{l'}$, $m_{l_1}$, $m_{l_2}$ are the  masses of the
initial- and  final-state charged leptons and $\Gamma_l$  is the decay
width of  the charged lepton $l$.   Here we should  emphasize that the
transition amplitudes  (\ref{Tgl3l}), (\ref{TZl3l}), (\ref{TBl3l}) and
the branching  ratios (\ref{Bl3l_1}) and (\ref{Bl3l_2})  have the most
general chiral  and formfactor structure to leading  order in external
masses and  momenta and so they  are applicable to most  models of New
Physics with CLFV. Finally, we  have checked that the branching ratios
(\ref{Bl3l_1})   and   (\ref{Bl3l_2})   go   over   to   the   results
of~\cite{IPNPB}, in the $\nu_R$SM limit of the theory.

\subsection{Coherent {\boldmath $\mu\to e$}
                       Conversion in a Nucleus}\label{sec:mueff} 

The coherent  $\mu\to e$  conversion in a  nucleus corresponds  to the
process  $J_\mu\to e^-J^+$, where  $J_\mu$ is  an atom  of nucleus~$J$
with  one  orbital  electron  replaced  by a  muon  and~$J^+$  is  the
corresponding ion without the muon. The transition amplitude for such a
CLFV process,
\begin{eqnarray}
\label{TmueJ}
{\cal T}^{\mu e;J} 
 &=& 
   \langle J^+ e^- | {\cal T}^{d\mu \to de} | J_\mu\rangle
 + \langle J^+ e^- | {\cal T}^{u\mu \to ue} | J_\mu\rangle\; ,
\end{eqnarray}  
depends on two effective box operators,
\begin{eqnarray}
  \label{Tdmude}
{\cal T}_{\rm box}^{d\mu \to de} 
 &=&
  - \frac{\alpha_w^2}{4 M_W^2} \sum_{X,Y=L,R}\ \sum_{A=V,S,T}
   B_{dA}^{XY}\: \overline{e}\, \Gamma^X_A \mu\ \bar{d}\, \Gamma^X_A d
\nonumber\\
 &=& 
 - \frac{\alpha_w^2}{2 M_W^2}\, 
 (d^\dagger d)\;\, \bar{e}\, ( V_d^R\, P_R + V_d^L\, P_L )\, \mu\; ,\\
  \label{Tumuue}
{\cal T}_{\rm box}^{u\mu \to ue} 
 &=&
 - \frac{\alpha_w^2}{4 M_W^2} \sum_{X,Y=L,R}\ \sum_{A=V,S,T}
   B_{u A}^{XY}\: \bar{e}\, \Gamma^X_A \mu\ \bar{u}\, \Gamma^X_A u
\nonumber\\
 &=&
 - \frac{\alpha_w^2}{2 M_W^2}\, 
 (u^\dagger u)\;\, \bar{e}\, (V_u^R\, P_R + V_u^L\, P_L)\, \mu\; . 
\end{eqnarray}
Here $\mu$  and $e$ are the  muon and electron wave  functions and $d$
and $u$  are field operators acting  on the $J_\mu$  and $J^+$ states.
The formfactors $B_{dA}^{XY}$ and  $B_{uA}^{XY}$ are given in Appendix
\ref{sec:olff}.   The   composite  formfactors  $V_d^{L}$,  $V_u^{L}$,
$V_d^{R}$, $V_u^{R}$ may conveniently be expressed as
\begin{eqnarray}
\label{V_ff}
V_d^L 
 &=& 
 - \frac{1}{3} s_w^2 \Big(F^L_\gamma - \frac{1}{m_\mu} G^R_\gamma\Big)
 + \Big(\frac{1}{4} - \frac{1}{3} s_w^2\Big) F_Z^L\,
 +\, \frac{1}{4}\Big(B_{d V}^{LL}+B_{d V}^{LR} + B_{d S}^{RR} + B_{d S}^{RL}\Big)\;, 
\nonumber\\
V_d^R 
 &=& 
 -  \frac{1}{3} s_w^2 \Big(F^R_\gamma - \frac{1}{m_\mu} G^L_\gamma\Big)
 + \Big(\frac{1}{4} - \frac{1}{3} s_w^2\Big) F_Z^R\,
 +\, \frac{1}{4}\Big(B_{d V}^{RR}+B_{d V}^{RL} + B_{d S}^{LL} + B_{d S}^{LR}\Big)\;,
\nonumber\\
V_u^L 
 &=& 
 \frac{2}{3} s_w^2 \Big(F^L_\gamma - \frac{1}{m_\mu} G^R_\gamma\Big)
 + \Big(- \frac{1}{4} + \frac{2}{3} s_w^2\Big) F_Z^L
 + \frac{1}{4}\Big(B_{u V}^{LL}+B_{u V}^{LR} + B_{u S}^{RR} + B_{u S}^{RL}\Big)\;,
\nonumber\\
V_u^R 
 &=& 
  \frac{2}{3} s_w^2 \Big(F^R_\gamma - \frac{1}{m_\mu} G^L_\gamma\Big)
 + \Big(- \frac{1}{4} + \frac{2}{3} s_w^2\Big) F_Z^R
 + \frac{1}{4}\Big(B_{u V}^{RR}+B_{u V}^{RL} + B_{u S}^{LL} + B_{u S}^{LR}\Big)\;,
\end{eqnarray}
where  $F_\gamma^L$, $F_\gamma^R$,  $F_Z^L$, $F_Z^R$  is  the shorthand
notation     for     $(F_\gamma^L)_{e\mu}$,     $(F_\gamma^R)_{e\mu}$,
$(F_Z^L)_{e\mu}$,  $(F_Z^R)_{e\mu}$.  

Our  next step  is to  determine the  nucleon matrix  elements  of the
operators $u^\dagger u$ and $d^\dagger d$. These are given by
\begin{eqnarray}
\langle J^+e^- |u^\dagger u| J_\mu \rangle &=& (2Z+N) F(-m_\mu^2)\; ,
\nonumber\\
\langle J^+e^- |d^\dagger d| J_\mu \rangle &=& (Z+2N) F(-m_\mu^2)\; ,
\end{eqnarray}
where  the formfactor $F(q^2)$  incorporates the  recoil of  the $J^+$
ion~\cite{COKFV}, and  the factors $2Z+N$ and $Z+2N$  count the number
of $u$  and $d$  quarks in the  nucleus $J$, respectively.  Hence, the
matrix element for $J_\mu\to J^+\mu^-$ can be written down as
\begin{eqnarray}
T^{J_\mu\to J^+ e^-} 
 &=&
 -\frac{\alpha_w^2}{2 M_W^2}\, F(-m_\mu^2)\: \bar{e} 
 \,(Q_W^L\, P_R + Q_W^R\, P_L )\, 
 \mu\; ,
\label{TJmue}
\end{eqnarray}
with
\begin{eqnarray}
\label{QWLR}
Q_W^L &=& (2Z+N) V_u^L +(Z+2N)V_d^L\; ,
\nonumber\\
Q_W^R &=& (2Z+N) V_u^R +(Z+2N)V_d^R\; .
\end{eqnarray}

Given the transition amplitude~(\ref{TJmue}), the decay rate $J_\mu\to
J^+e^-$ is found to be
\begin{eqnarray}
\label{RmueJ}
R^J_{\mu    e}     &=&    \frac{\alpha^3\alpha_w^4    m_\mu^5}{16\pi^2
  M_W^4\Gamma_{{\rm  capture}}} \frac{Z^4_{\rm  eff}}{Z} |F(-m_\mu^2)|^2
\: \Big(\,|Q_W^L|^2+|Q_W^R|^2\,\Big)\; ,
\end{eqnarray}
where $\Gamma_{{\rm capture}}$ is the  capture rate of the muon by the
nucleus, and  $Z_{\rm eff}$ is  the effective charge which  takes into
account coherent effects that can occur  in the nucleus $J$ due to its
finite  size. In  our analysis,  we use  the values  of  $Z_{\rm eff}$
quoted    in~\cite{KKO}.    We    reiterate    that   the    branching
ratio~(\ref{RmueJ}) possesses the most general formfactor structure to
leading order in  external masses and momenta and  is relevant to most
models of New  Physics with CLFV.  Finally, we  have verified that our
analytical              results             are             consistent
with~\cite{IPPRD,Ilakovac:1995km,Alonso:2012ji} in the $\nu_R$SM limit
of the theory.

\setcounter{equation}{0}
\section{Numerical Results}\label{numerics}

In this section,  we present a numerical analysis  of CLFV observables
in  the $\nu_R$MSSM.   In order  to reduce  the number  of independent
parameters, we adopt the  constrained framework of mSUGRA.  In detail,
our model parameters  are: (i)~the usual SM parameters,  such as gauge
coupling  constants,  the  quark  and charged-lepton  Yukawa  matrices
inputted at  the scale $M_Z$,  (ii)~the heavy neutrino mass  $m_N$ and
the neutrino Yukawa matrix ${\bf h}_\nu$ evaluated at $m_N$, (iii)~the
universal mSUGRA parameters $m_0$,  $M_{1/2}$ and $A_0$ inputted at the
GUT scale,  and (iv)~the ratio $\tan\beta$  of the Higgs  VEVs and the
sign of the superpotential Higgs-mixing parameter $\mu$.

The  allowed  ranges  of  the  soft  SUSY-breaking  parameters  $m_0$,
$M_{1/2}$, $A_0$ and $\tan\beta$  are strongly constrained by a number
of              accelerator              and              cosmological
data~\cite{m-gtqt,HdisATL,HdisCMS,ElOl12}.    For   definiteness,   we
consider the following set of input parameters:
\begin{equation}
  \label{mSUGRA}
\tan\beta = 10\,,\qquad  m_0=1000~{\rm GeV}\,,\qquad 
A_0 = -3000~{\rm GeV}\,,\qquad M_{1/2} = 1000~{\rm GeV}\; .
\end{equation}
Here we take  the $\mu$ parameter to be  positive, whilst its absolute
value $|\mu |$ is derived form the minimization of the Higgs potential
at         the          scale         $M_Z$.          With         aid
of~\cite{CHHHWW12,CPsuperH,Heinemeyer:2010eg},  we   verify  that  the
parameter  set~(\ref{mSUGRA})  predicts  a  SM-like Higgs  boson  with
$m_H\approx 125$~GeV,  in agreement with  the recent discovery  at the
LHC~\cite{HdisATL,HdisCMS}, and  is compatible with  the current lower
limits   on    gluino   and   squark    masses   \cite{m-gtqt}.    The
set~(\ref{mSUGRA})  is  also in  agreement  with  having the  lightest
neutralino as the Dark Matter in the Universe~\cite{ElOl12}.

We employ  the one-loop  RG equations of~\cite{Cha02,Pet04}  to evolve
the  the gauge  coupling constants  and the  quark and  charged lepton
Yukawa matrices from $M_Z$ to  the GUT scale, while the heavy neutrino
mass matrix ${\bf  m}_M$ and the neutrino Yukawa  matrix ${\bf h}_\nu$
are evolved  from the heavy neutrino  mass threshold $m_N$  to the GUT
scale.   Furthermore,  we  assume  that the  heavy  neutrino-sneutrino
sector is  supersymmetric above $m_N$.  For purposes  of RG evolution,
this is  a good approximation for  $m_N$ larger than  the typical soft
SUSY-breaking  scale~\cite{IPPRD}.   At  the  GUT  scale,  the  mSUGRA
universality  conditions are  used to  express the  soft SUSY-breaking
masses,  in terms  of $m_0$,  $M_{1/2}$ and  $A_0$. Hence,  all scalar
masses  receive  a soft  SUSY-breaking  mass  $m_0$,  all gaugino  are
mass-degenerate to  $M_{1/2}$, and all scalar  trilinear couplings are
of the form ${\bf h}_x A_0$, with $x=u,d,l,\nu$, where ${\bf h}_x$ are
the  Yukawa matrices  at the  GUT  scale.  The  sneutrino mass  matrix
acquires additional contributions from the heavy neutrino mass matrix.
The sparticle  mass matrices and trilinear couplings  are evolved from
the  GUT scale to  $M_Z$, except  for the  sneutrino masses  which are
evolved to  the heavy neutrino  threshold $m_N$. Having  thus obtained
all sparticle and sneutrino mass matrices, we can numerically evaluate
all  particle  masses and  interaction  vertices  in the  $\nu_R$MSSM,
without approximations.

To  simplify our  numerical analysis,  we consider  two representative
scenarios  of  Yukawa  textures discussed  in  Section~\ref{lowscale}.
Specifically,  the first scenario  realizes the  U(1)-symmetric Yukawa
texture in~(\ref{YU1}), for  which we take either $a=b$  and $c=0$, or
$a=c$  and  $b=0$,  or $b=c$  and  $a=0$,  thus  giving rise  to  CLFV
processes $\mu\to e X$, $\tau\to eX$ and $\tau\to\mu X$, respectively.
Here   $X$   represents   the  lepton-flavour   conserving   state(s),
e.g.~$X=\gamma,\, e^+e^-,\,  \mu^+\mu^-,\, q\overline{q}$.  The~second
scenario  is  motivated  by  the  $A_4$  group  and  uses  the  Yukawa
texture~(\ref{YA4}), where  the parameters $a$, $b$  and~$c$ are taken
to be all equal, i.e.~$a=b=c$.

The heavy  neutrino mass scale $m_N$  strongly depends on  the size of
the symmetry-breaking  terms in the  Yukawa matrix ${\bf  h}_\nu$. For
instance,  for  the  model  (\ref{YU1}),  the typical  values  of  the
$U(1)$-lepton-symmetry-breaking    parameters    $\epsilon_l    \equiv
\epsilon_{e,\mu,\tau}$ consistent with low-scale resonant leptogenesis
is $\epsilon \stackrel{<}{{}_\sim} 10^{-5}$ \cite{APRLtau}, leading to
light-neutrino masses
\begin{eqnarray}
  \label{mNmax}
m_\nu\ \sim\ \frac{\epsilon^2_l v^2}{m_N} &\sim& 10^{-2}\, \textrm{eV} 
 \Big(\frac{\epsilon_l}{10^{-6}}\Big)^2\
                            \Big(\frac{1\,\textrm{TeV}}{m_N}\Big)\ .  
\end{eqnarray}
Taking  into  account   the  constraint  $m_\nu  \stackrel{>}{{}_\sim}
10^{-1}$~eV generically derived from neutrino oscillation data, we may
estimate  that  the  heavy  neutrino  mass scale  $m_N$  is  typically
restricted to be  less than $10$~TeV, for $\epsilon_l  = 10^{-5}$.  If
the assumption  of successful  low-scale leptogenesis is  relaxed, the
symmetry-breaking  parameters $\epsilon_l$  has only  to be  couple of
orders in  magnitude smaller than  the Yukawa parameters $a$,  $b$ and
$c$, with $a,\,b,\,c \stackrel{<}{{}_\sim} 10$.  Thus, for $\epsilon_l
<10^{-3}-10^{-2}$, the heavy neutrino mass scale $m_N$ may be as large
as $10^7-10^9$~TeV, leading to  the decoupling of heavy neutrinos from
low-energy observables.   As our interest is in  the interplay between
heavy neutrino, sneutrino and soft SUSY-breaking contributions to CLFV
observables, we will only study here the parameter space in which $m_N
< 10$~TeV.

In  the present  analysis,  we consider  that the  symmetry-preserving
Yukawa  parameters   $a$,  $b$  and   $c$  are  limited   through  the
perturbativity condition: $\mbox{Tr}\,  {\bf h}^\dagger_\nu {\bf h}_\nu
< 4\pi$, which we require to  hold true for the entire interval of the
RG     evolution:    $\ln(M_Z/\textrm{TeV})<     t     <    \ln(M_{\rm
  GUT}/\textrm{TeV})$.  For  the model in~(\ref{YU1}),  this condition
translates  into   the  constraint:   $a<0.34$,  and  for   the  model
in~(\ref{YA4}), to $a<0.23$.   As a consequence, we do  not display in
plots numerical  values for points  in parameter space, for  which the
aforementioned perturbativity condition gets violated.

%%%%%%%%%%%%%%%%%%%%%%%%%%%%%%%%%%%%%%%%%%%%%%%%%%%%%%%%%%%%%%%%%%%%%%%
\begin{figure}[!ht]
 \centering
 \includegraphics[clip,width=0.30\textwidth]{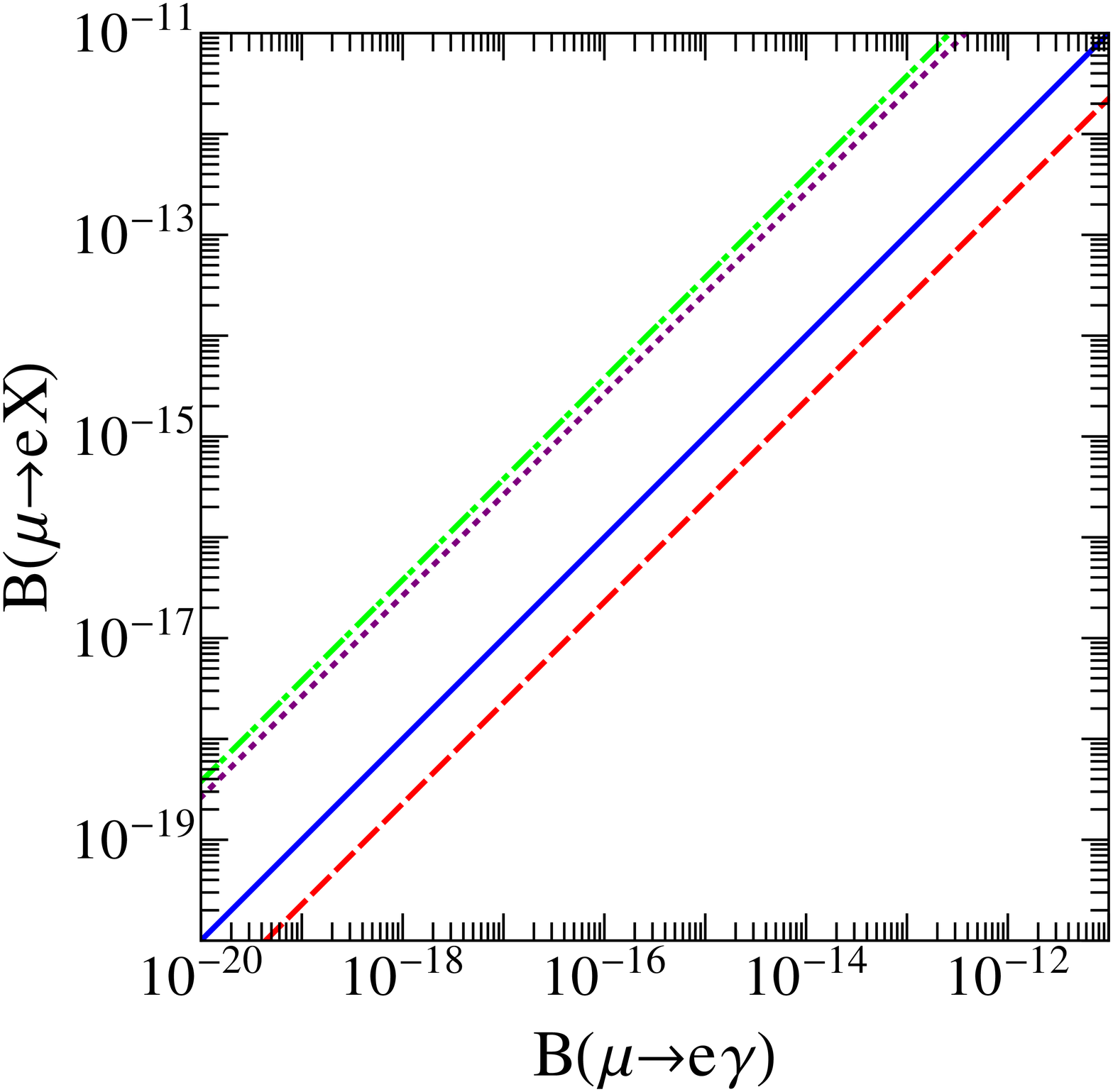}\hspace{1cm}
 \includegraphics[clip,width=0.285\textwidth]{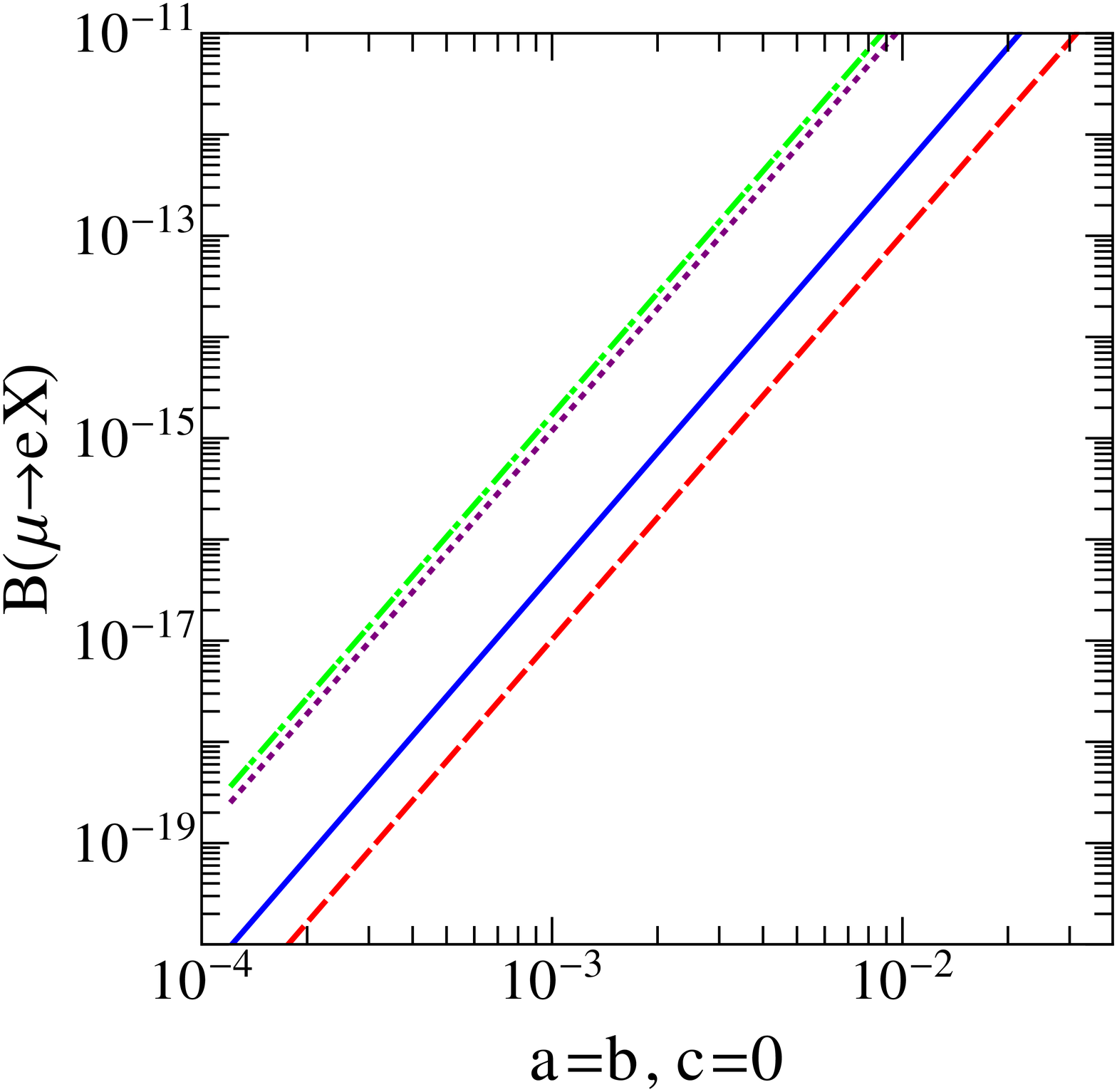}
 \\[.02\textwidth]
 \includegraphics[clip,width=0.30\textwidth]{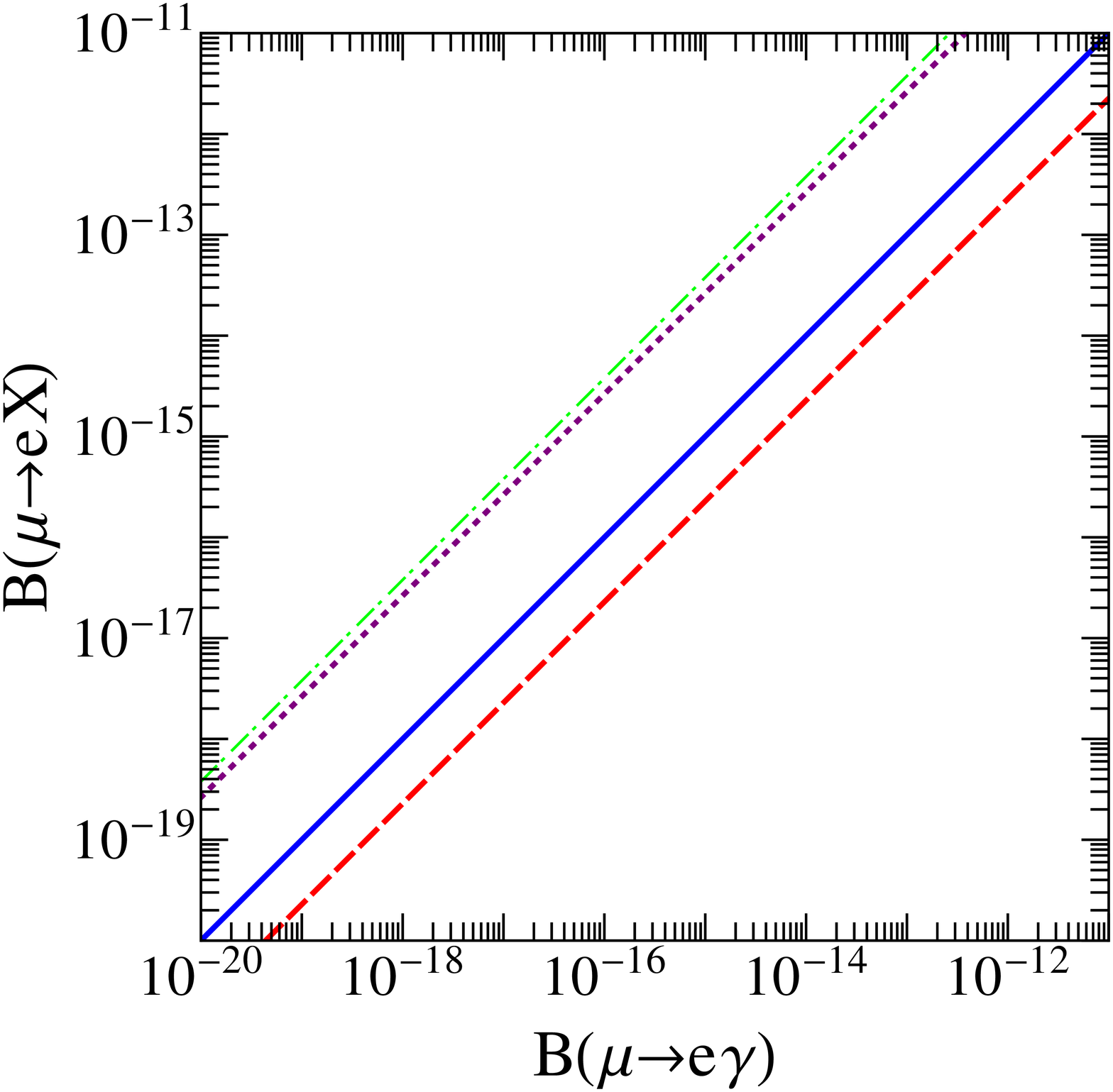}\hspace{1cm}
 \includegraphics[clip,width=0.285\textwidth]{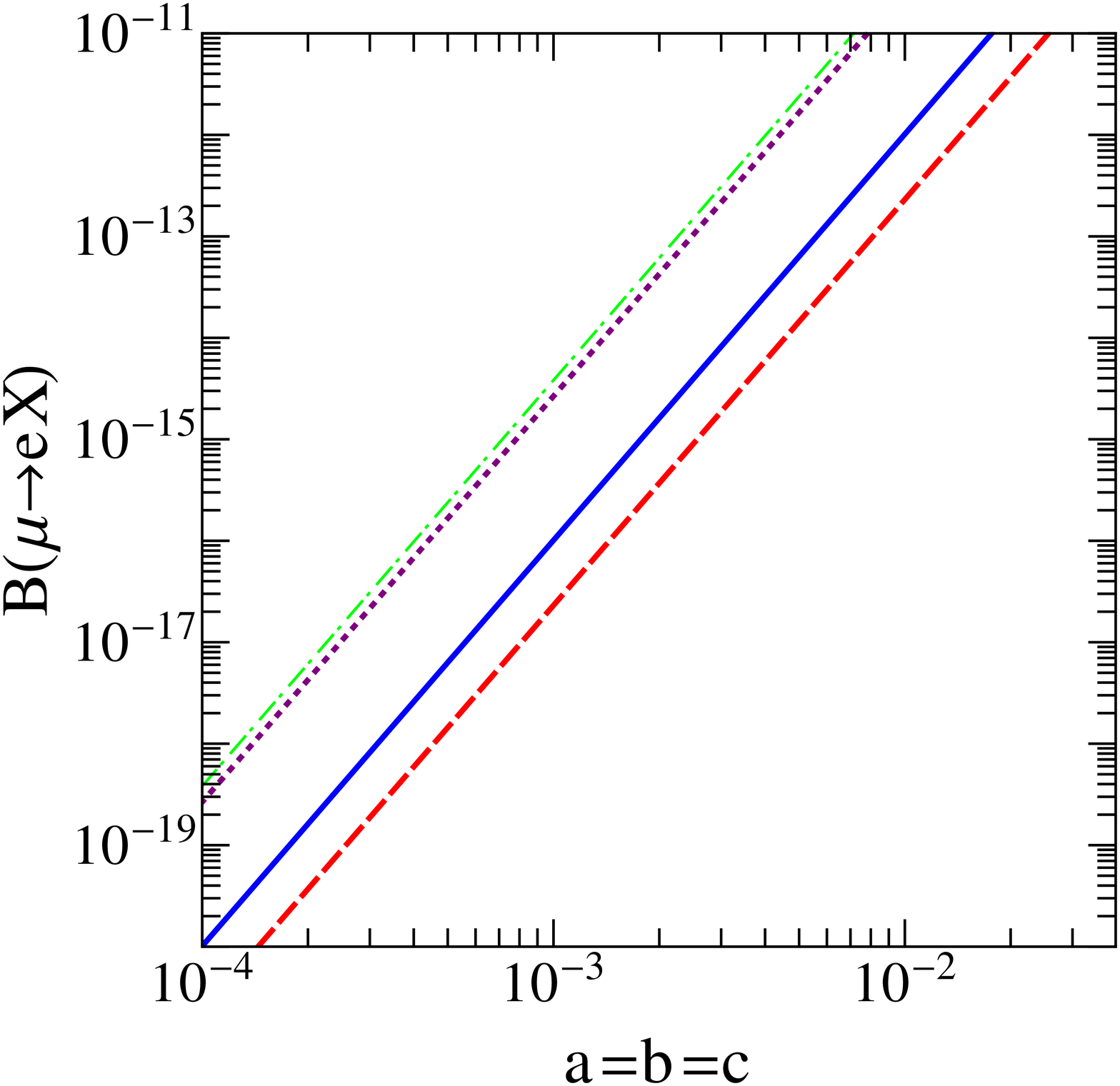}
\caption{Numerical  estimates of  $B(\mu\to e\gamma)$  [blue (solid)],
  $B(\mu\to  eee)$   [red  (dashed)],  $R^{\rm   Ti}_{\mu  e}$~[violet
    (dotted)]  and  $R^{\rm  Au}_{\mu  e}$~[green  (dash-dotted)],  as
  functions  of  $B(\mu\to e\gamma)$  (left  pannels)  and the  Yukawa
  parameter  $a$  (right pannels),  for  $m_N=400$~GeV and  $\tan\beta
  =10$.   The  upper two  pannels  correspond  to  the Yukawa  texture
  (\ref{YU1}), with $a=b$ and $c=0$,  and the lower two pannels to the
  Yukawa texture (\ref{YA4}), with $a=b=c$.}
\label{Fig3}
\end{figure}

\begin{figure}[!ht]
 \centering
 \includegraphics[clip,width=0.30\textwidth]{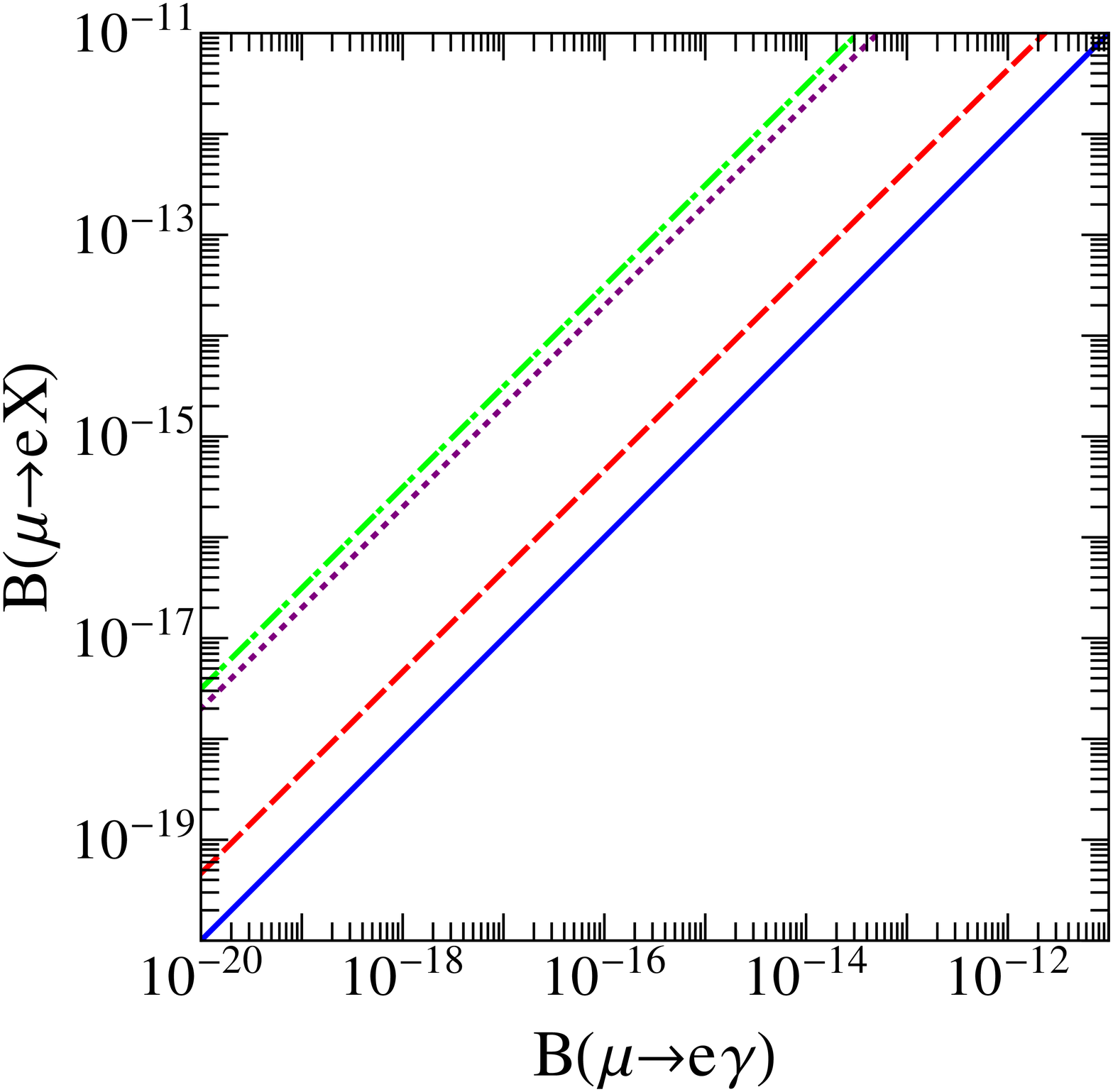}\hspace{1cm}
 \includegraphics[clip,width=0.285\textwidth]{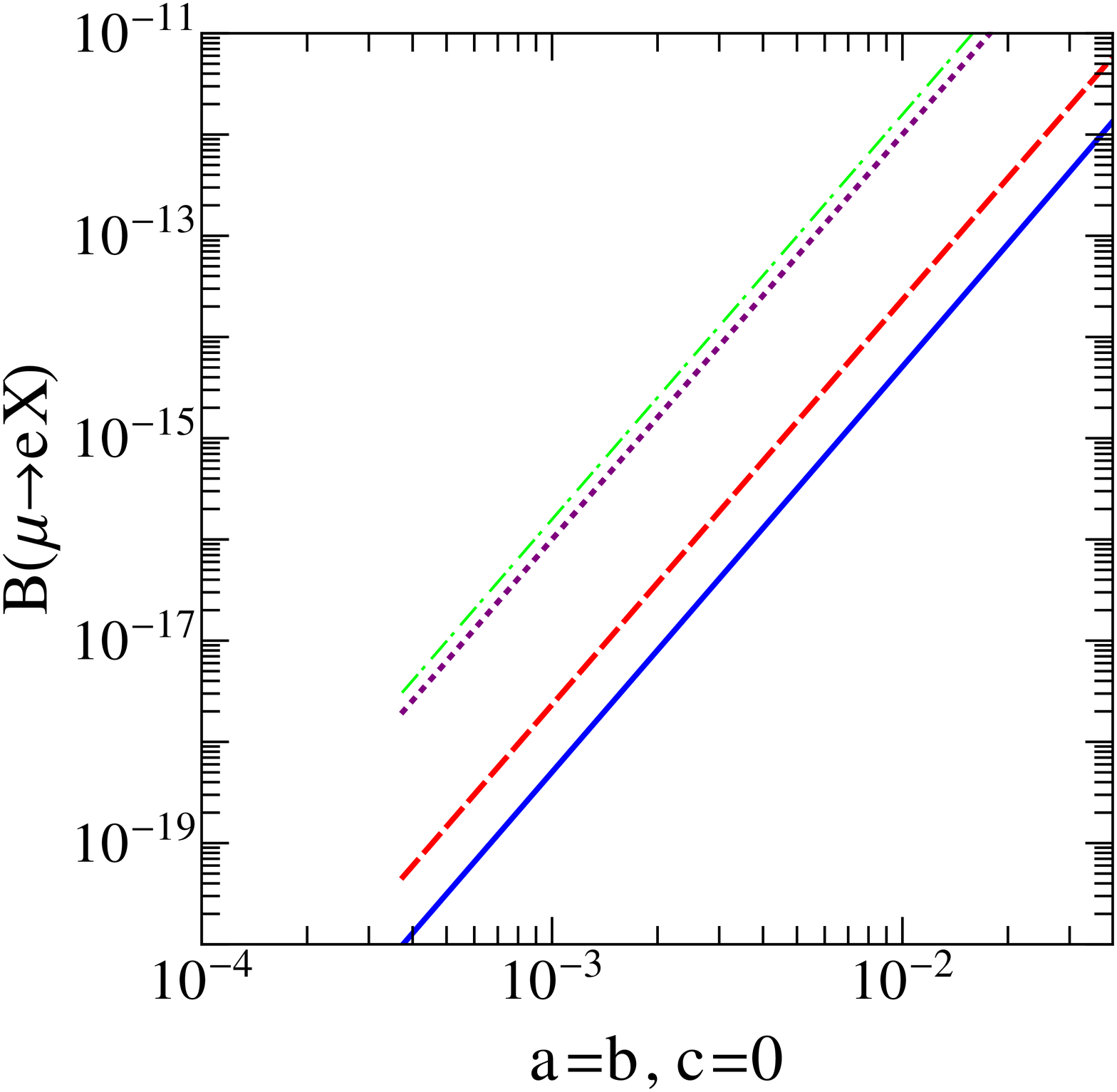}
 \\[.02\textwidth]
 \includegraphics[clip,width=0.30\textwidth]{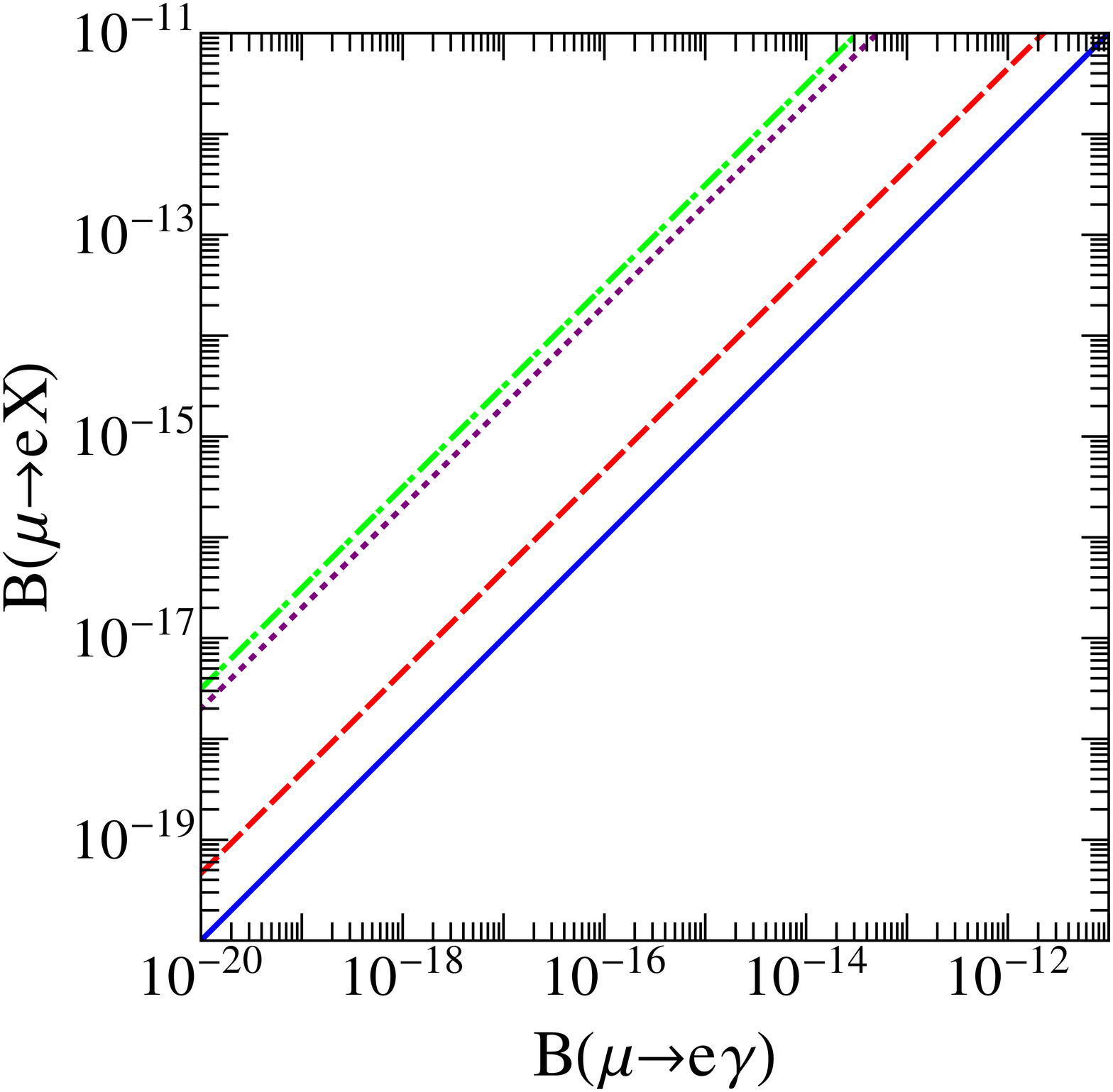}\hspace{1cm}
 \includegraphics[clip,width=0.285\textwidth]{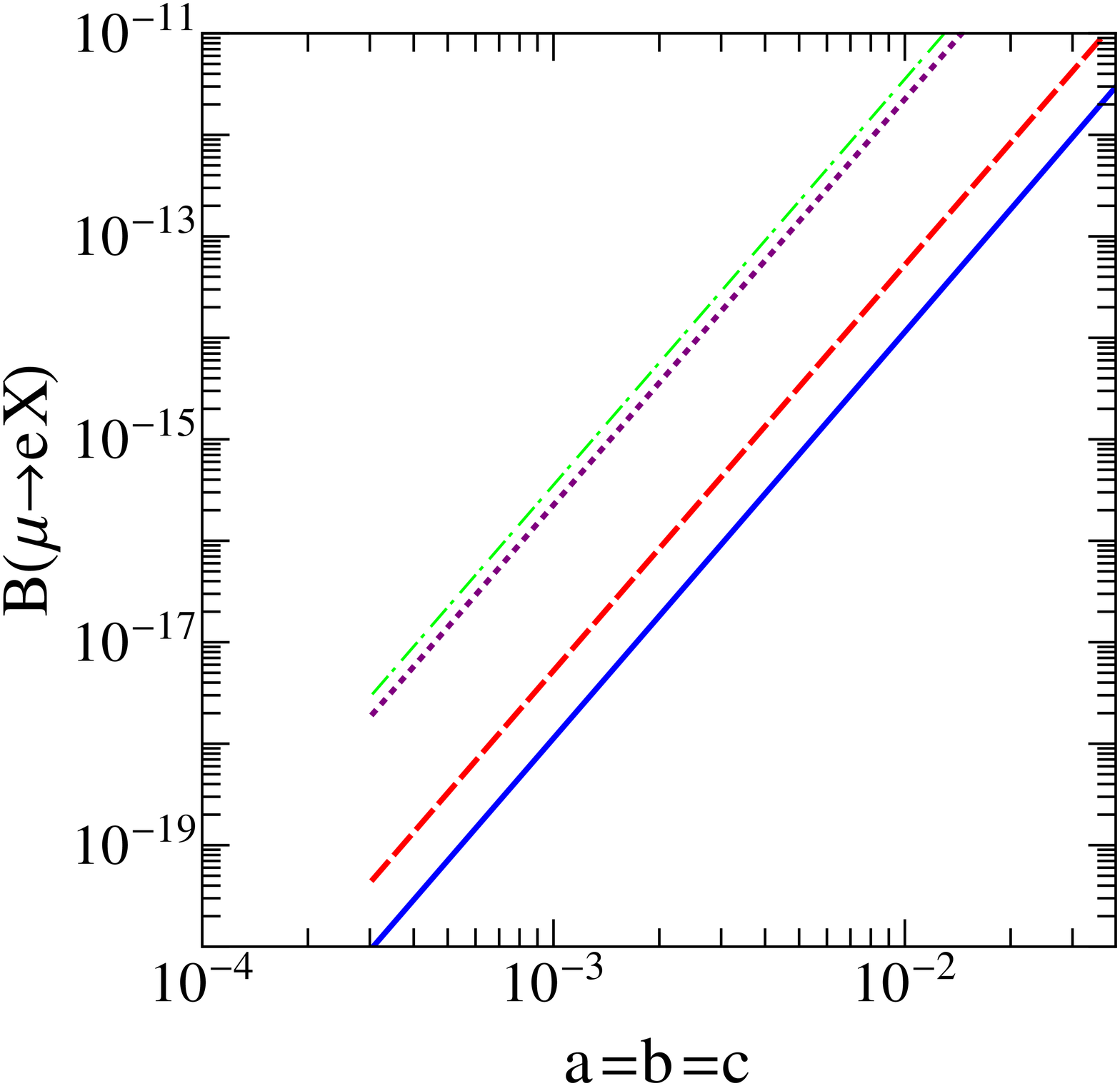}
\caption{The same as in Fig.~\ref{Fig3}, but for $m_N=1$~TeV.}
\label{Fig4}
\end{figure}
%%%%%%%%%%%%%%%%%%%%%%%%%%%%%%%%%%%%%%%%%%%%%%%%%%%%%%%%%%%%%%%%%%%%%%%

In Fig.~\ref{Fig3}, we display numerical predictions for the $\mu$-LFV
observables $B(\mu \to e X)$: $B(\mu\to e\gamma)$ [blue (solid) line],
$B(\mu\to  eee)$  [red (dashed)  line],  $R^{\rm Ti}_{\mu  e}$~[violet
  (dotted) line] and  $R^{\rm Au}_{\mu e}$~[green (dash-dotted) line],
as  functions of  $B(\mu\to e\gamma)$  (left pannels)  and  the Yukawa
parameter $a$ (right pannels),  for $m_N=400$~GeV and $\tan\beta =10$.
The upper  two pannels assume the Yukawa  texture in~(\ref{YU1}), with
$a=b$ and $c=0$, whilst the lower two pannels correspond to the Yukawa
texture  in~(\ref{YA4}), with  $a=b=c$.  In  Fig.~\ref{Fig4},  we give
numerical estimates for the same set of $\mu$-LFV observables, but for
$m_N  =  1$~TeV.    In  Figs.~\ref{Fig3}  and~\ref{Fig4},  the  Yukawa
parameter   $a$  has   been  chosen,   such   that  $10^{-20}<B(\mu\to
e\gamma)<10^{-10}$. Such  a range of values includes  both the present
\cite{MEG11,SINDRUM88,Titanium,Gold,tau-8a,tau-8b}      and     future
\cite{MEG,PRISM,Mu2e,mueAl-16,mueTi-18a,mueTi-18b,Bon07}   experimental
limits.   As we  see from  Figs.~\ref{Fig3} and  \ref{Fig4},  the CLFV
observables under  study depend quadratically on  the Yukawa parameter
$a$,  namely they  are proportional  to $a^2$.   Instead,  the quartic
Yukawa terms  proportional to $a^4$~\cite{IPPRD}  remain always small,
which  is  a consequence  of  the  imposed perturbativity  constraint:
$\textrm{Tr}({\bf  h}_\nu^\dagger{\bf  h}_\nu)<4\pi$,  up to  the  GUT
scale.

%%%%%%%%%%%%%%%%%%%%%%%%%%%%%%%%%%%%%%%%%%%%%%%%%%%%%%%%%%%%%%%%%%%%%%
\begin{figure}[!ht]
 \centering
 \includegraphics[clip,width=0.29\textwidth]{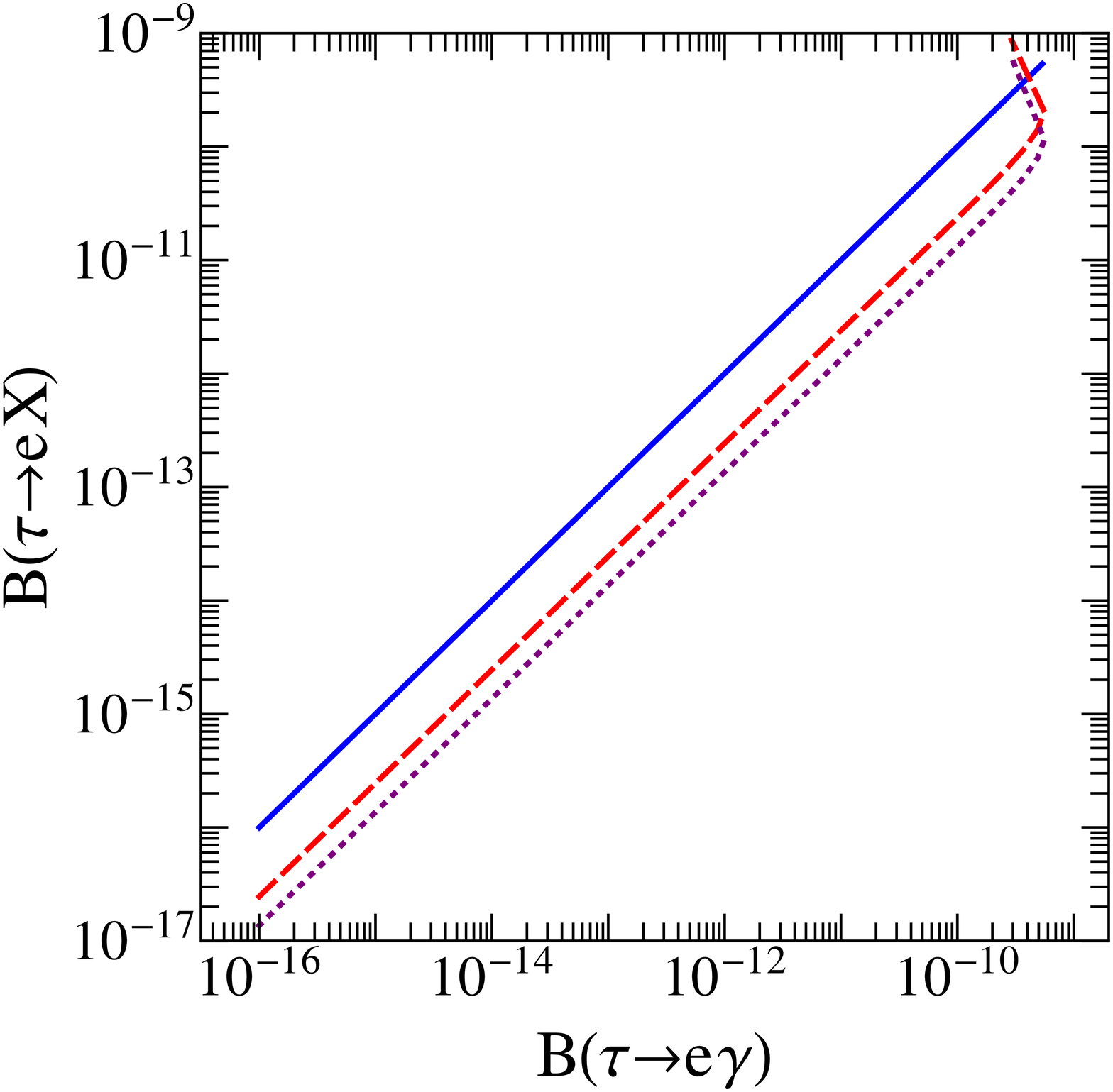}\hspace{1cm}
 \includegraphics[clip,width=0.29\textwidth]{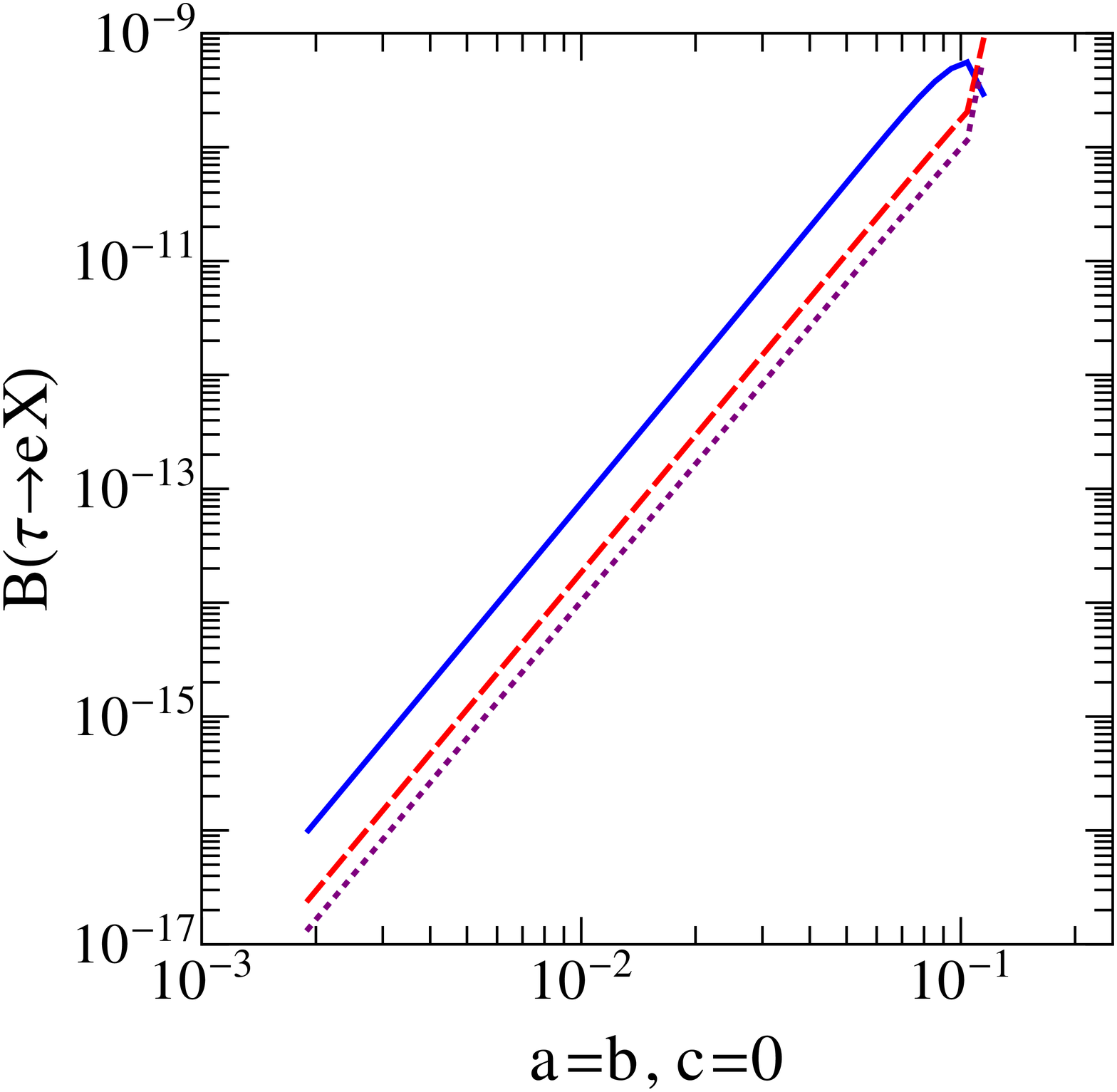}
 \\[.02\textwidth]
 \includegraphics[clip,width=0.29\textwidth]{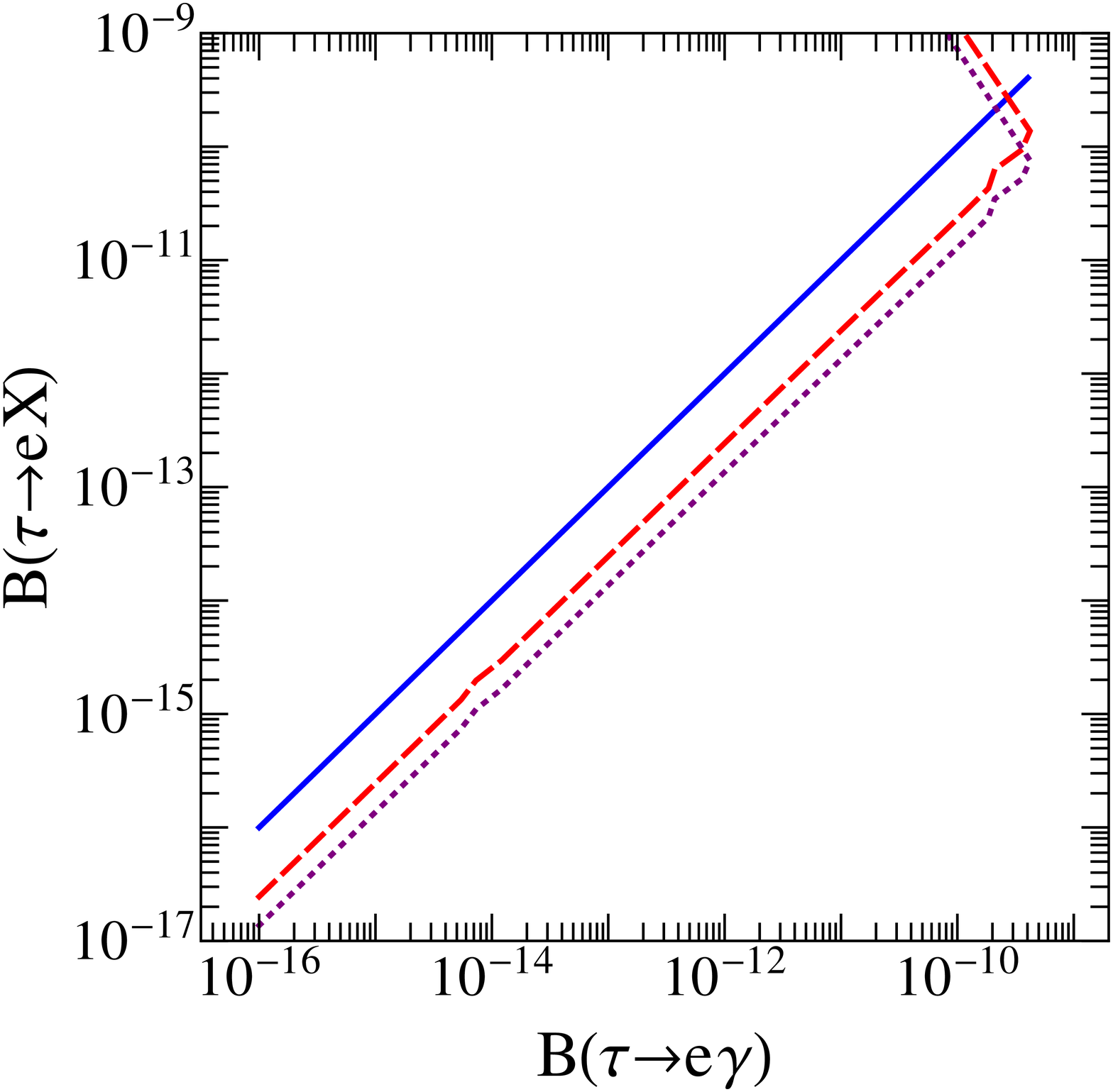}\hspace{1cm}
 \includegraphics[clip,width=0.29\textwidth]{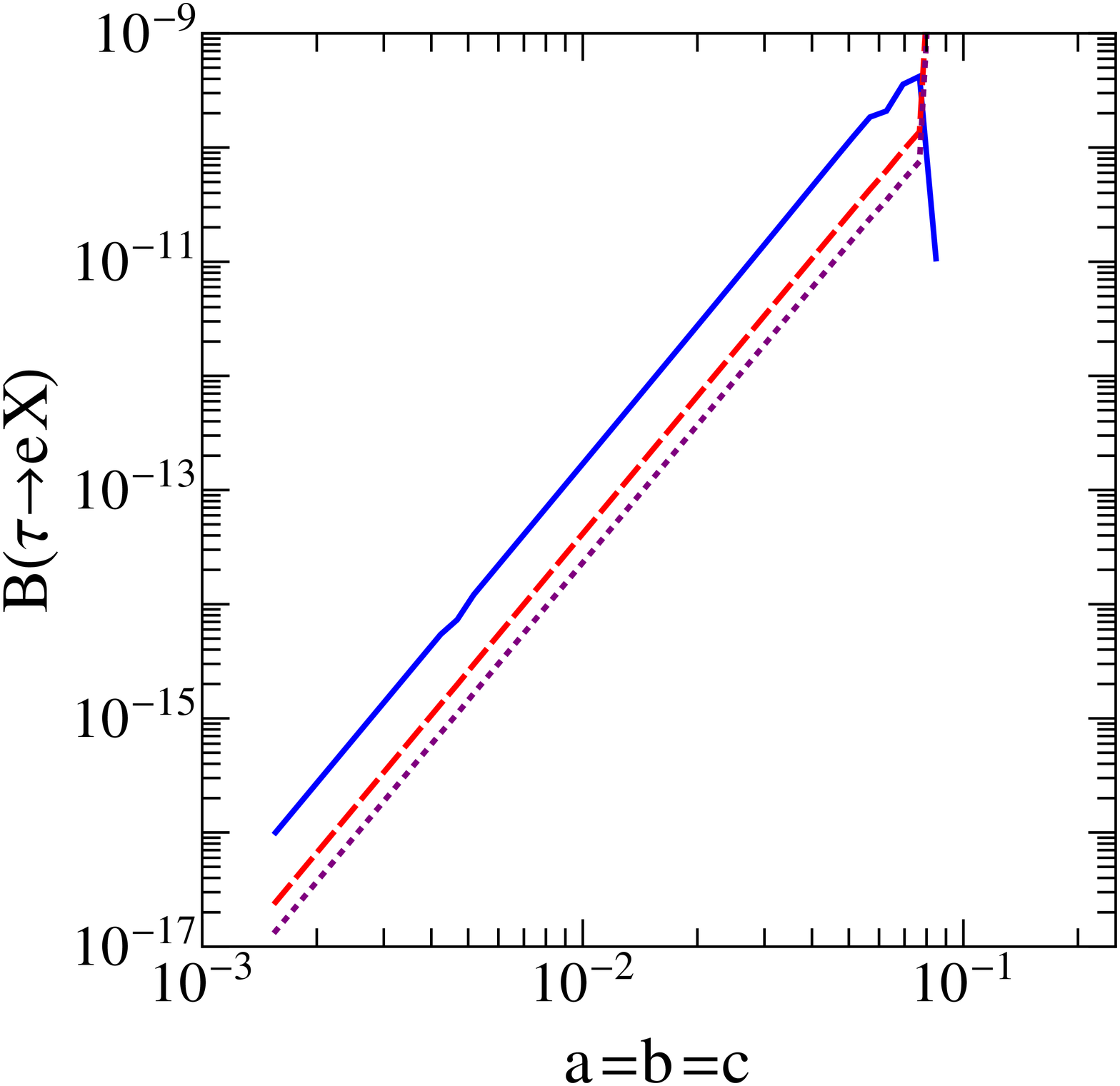}
\caption{Numerical estimates of $B(\tau\to e\gamma)$ [blue (solid)],
  $B(\tau\to eee)$ [red (dashed)] and $B(\tau\to e\mu\mu)$ [violet
    (dotted)], as functions of $B(\tau\to e\gamma)$ (left pannels) and
  the Yukawa parameter $a$ (right pannels), for $m_N=400$~GeV and
  $\tan\beta=10$.  The upper pannels present predictions for the Yukawa
  texture (\ref{YU1}), with $a=c$ and $b=0$, and the lower pannels for
  the Yukawa texture (\ref{YA4}), with $a=b=c$.}
\label{Fig5}
\end{figure}

\begin{figure}[!ht]
 \centering
 \includegraphics[clip,width=0.29\textwidth]{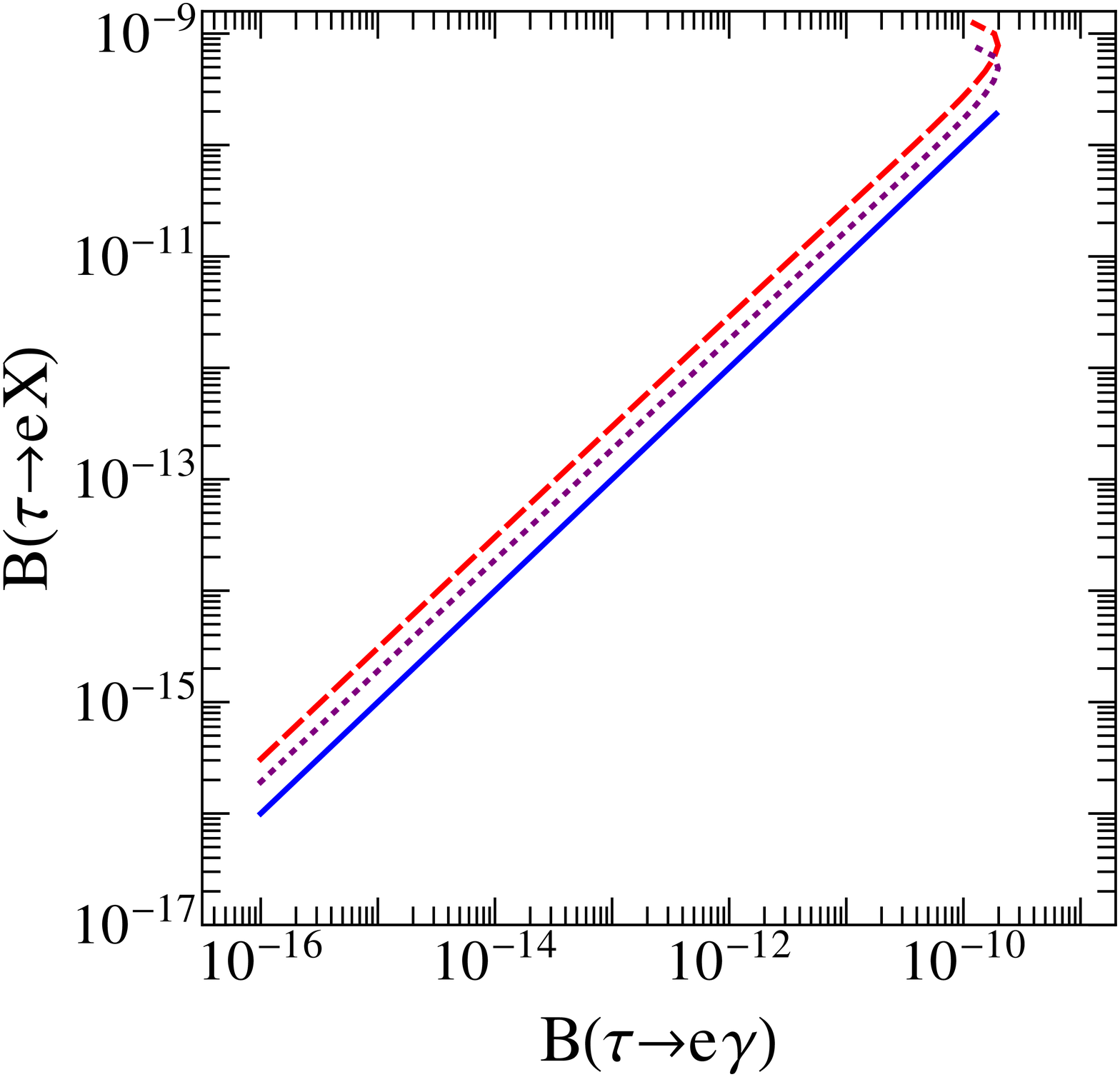}\hspace{1cm}
 \includegraphics[clip,width=0.29\textwidth]{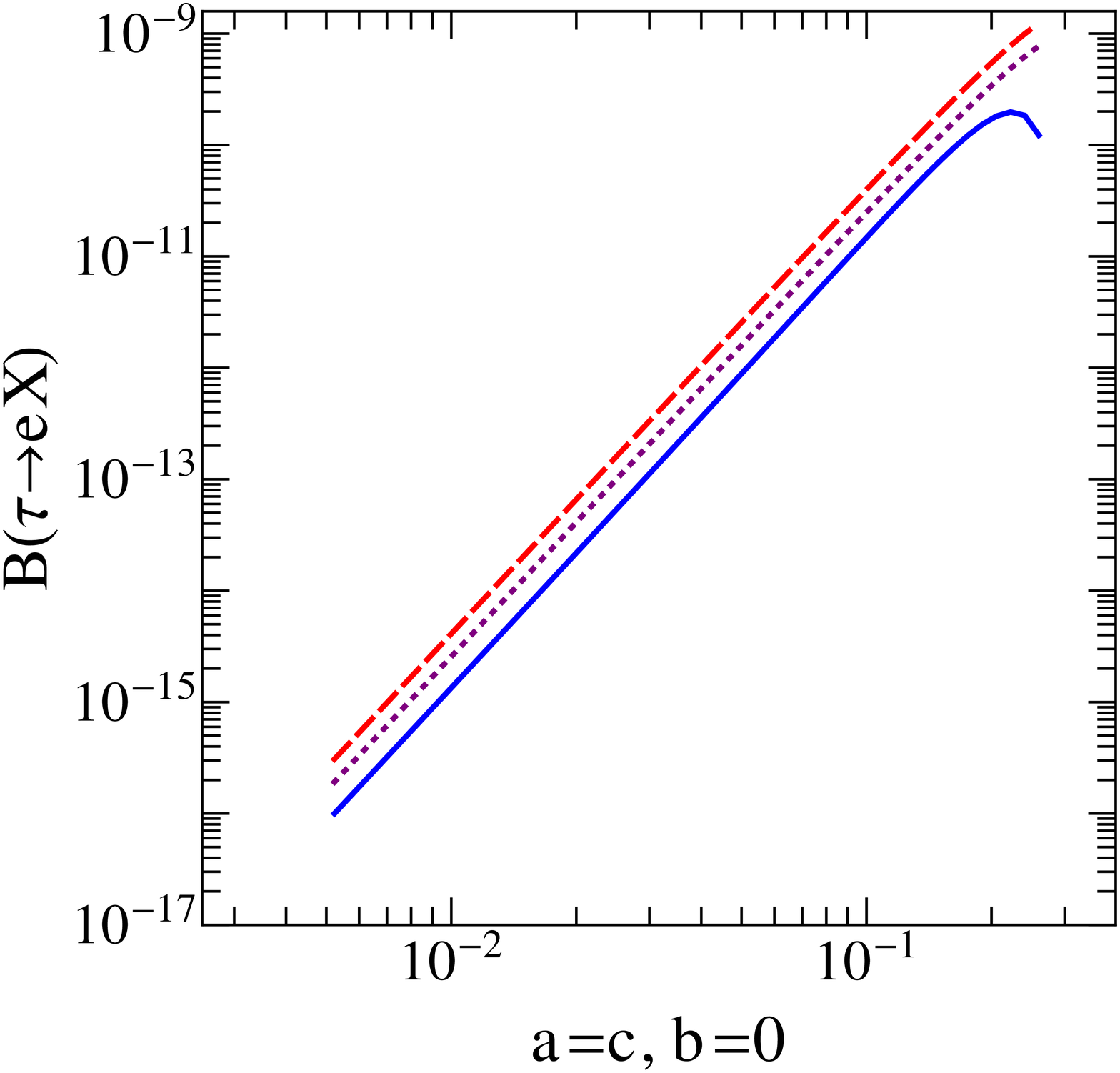}
 \\[.02\textwidth]
 \includegraphics[clip,width=0.29\textwidth]{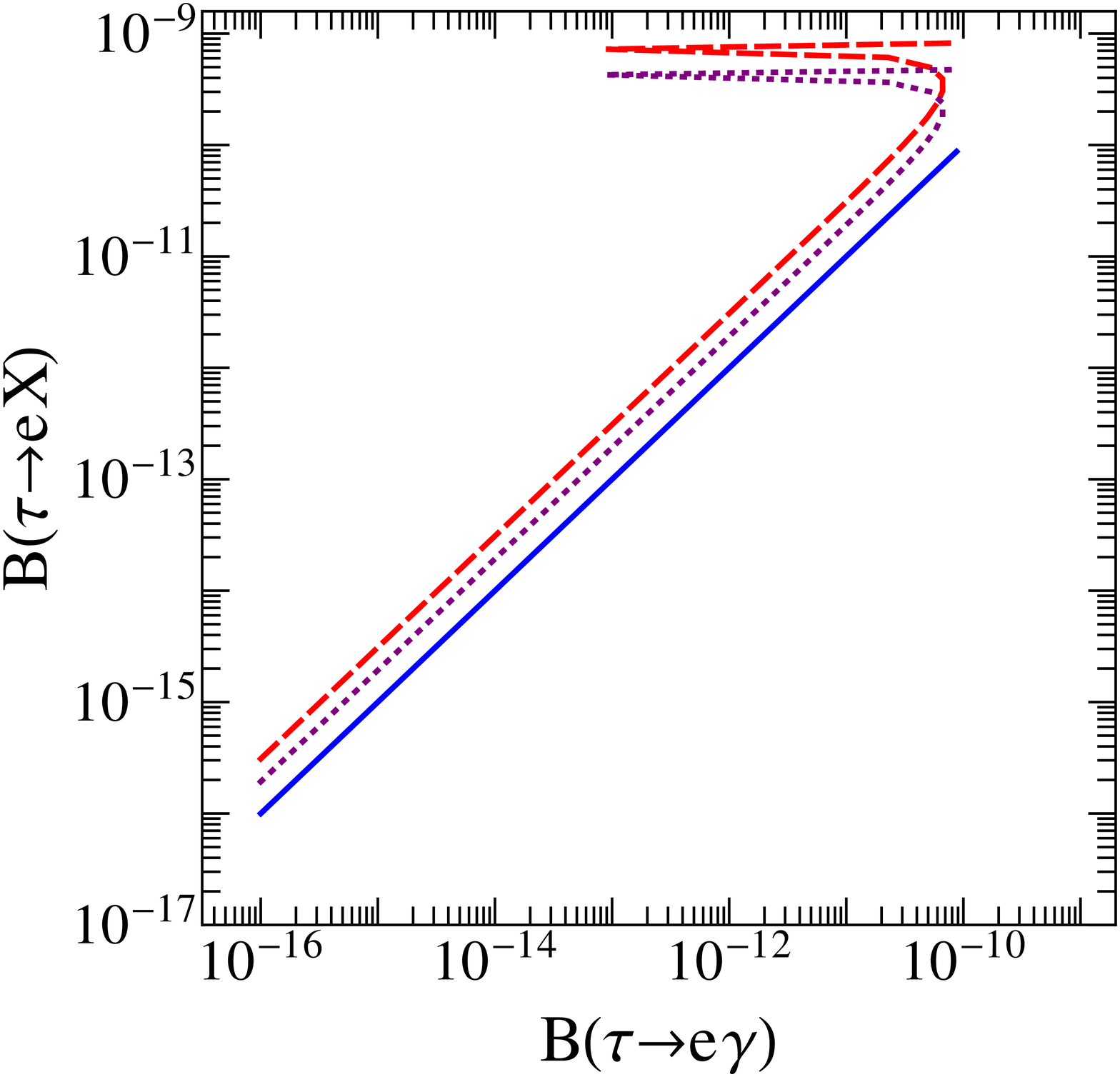}\hspace{1cm}
 \includegraphics[clip,width=0.29\textwidth]{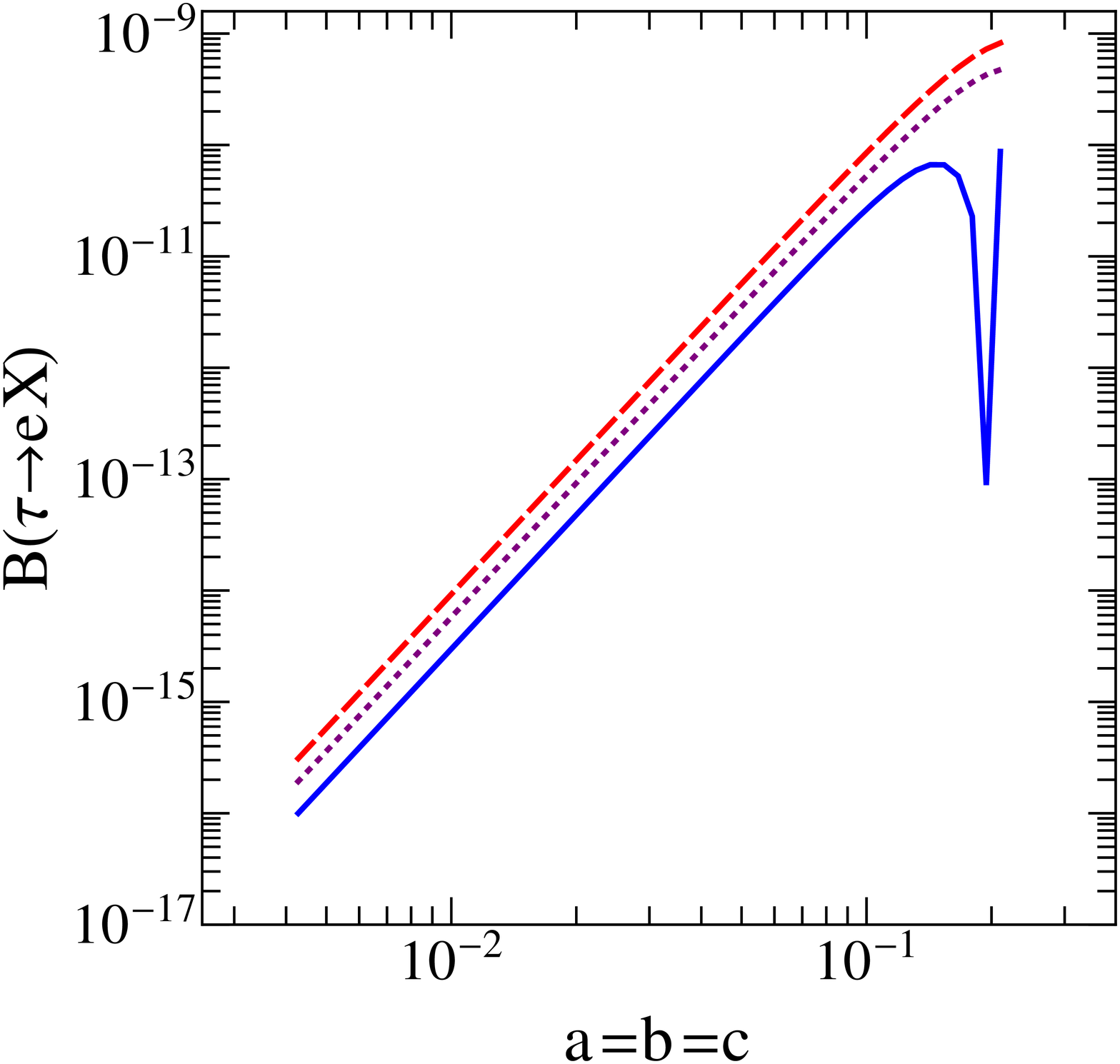}
\caption{The same as in Fig.~\ref{Fig5}, but for $m_N=1$~TeV.}
\label{Fig6}
\end{figure}
%%%%%%%%%%%%%%%%%%%%%%%%%%%%%%%%%%%%%%%%%%%%%%%%%%%%%%%%%%%%%%%%%%%%%%

By   analogy,   Figs.~\ref{Fig5}   and~\ref{Fig6}  present   numerical
estimates of the $\tau$-LFV  observables $B(\tau \to e X)$: $B(\tau\to
e\gamma)$ [blue (solid) lines],  $B(\tau\to eee)$ [red (dashed) lines]
and  $B(\tau\to e\mu\mu)$  [violet  (dotted) lines],  as functions  of
$B(\tau\to  e\gamma)$  (left pannels)  and  the  Yukawa parameter  $a$
(right pannels),  for $m_N=400$~GeV  and $m_N =  1$~TeV, respectively.
Note  that we  do not  show  predictions for  the fully  complementary
observables  $B(\tau \to \mu  X)$: $B(\tau\to  \mu\gamma)$, $B(\tau\to
\mu\mu\mu)$  and $B(\tau\to  \mu  ee)$.  The  upper  pannels give  our
predictions for the Yukawa  texture (\ref{YU1}), with $a=c$ and $b=0$,
and  the  lower  pannels  for  the Yukawa  texture  (\ref{YA4}),  with
$a=b=c$.    In  both   Figs.~\ref{Fig5}  and~\ref{Fig6},   the  Yukawa
parameter   $a$  has  been   chosen,  such   that  $10^{-16}<B(\tau\to
e\gamma)<10^{-7}$.    As  can  be   seen  from   Figs.~\ref{Fig5}  and
\ref{Fig6}, all observables $B(\tau \to e X)$ of $\tau$-LFV (with $X =
\gamma,\, ee,\,  \mu\mu$) exhibit similar quadratic  dependence on the
small   Yukawa   parameter~$a$.   However,   close   to  the   largest
perturbatively  allowed  values  of $a$,  i.e.~$a\stackrel{<}{{}_\sim}
0.34$ for the  model~(\ref{YU1}) and $a\stackrel{<}{{}_\sim} 0.23$ for
the model~(\ref{YA4}),  some of the observables  of $\tau$-LFV exhibit
either  numerical  instability, or  the  existence  of a  cancellation
region in parameter space, as we will see below.

%%%%%%%%%%%%%%%%%%%%%%%%%%%%%%%%%%%%%%%%%%%%%%%%%%%%%%%%%%%%%%%%%%%%%%
\begin{figure}[!ht]
 \centering
 \includegraphics[clip,width=0.355\textwidth]{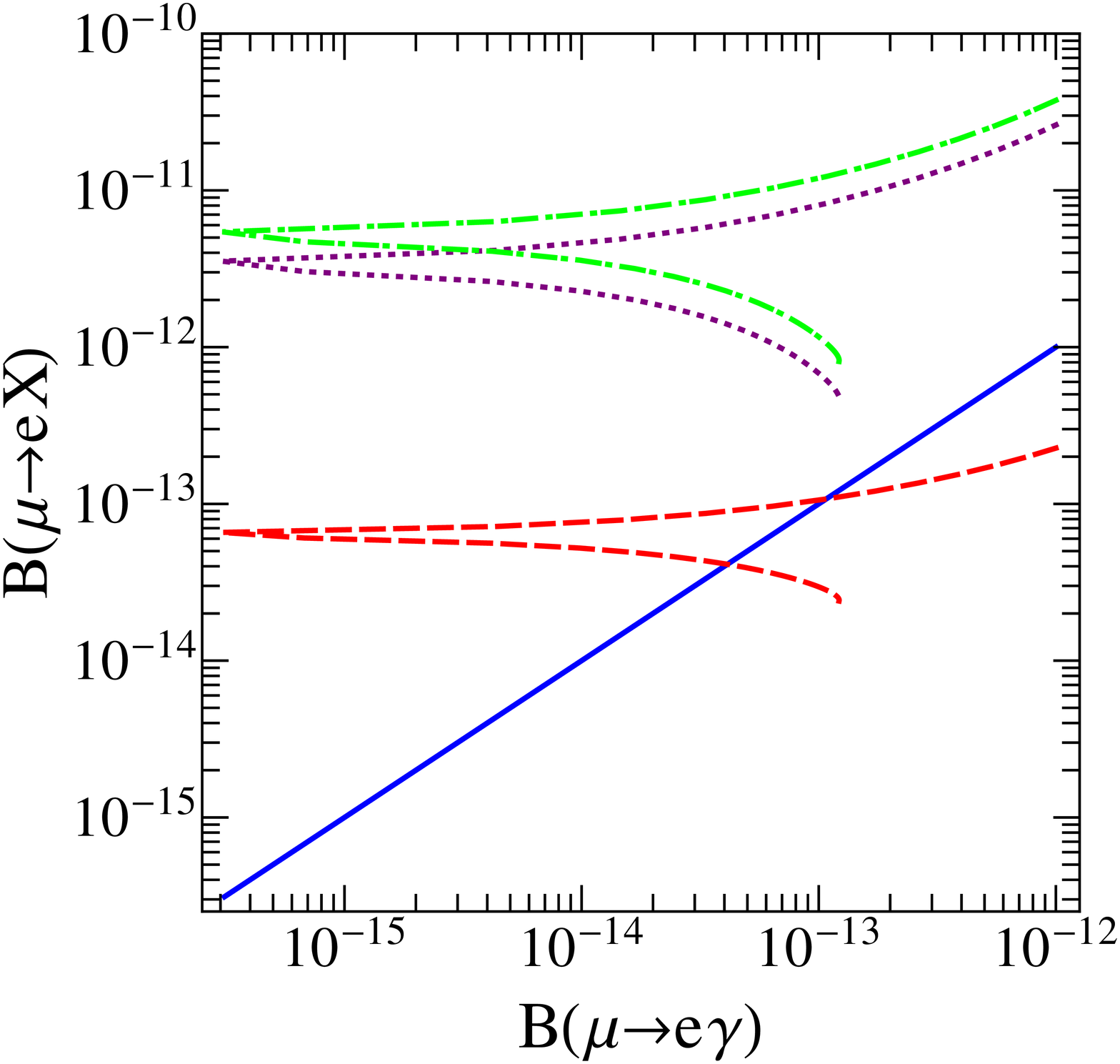}\hspace{1cm}
 \includegraphics[clip,width=0.355\textwidth]{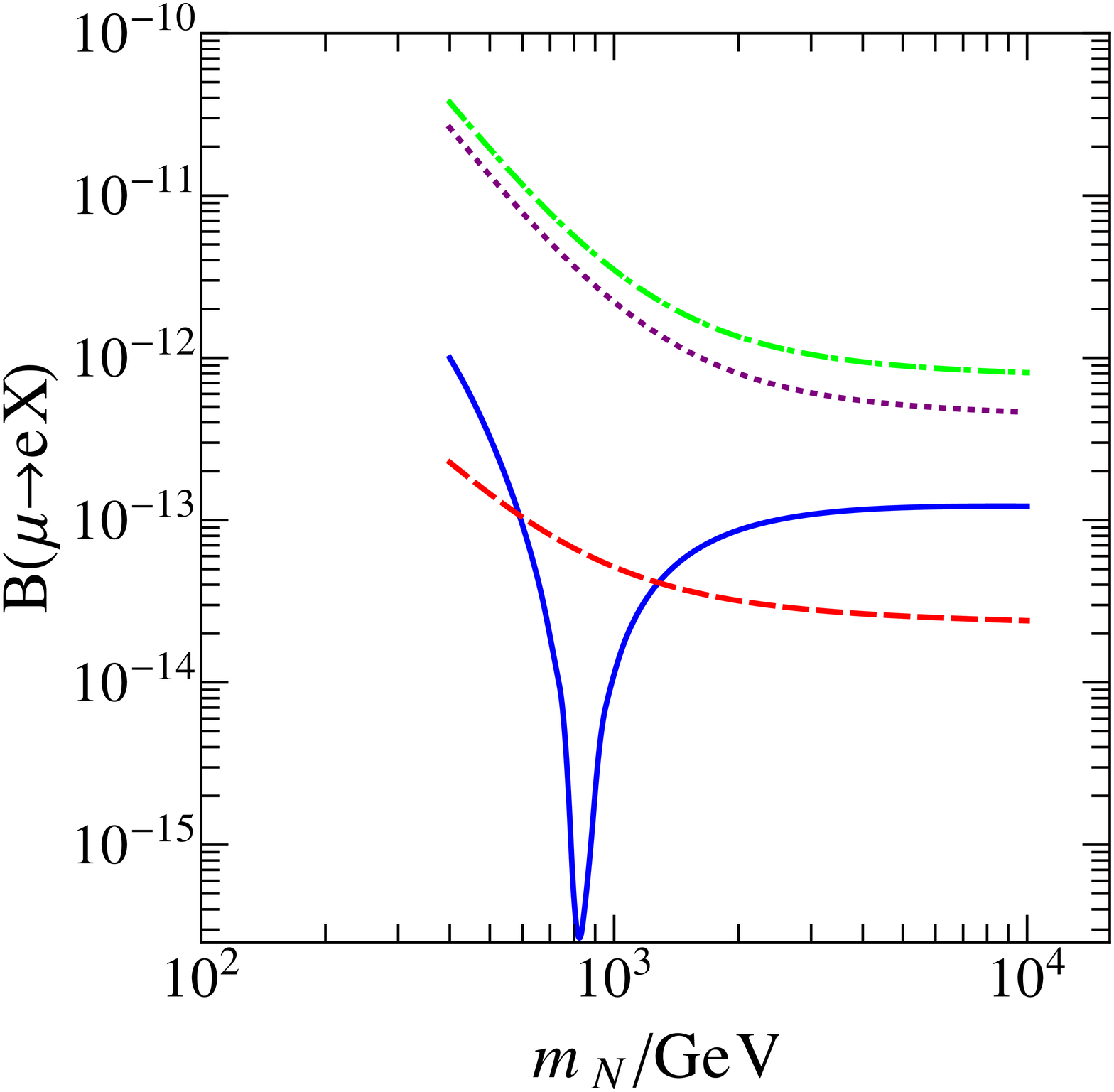}
 \\[.02\textwidth]
 \includegraphics[clip,width=0.355\textwidth]{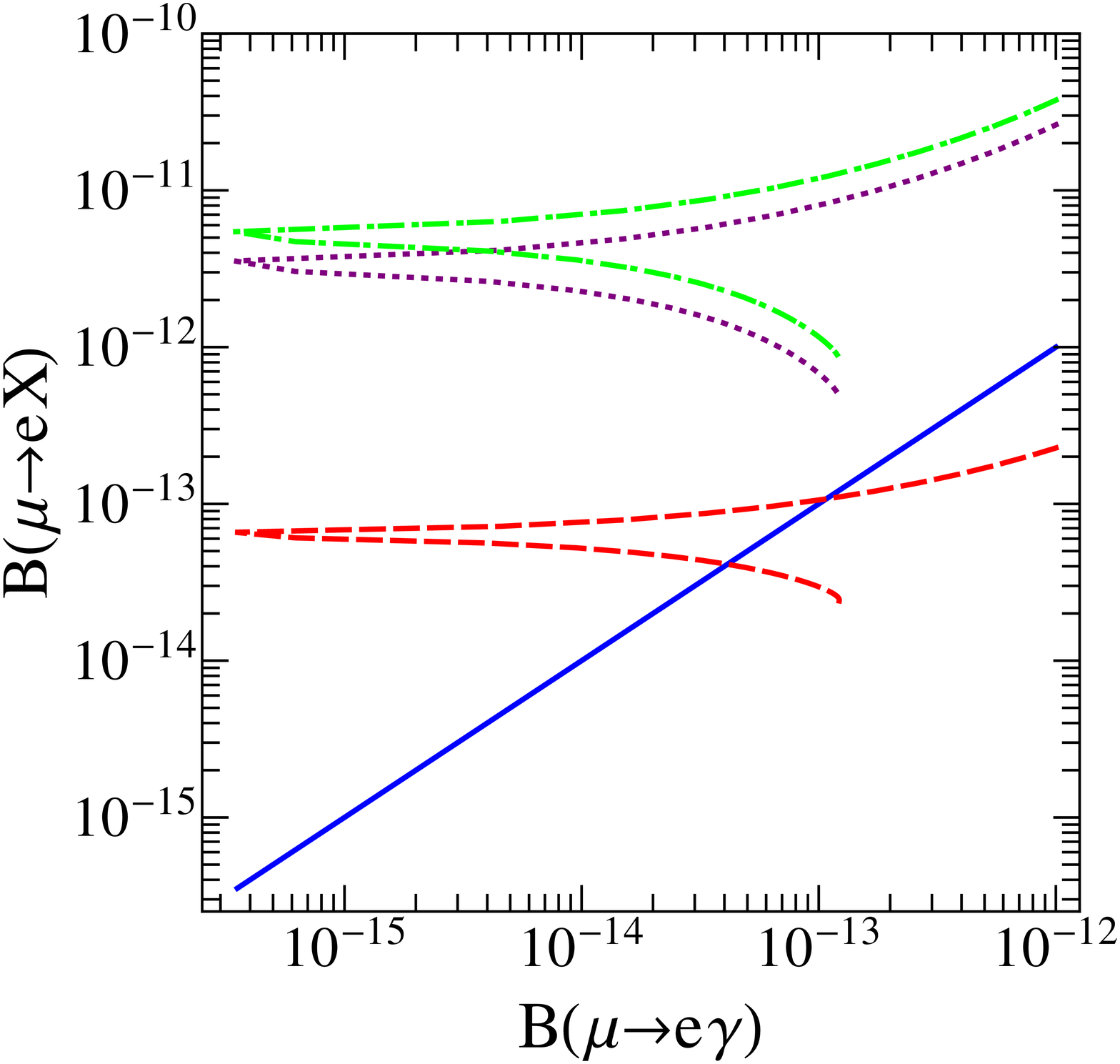}\hspace{1cm}
 \includegraphics[clip,width=0.355\textwidth]{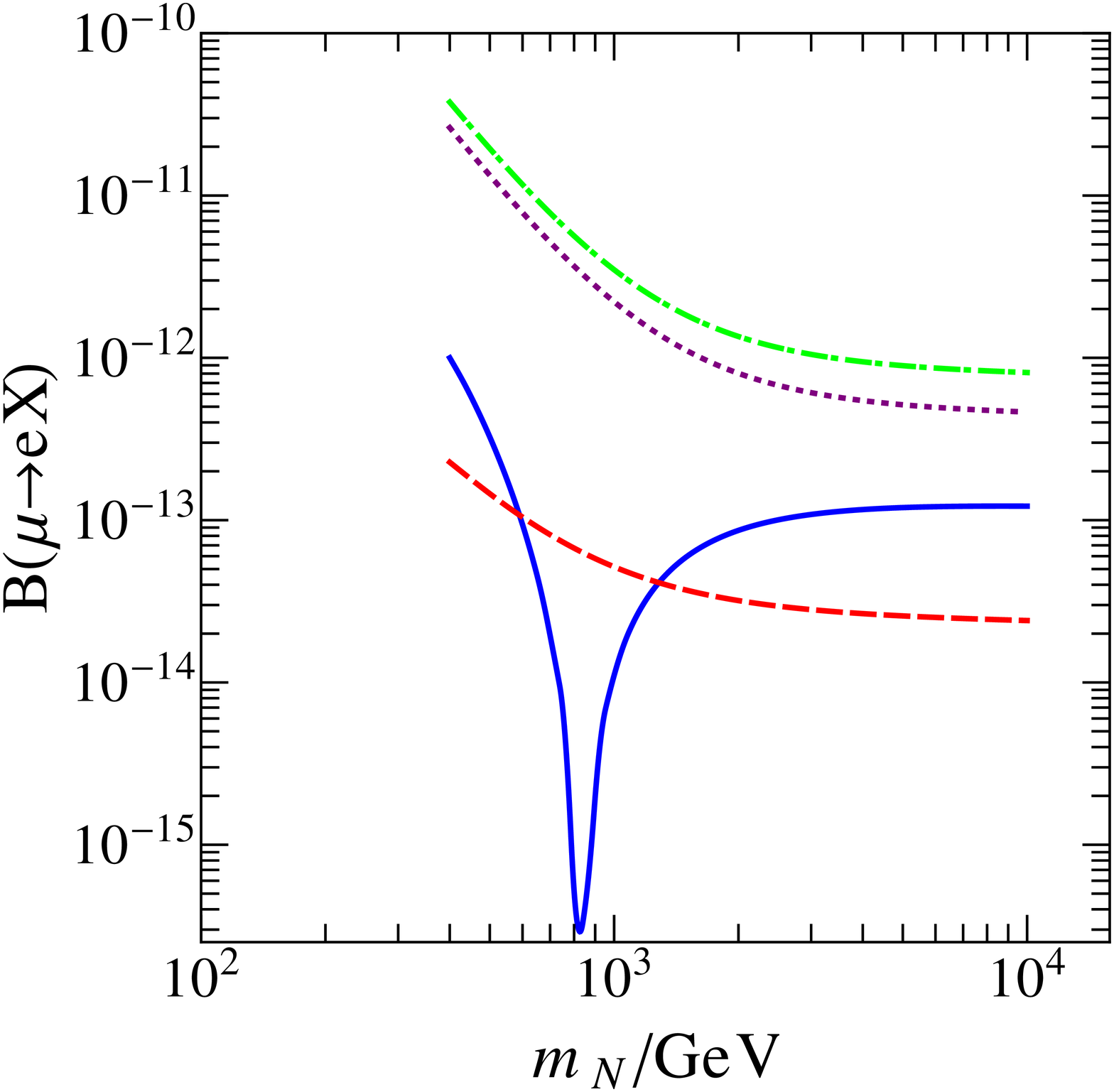}
\caption{Numerical  estimates of  $B(\mu\to e\gamma)$  [blue (solid)],
  $B(\mu\to  eee)$   [red  (dashed)],  $R^{\rm   Ti}_{\mu  e}$~[violet
    (dotted)]  and  $R^{\rm  Au}_{\mu  e}$~[green  (dash-dotted)],  as
  functions  of  $B(\mu\to  e\gamma)$  (left pannels)  and  the  heavy
  neutrino  mass scale  $m_N$ (right  pannels).  In  all  pannels, the
  Yukawa parameter  $a$ was  kept fixed by  the condition  $B(\mu\to e
  \gamma)=10^{-12}$  for $m_N  = 400$~GeV,  and $\tan\beta  =  10$ was
  used.   The upper pannels  display numerical  values for  the Yukawa
  texture~(\ref{YU1}), with $a=b$ and $c=0$, and the lower pannels for
  the Yukawa texture (\ref{YA4}), with $a=b=c$.}

\label{Fig7}
\end{figure}
%%%%%%%%%%%%%%%%%%%%%%%%%%%%%%%%%%%%%%%%%%%%%%%%%%%%%%%%%%%%%%%%%%%%%%

Figure~\ref{Fig7} presents numerical  estimates of $B(\mu\to e\gamma)$
[blue  (solid) line],  $B(\mu\to  eee)$ [red  (dashed) line],  $R^{\rm
  Ti}_{\mu e}$~[violet (dotted)  line] and $R^{\rm Au}_{\mu e}$~[green
  (dash-dotted)  line],  as  functions  of $B(\mu\to  e\gamma)$  (left
pannels) and the heavy neutrino  mass scale $m_N$ (right pannels).  In
all pannels, we  keep the Yukawa parameter $a$  fixed by the condition
$B(\mu\to e \gamma)=10^{-12}$ for $m_N = 400$~GeV, using the benchmark
value: $\tan\beta  = 10$.  The upper pannels  display numerical values
for  the Yukawa  texture~(\ref{YU1}), with  $a=b$ and  $c=0$,  and the
lower pannels  for the Yukawa texture (\ref{YA4}),  with $a=b=c$.  The
heavy  neutrino  mass  is  varied  within the  LHC  explorable  range:
$400\textrm{ GeV}<m_N<10\textrm{ TeV}$.   All observables $B(\mu \to
e X)$ of  $\mu$-LFV (with $X = \gamma,\, ee,\,  {\rm Ti},\, {\rm Au}$)
exhibit  a  non-trivial  dependence  on $m_N$.   The  branching  ratio
$B(\mu\to e\gamma)$  shows a dip  at $m_N\approx 800\textrm{  GeV}$ in
both models (\ref{YU1}) and (\ref{YA4}), signifying the existence of a
cancellation region  in parameter space,  due to loops  involving heavy
neutrino,   sneutrino  and   soft  SUSY-breaking   terms.    For  $m_N
\stackrel{>}{{}_\sim}  3$~TeV,  all  observables  tend to  a  constant
value,  as  a  result  of  the dominance  of  the  soft  SUSY-breaking
contributions.

%%%%%%%%%%%%%%%%%%%%%%%%%%%%%%%%%%%%%%%%%%%%%%%%%%%%%%%%%%%%%%%%%%%%%%
\begin{figure}[!ht]
 \centering
 \includegraphics[clip,width=0.355\textwidth]{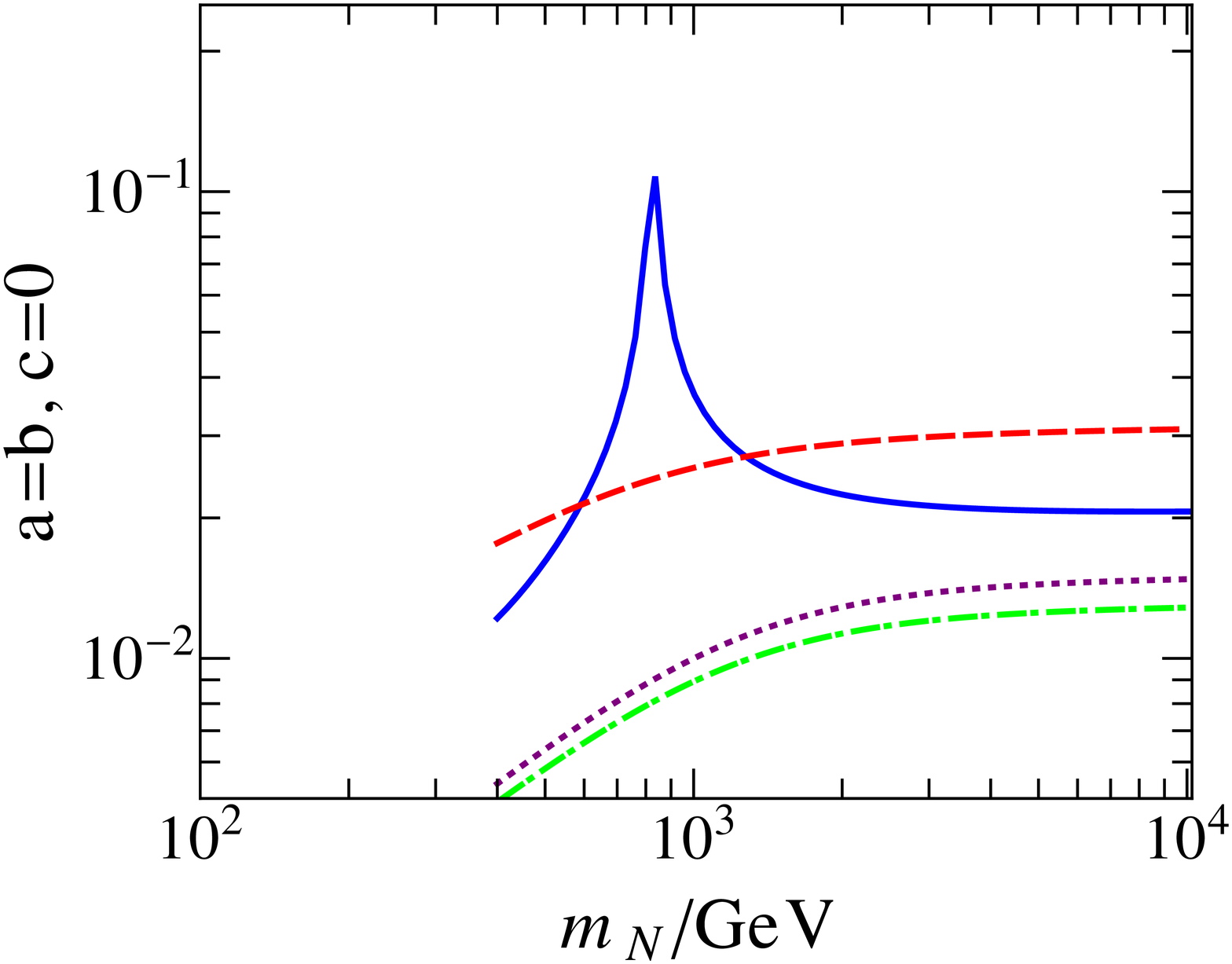}\hspace{1cm}
 \includegraphics[clip,width=0.355\textwidth]{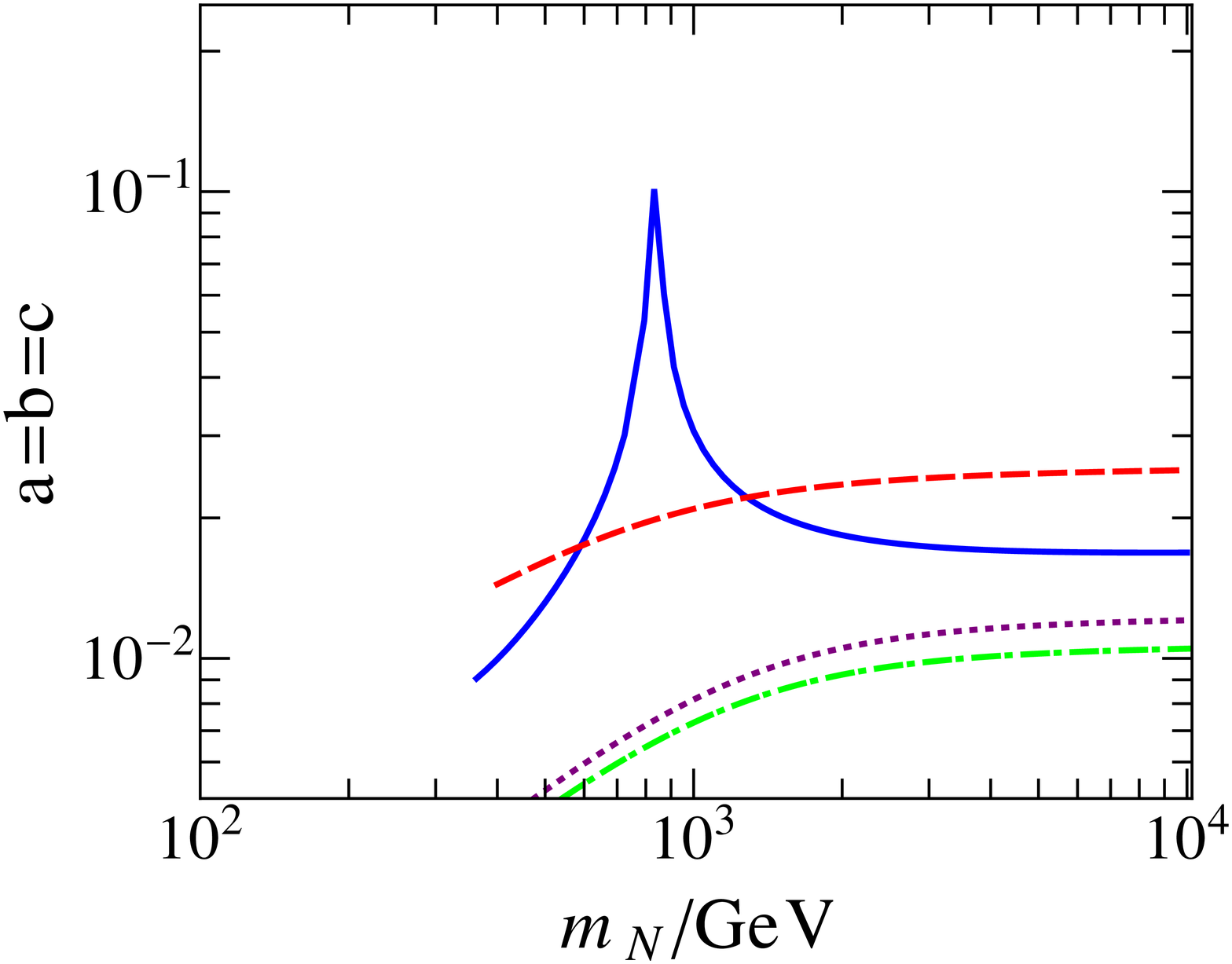}
\caption{Contours of the Yukawa  parameters $(a,b,c)$ versus $m_N$, for
  $B(\mu\to e\gamma)$ [blue  (solid)], $B(\mu\to eee)$ [red (dashed)],
  $R^{\rm   Ti}_{\mu  e}$~[violet   (dotted)]  and   $R^{\rm  Au}_{\mu
    e}$~[green (dash-dotted)],  where $a$ and $m_N$  are determined by
  the  condition  $B(\mu\to e\gamma)  =  10^{-12}$.  All contours  are
  evaluated with $\tan\beta=10$ and  for different Yukawa textures, as
  indicated by the vertical axes labels.}
\label{Fig8}
\end{figure}
%%%%%%%%%%%%%%%%%%%%%%%%%%%%%%%%%%%%%%%%%%%%%%%%%%%%%%%%%%%%%%%%%%%%%%

In Fig.~\ref{Fig8} we show contours of the Yukawa parameters $(a,b,c)$
versus the  heavy neutrino mass  scale $m_N$, for  $B(\mu\to e\gamma)$
[blue  (solid) line],  $B(\mu\to  eee)$ [red  (dashed) line],  $R^{\rm
  Ti}_{\mu e}$~[violet (dotted)  line] and $R^{\rm Au}_{\mu e}$~[green
  (dash-dotted)  line].   The  Yukawa  parameter  $a$  and  $m_N$  are
determined  by  the condition  $B(\mu\to  e\gamma)  = 10^{-12}$.   The
labels  in  the  vertical   axes  indicate  the  two  Yukawa  textures
in~(\ref{YU1}) and~(\ref{YA4}), which we have adopted in our analysis.
The  contours  for $B(\mu\to  e\gamma)$  display  a  maximum for  $m_N
\approx  800$~GeV, as  a  consequence of  cancellations between  heavy
neutrino,    sneutrino    and    soft   SUSY-breaking    contributions
(cf.~Fig.~\ref{Fig7}).

%%%%%%%%%%%%%%%%%%%%%%%%%%%%%%%%%%%%%%%%%%%%%%%%%%%%%%%%%%%%%%%%%%%%%%

\begin{figure}[!ht]
 \centering
 \includegraphics[clip,width=0.33\textwidth]{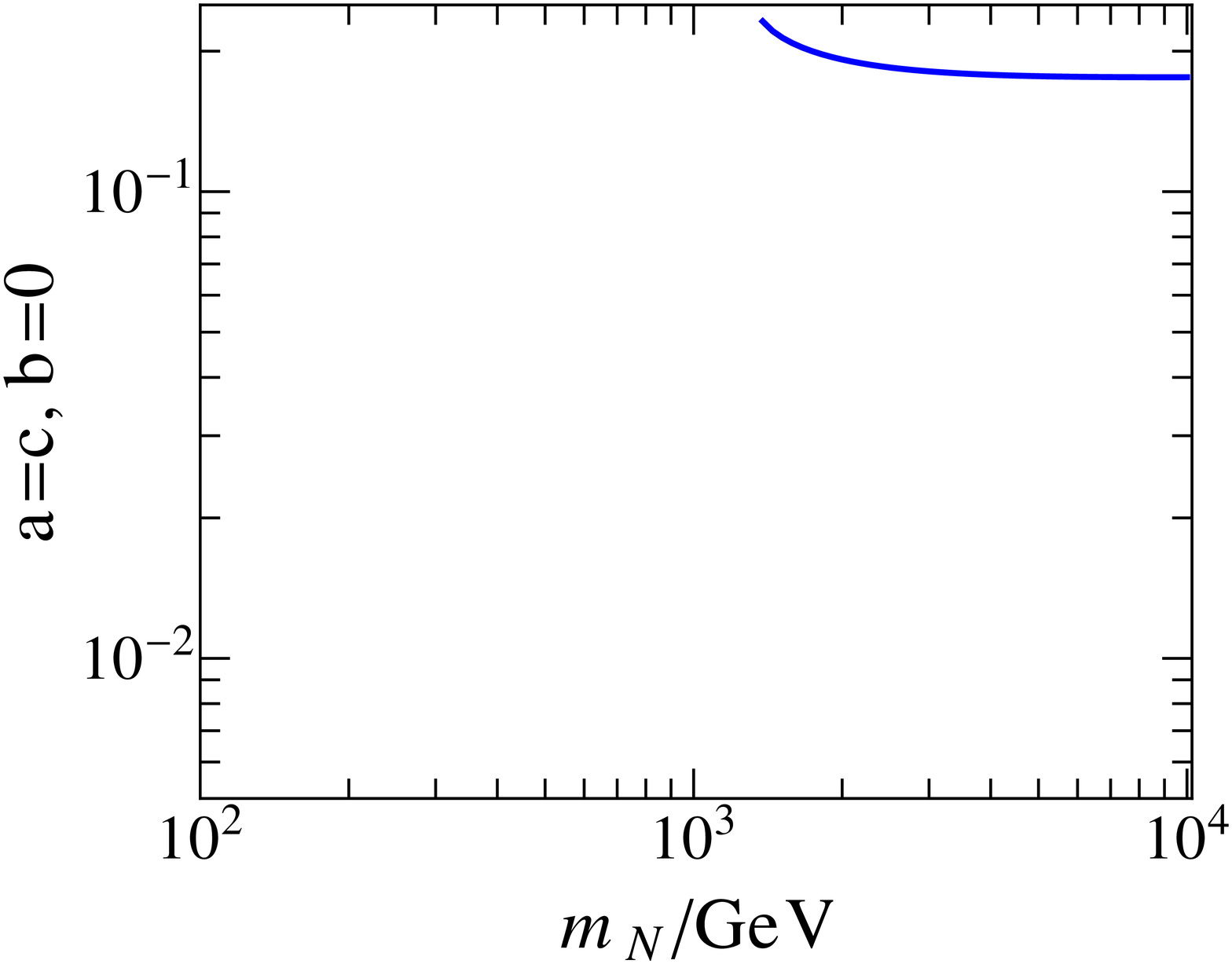}\hspace{1cm}
 \includegraphics[clip,width=0.33\textwidth]{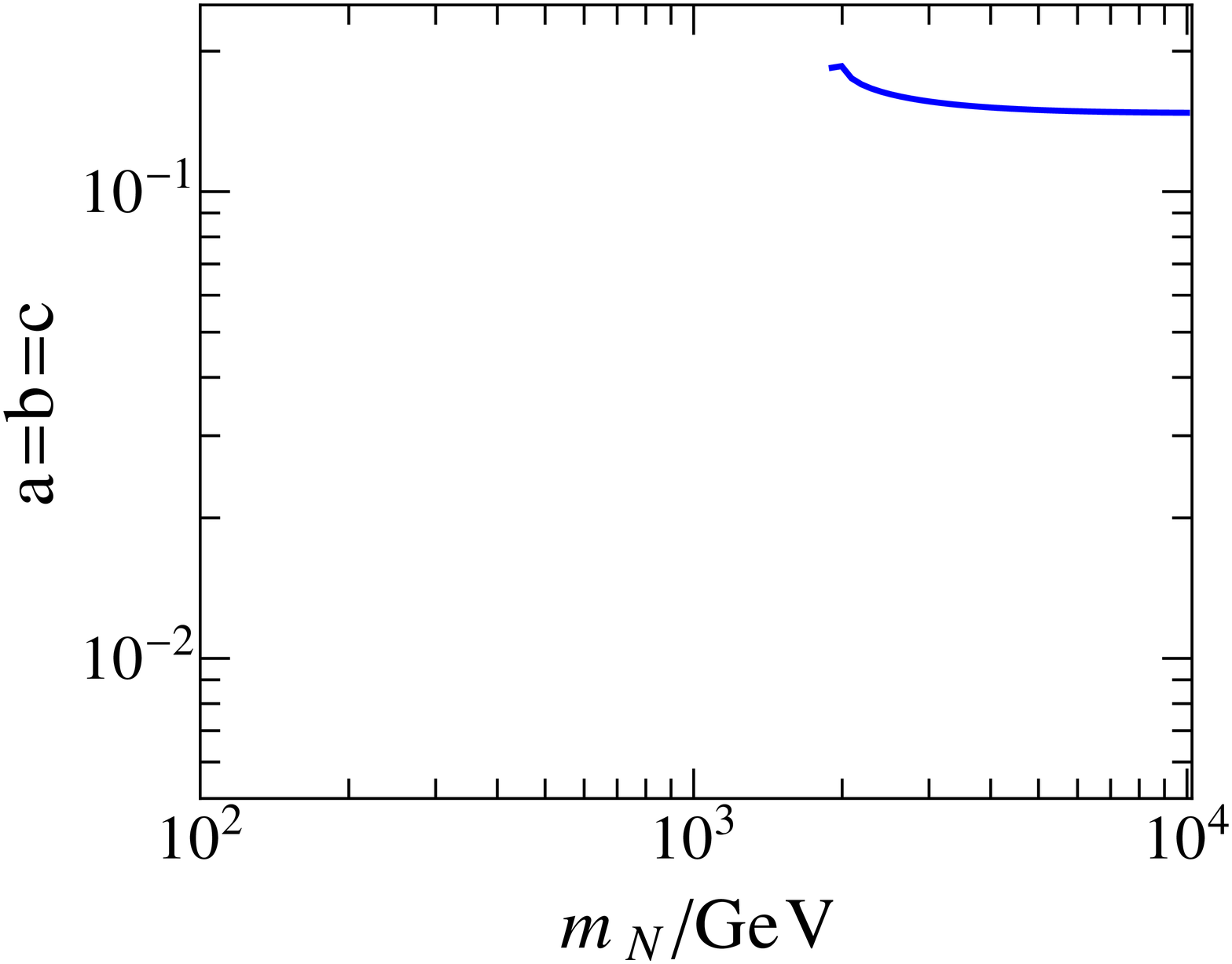}
\caption{Contours of the Yukawa parameters $(a,b,c)$ versus $m_N$, for
  $B(\tau\to e\gamma)$ [blue (solid)], 
%and $B(\tau\to \mu \gamma)$ [red (dashed)], 
where $\tan\beta=10$ and $a$ and $m_N$ are determined by
  the condition $B(\tau\to e\gamma)=10^{-9}$. No solutions have been
  found for $B(\tau\to eee)$ and $B(\tau\to e\mu\mu)$. }
\label{Fig9}
\end{figure}
%%%%%%%%%%%%%%%%%%%%%%%%%%%%%%%%%%%%%%%%%%%%%%%%%%%%%%%%%%%%%%%%%%%%%%

Figure~\ref{Fig9} shows  contours of the  Yukawa parameters $(a,b,c)$,
as functions  of $m_N$, for $B(\tau\to e\gamma)$  [blue (solid) line],
where the  parameters $a$  and $m_N$ are  determined by  the condition
$B(\tau\to e\gamma)=10^{-9}$.   We do  not give numerical  results for
$B(\tau\to \mu \gamma)$, as these  are fully complementary to the ones
given for $B(\tau\to e\gamma)$.  Notice that given the above condition
on $B(\tau\to  e\gamma)$, no solution exists for  $B(\tau\to eee)$ and
$B(\tau\to e\mu\mu)$.

%%%%%%%%%%%%%%%%%%%%%%%%%%%%%%%%%%%%%%%%%%%%%%%%%%%%%%%%%%%%%%%%%%%%%%
\begin{figure}
 \centering
 \includegraphics[clip,width=0.335\textwidth]{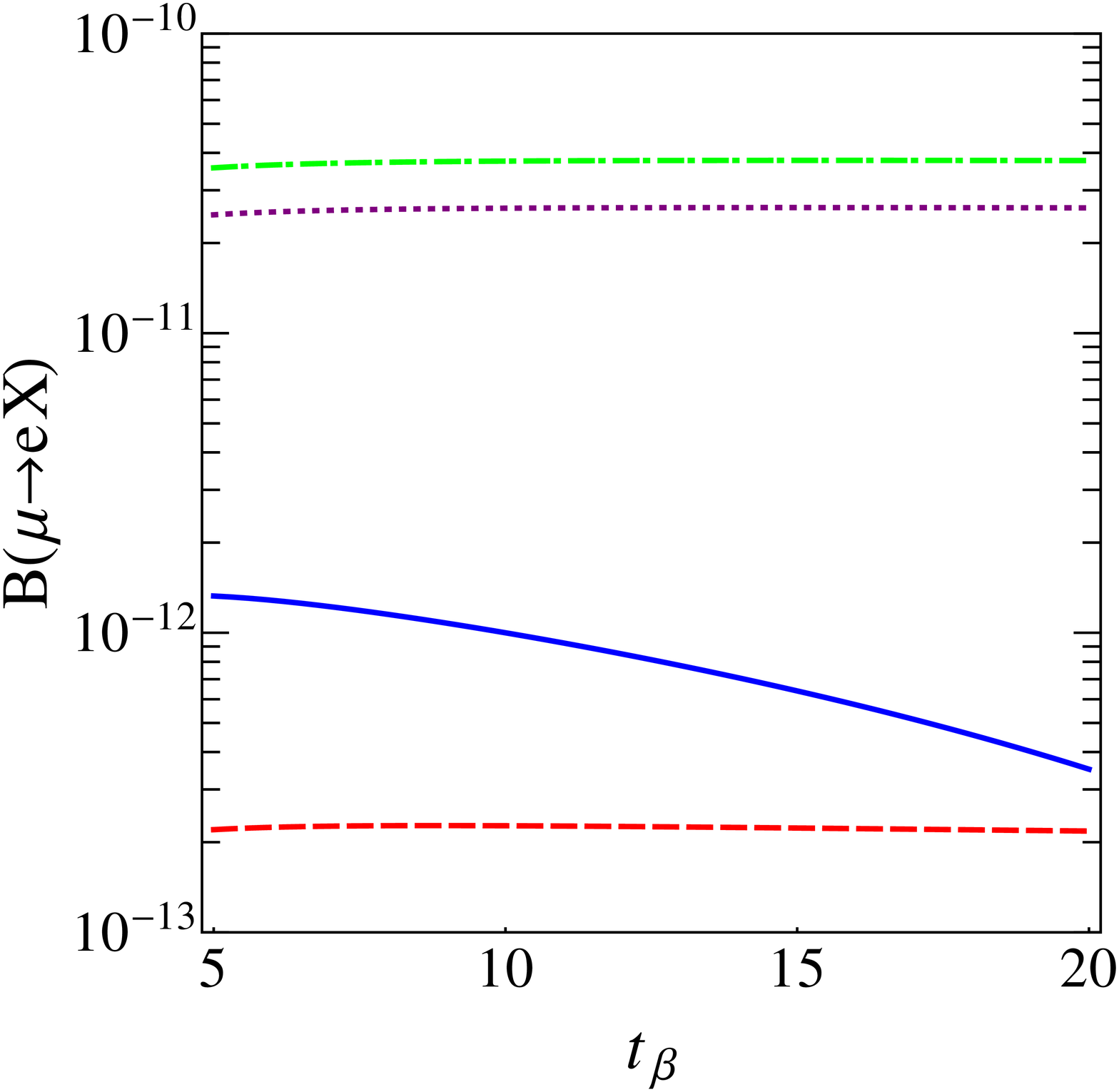}\hspace{1cm}
 \includegraphics[clip,width=0.335\textwidth]{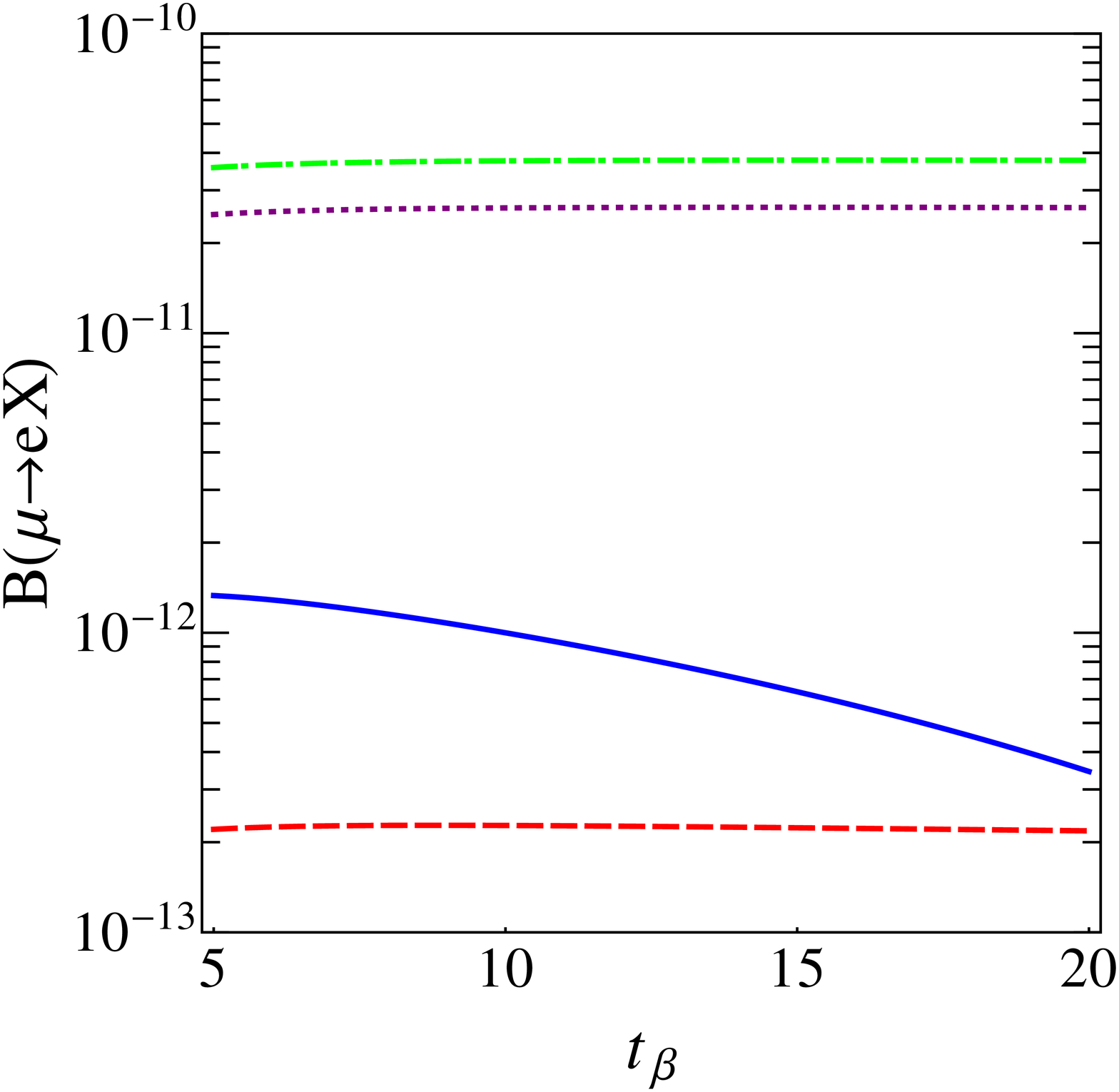}
 \\[.02\textwidth]
 \includegraphics[clip,width=0.335\textwidth]{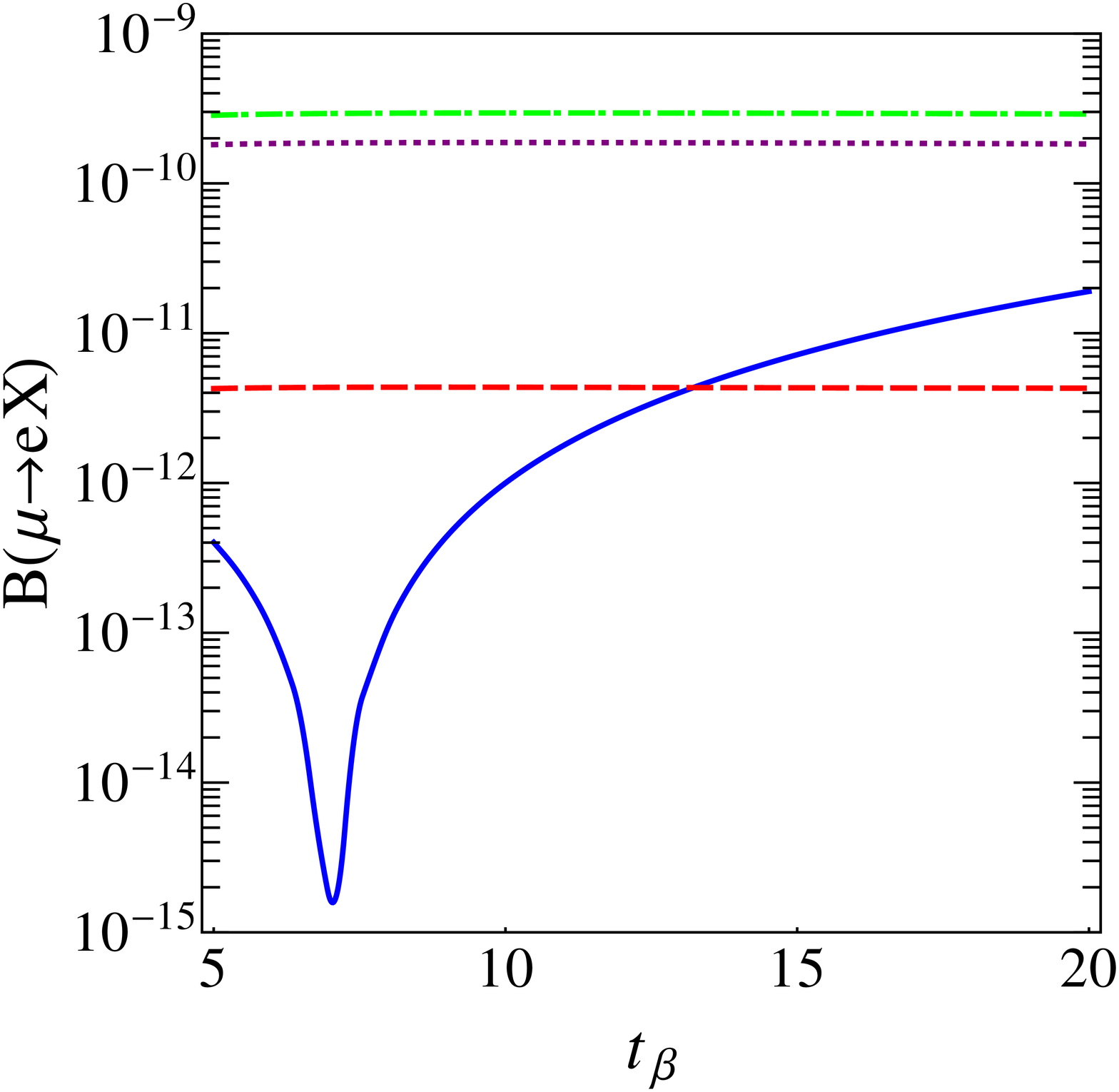}\hspace{1cm}
 \includegraphics[clip,width=0.335\textwidth]{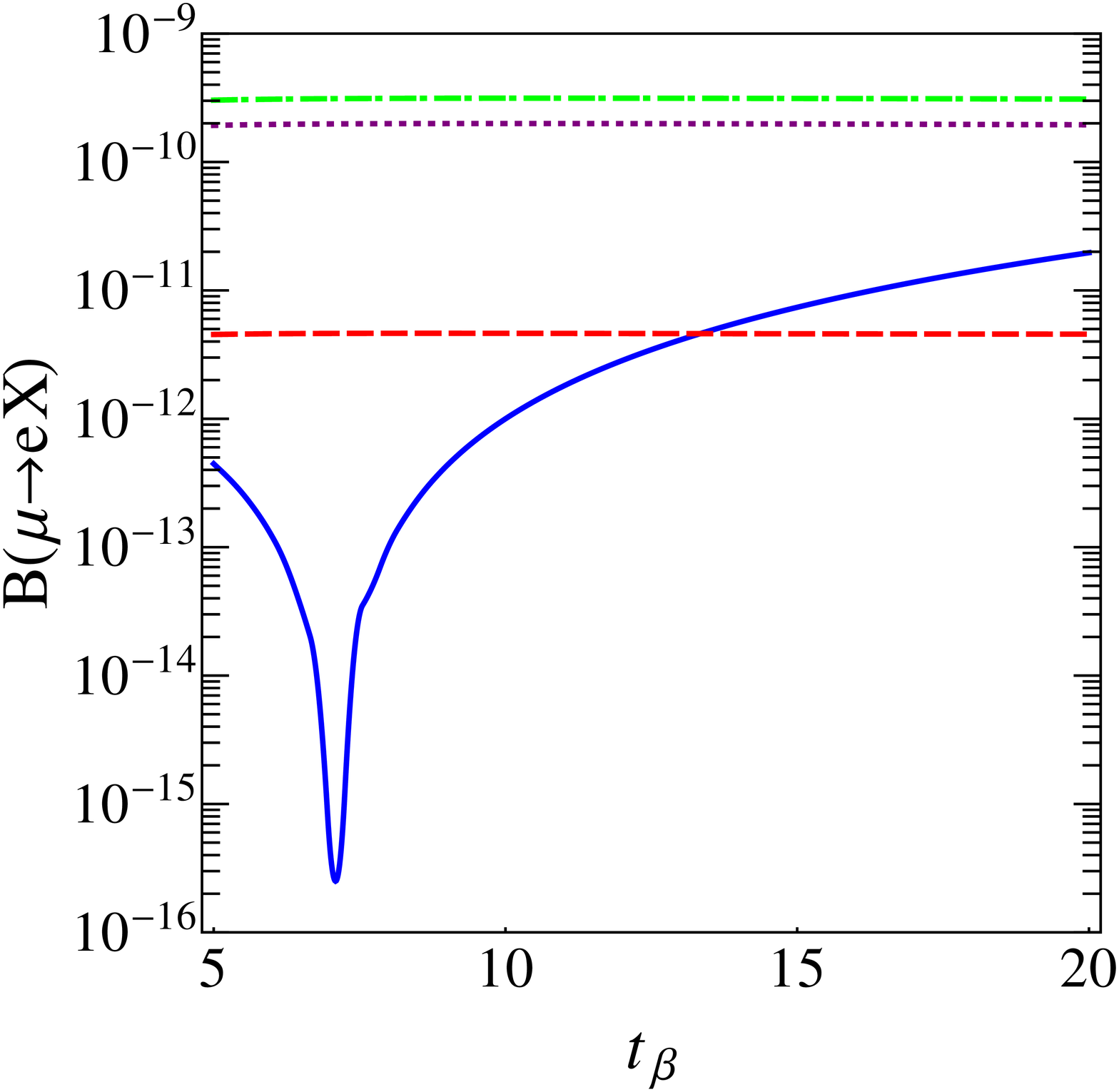}
\caption{Numerical estimates  of $B(\mu\to e  \gamma)$ [blue (solid)],
  $B(\mu\to  eee)$   [red  (dashed)],  $R^{\rm   Ti}_{\mu  e}$~[violet
    (dotted)]  and  $R^{\rm  Au}_{\mu  e}$~[green  (dash-dotted)],  as
  functions  of  $\tan\beta$.   The  upper pannels  are  obtained  for
  $m_N=400$~GeV  and  the  lower  pannels for~$m_N=1$~TeV.   The  left
  pannels use  the Yukawa texture  (\ref{YU1}), with $a=b$  and $c=0$,
  and the right pannels  the Yukawa texture (\ref{YA4}), with $a=b=c$.
  In  all pannels,  the  Yukawa  parameter $a$  is  determined by  the
  condition $B(\mu\to e \gamma)=10^{-12}$.}
\label{Fig10}
\end{figure}
%%%%%%%%%%%%%%%%%%%%%%%%%%%%%%%%%%%%%%%%%%%%%%%%%%%%%%%%%%%%%%%%%%%%%%

In  our  numerical  analysis  so  far,  we  have  kept  the  value  of
$\tan\beta$  fixed  to its  benchmark  value given  in~(\ref{mSUGRA}):
$\tan\beta  = 10$.   In  Fig.~\ref{Fig10}, we  relax this  assumption,
varying $\tan\beta$ in the interval $5 \stackrel{<}{{}_\sim} \tan\beta
\stackrel{<}{{}_\sim} 20$, while  maintaining agreement with a SM-like
Higgs boson  mass $M_H \approx  125$~GeV and taking into  account that
the combined experimental and  theoretical errors are currently of the
order of~5--6~GeV.   Specifically, in Fig.~\ref{Fig10}  we display the
dependence  of $B(\mu\to  e  \gamma)$ [blue  (solid) line],  $B(\mu\to
eee)$ [red (dashed) line], $R^{\rm Ti}_{\mu e}$~[violet (dotted) line]
and  $R^{\rm Au}_{\mu e}$~[green  (dash-dotted) line]  on $\tan\beta$.
In  all  pannels,  the  Yukawa  parameter $a$  is  determined  by  the
condition  $B(\mu\to  e   \gamma)=10^{-12}$.   The  upper  pannels  in
Fig.~\ref{Fig10} show  numerical results for  $m_N=400$~GeV, while the
lower pannels for~$m_N=1$~TeV.  The  left pannels give our predictions
for  the Yukawa  texture~(\ref{YU1}), with  $a=b$ and  $c=0$,  and the
right pannels  for the  Yukawa texture~(\ref{YA4}), with  $a=b=c$.  In
the lower pannels,  we observe a suppression of  $B(\mu\to e \gamma)$,
for values  $\tan\beta \approx 7$,  due to cancellation  between heavy
neutrino, sneutrino and soft SUSY-breaking effects.

%%%%%%%%%%%%%%%%%%%%%%%%%%%%%%%%%%%%%%%%%%%%%%%%%%%%%%%%%%%%%%%%%%%%%%
\begin{figure}
 \centering
 \includegraphics[clip,width=0.38\textwidth]{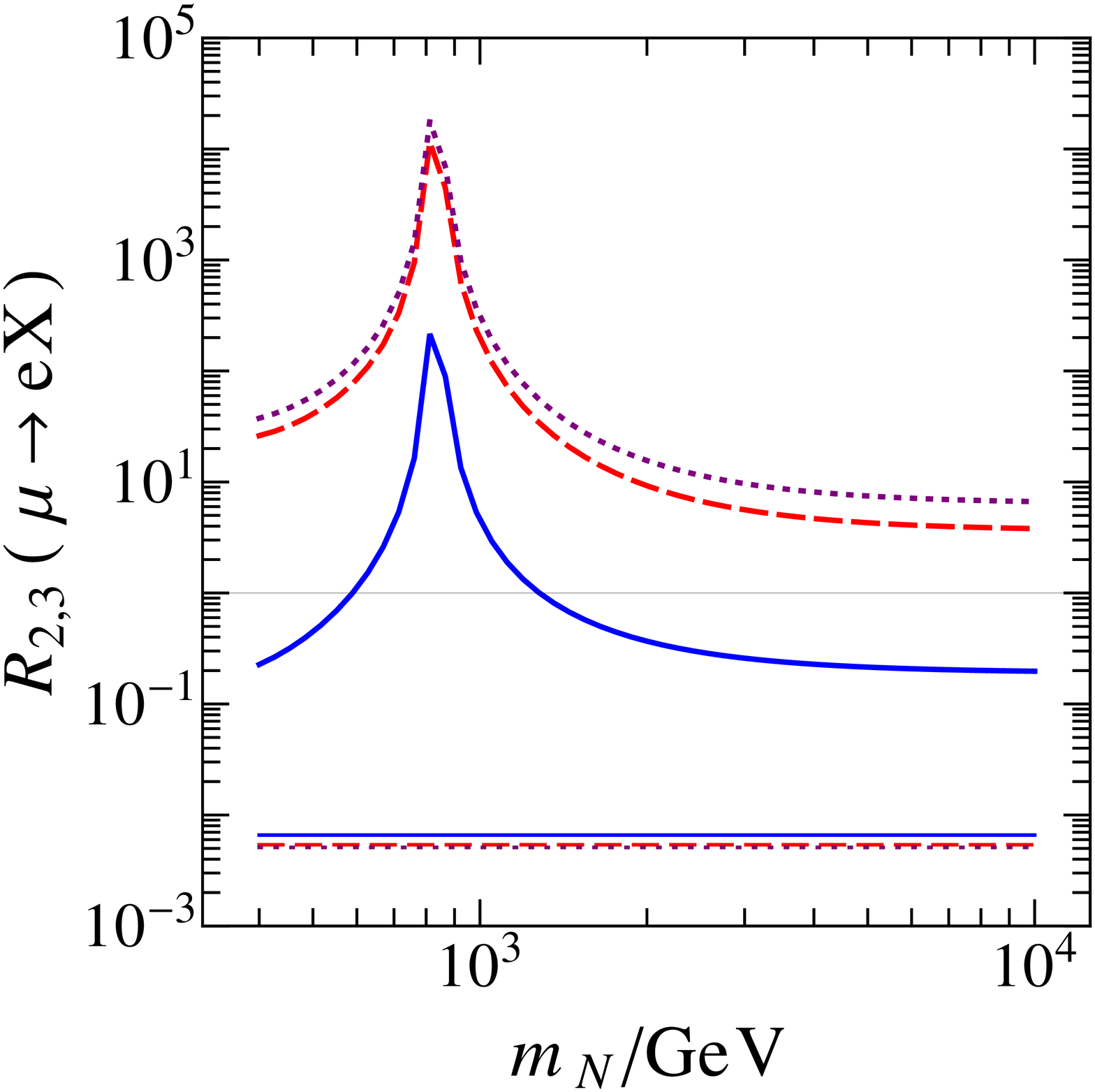} \\
 \includegraphics[clip,width=0.35\textwidth]{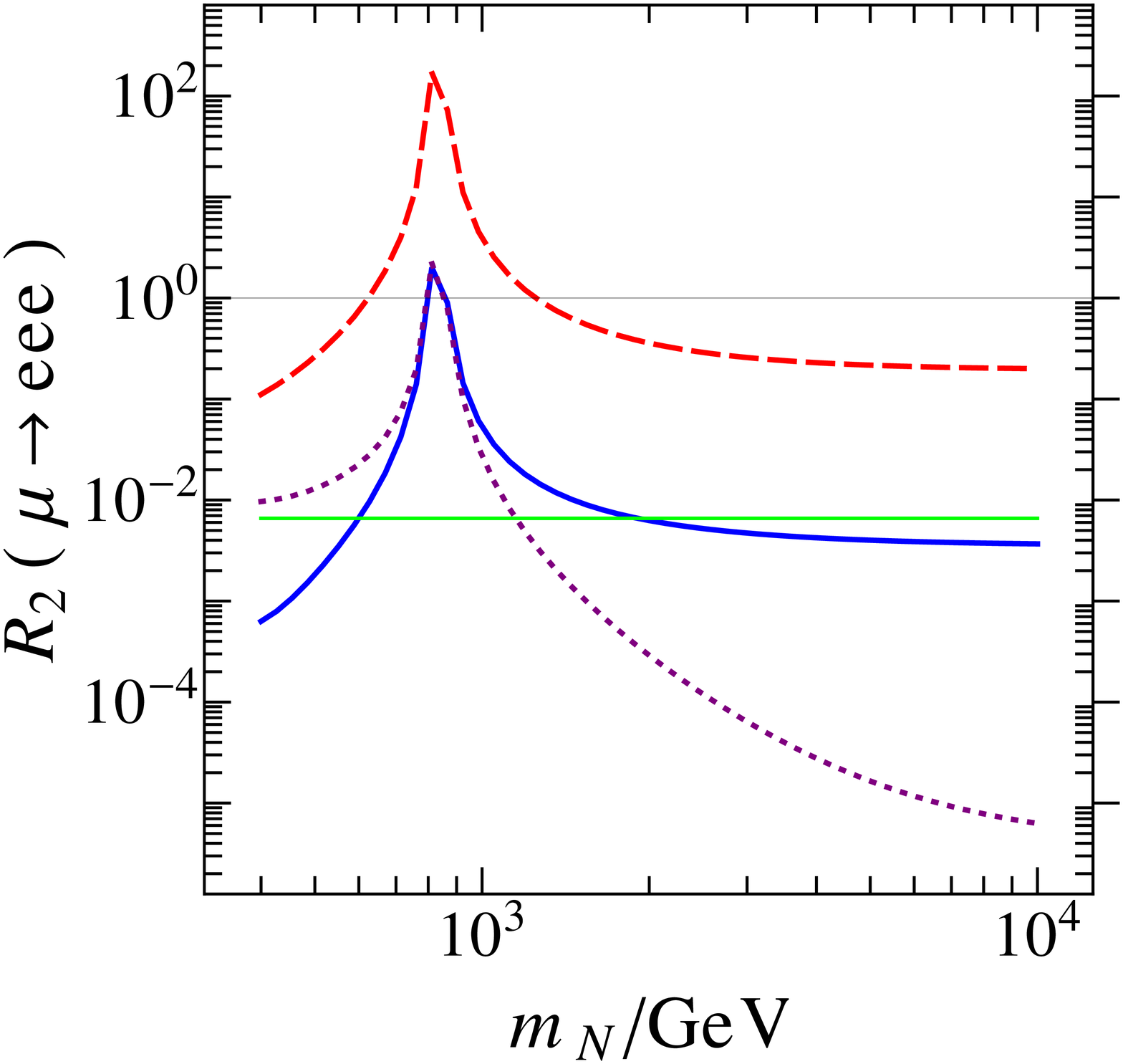}\hspace{1cm}
 \includegraphics[clip,width=0.35\textwidth]{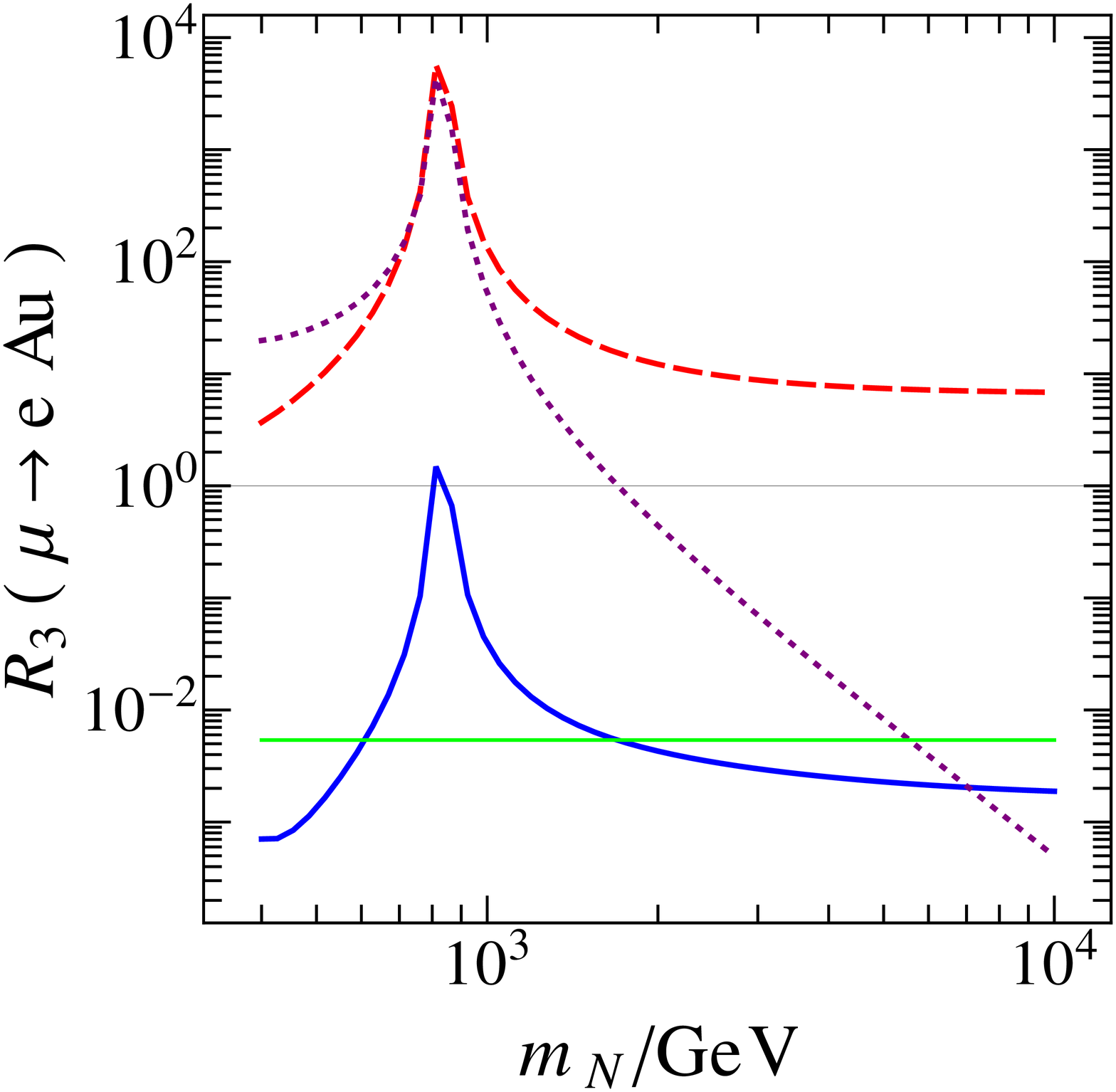} \\
 \includegraphics[clip,width=0.35\textwidth]{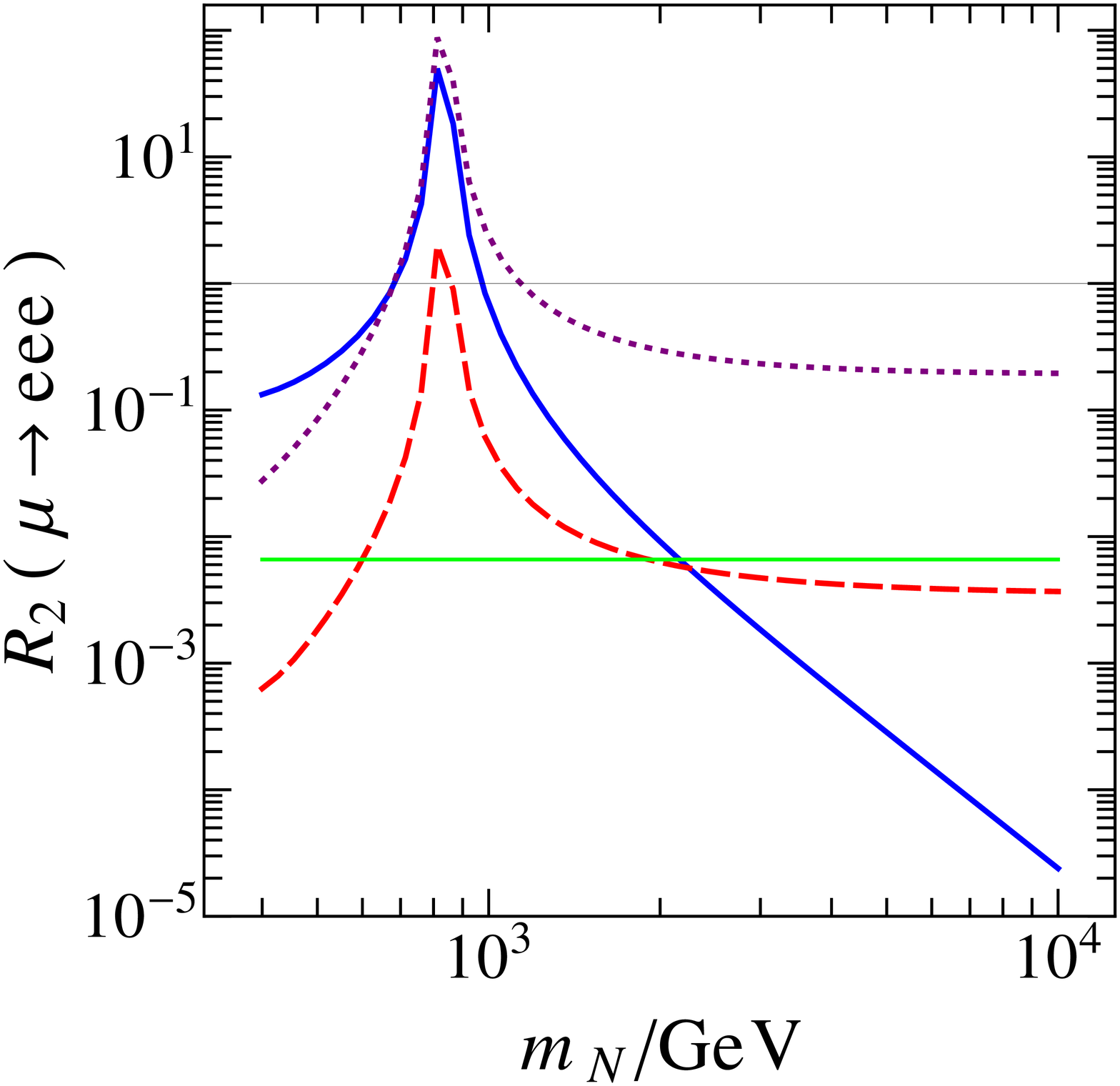}\hspace{1cm}
 \includegraphics[clip,width=0.35\textwidth]{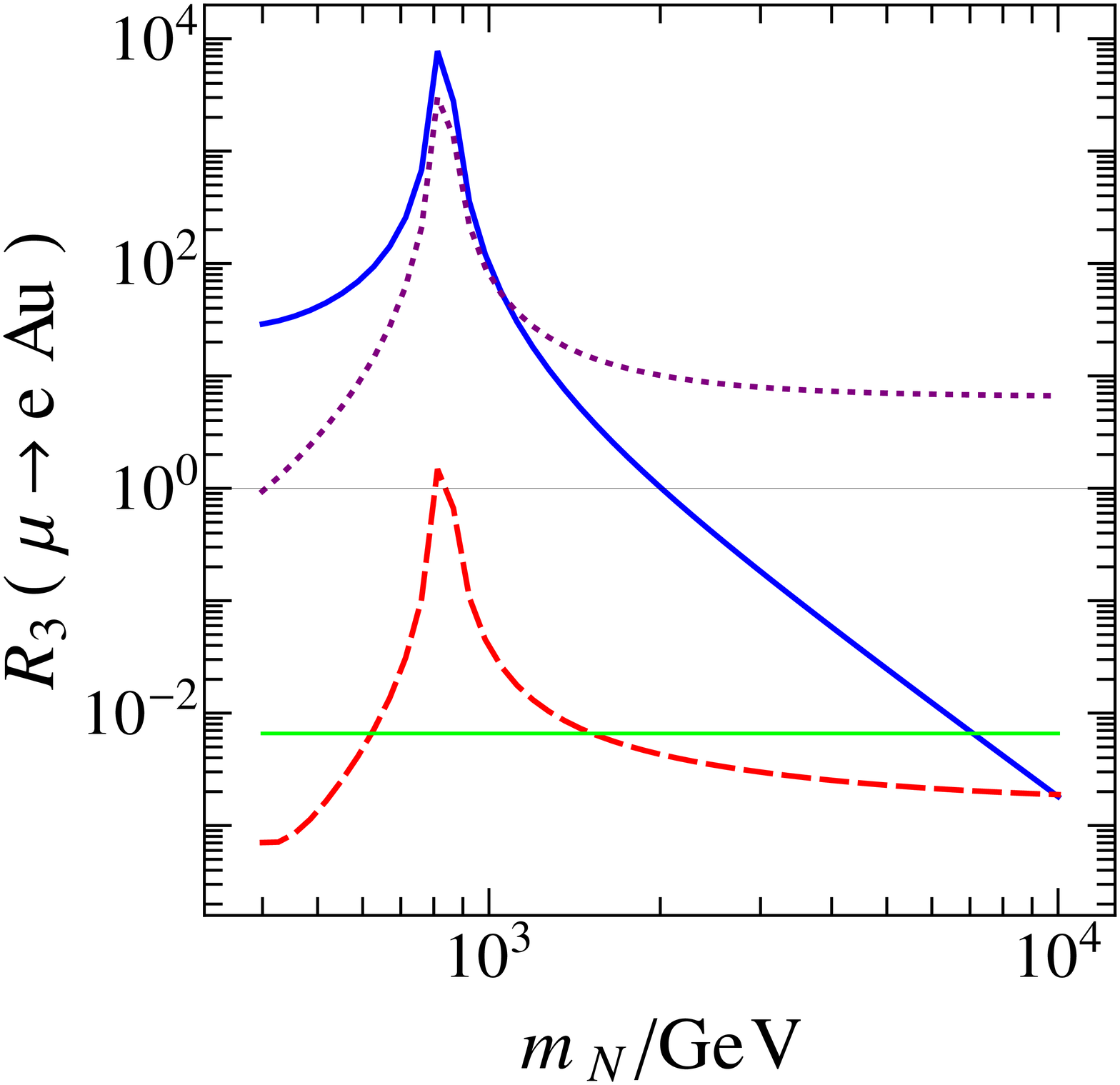}
\caption{Numerical estimates of the ratios $R_2(\mu \to eee)$, $R^{\rm
    Ti}_3$  and $R^{\rm  Au}_3$,  as functions  of  $m_N$. The  Yukawa
  parameter  $a$   is  fixed   by  the  condition   $B(\mu\to  e\gamma
  )=10^{-12}$, for  $m_N = 400$~GeV and $\tan\beta=10$.   In the upper
  pannel,  thick lines give  the complete  evaluation of  $R_2(\mu \to
  eee)$  [blue (solid)],  $R^{\rm  Ti}_3$ [red  (dashed)] and  $R^{\rm
    Au}_3$  [violet (dotted)],  while  the respective  thin lines  are
  evaluated keeping only  the magnetic dipole formfactors $G_\gamma^L$
  and  $G_\gamma^R$.   The two  middle  pannels  provide a  formfactor
  analysis  of $R_2(\mu  \to eee)$  and  $R^{\rm Au}_3$,  in terms  of
  contributions  due  to  $G_\gamma$  and $F_\gamma$  [blue  (solid)],
  $F_Z$~[red~(dashed)]  and box  formfactors  [violet (dotted)].   The
  lower  two pannels  show  the separate  contributions  due to  heavy
  neutrinos~$N$                                         [blue~(solid)],
  sneutrinos~$\widetilde{N}$~[red~(dashed)] and soft SUSY-breaking LFV
  terms~[violet~(dotted)].  The green (horizontal) lines in the middle
  and  lower pannels give  the predicted  values obtained  by assuming
  that  only  the   $G^{L,R}_\gamma$  formfactors  contribute  to  the
  amplitudes.}
  \label{Fig11}
\end{figure}
%%%%%%%%%%%%%%%%%%%%%%%%%%%%%%%%%%%%%%%%%%%%%%%%%%%%%%%%%%%%%%%%%%%%%%

It is interesting to compare  the contributions of the magnetic dipole
formfactors to  the CLFV observables, with those  originating from the
remaining  formfactors. Specifically,  if  one assumes  that only  the
magnetic      dipole      formfactors~$G^{L,R}_\gamma$      contribute
in~(\ref{Bl3l_1}),   (\ref{Bl3l_2})   and   (\ref{RmueJ}),  then   the
following analytical results are obtained for the ratios:
\begin{eqnarray}
R_1 &\equiv& \frac{B(l\to l'l_1l_1^c)}{B(l\to l'\gamma)}
 \ =\ 
 \frac{\alpha}{3\pi}\Big(\ln\frac{m_l^2}{m_{l'}^2} - 3\Big)
\label{RBR1}
\\
R_2 &\equiv&
\frac{B(l\to l'l'l'^c)}{B(l\to l'\gamma)}
 \ =\
 \frac{\alpha}{3\pi}\Big(\ln\frac{m_l^2}{m_{l'}^2} - \frac{11}{4}\Big)
\label{RBR2}
\\
R_3 &\equiv& 
 \frac{R^J_{\mu e}}{B(\mu\to e\gamma)} 
 \ =\ 
 16\alpha^4 \frac{\Gamma_\mu}{\Gamma_{\rm capture}} Z Z_{eff}^4 |F(-\mu^2)|^2\  .
\label{RBR3}
\end{eqnarray}
According  to the  formulae (\ref{RBR1})--(\ref{RBR3}),  the predicted
$R_1$ values for $\tau\to e\mu\mu$  and $\tau\to\mu ee$ are $1/90$ and
$1/419$ respectively,  the predicted $R_2$  values for $\mu\to  e ee$,
$\tau\to  eee$ and  the $\tau\to  \mu\mu\mu$ are  $1/159$,  $1/91$ and
$1/460$ respectively, and the predicted $R_3$ values for Ti and Au are
$1/198$  and  $1/188$  respectively. 

In  Fig.~\ref{Fig11},  we  give  numerical  estimates  of  the  ratios
$R_2(\mu \to eee)$, $R^{\rm Ti}_3$ and $R^{\rm Au}_3$, as functions of
$m_N$. The  Yukawa parameter $a$  is fixed by the  condition $B(\mu\to
e\gamma )=10^{-12}$,  for $m_N = 400$~GeV and  $\tan\beta=10$.  In the
upper  pannel, thick  lines show  the predicted  values obtained  by a
complete evaluation of $R_2(\mu \to eee)$ [blue (solid) line], $R^{\rm
  Ti}_3$  [red  (dashed) line]  and  $R^{\rm  Au}_3$ [violet  (dotted)
  line], while the respective thin  lines are obtained by keeping only
the magnetic dipole  formfactors $G_\gamma^L$ and $G_\gamma^R$. Hence,
we see that going beyond  the magnetic dipole moment approximation may
enhance the ratios $R_{2,3}$ by more than two orders of magnitude.

The  two  middle  pannels  of Fig.~\ref{Fig11}  provide  a  formfactor
analysis  of $R_2(\mu  \to eee)$  and $R^{\rm  Au}_3$,  by considering
separately the  contributions due  to $G_\gamma$ and  $F_\gamma$ [blue
  (solid) line], $F_Z$~[red~(dashed) line] and box formfactors [violet
  (dotted)  line].   In particular,  we  observe  that heavy  neutrino
contributions  to the box  formfactors become  comparable to  and even
larger than the $Z$-boson mediated graphs in $\mu \to e$ conversion in
Gold, for heavy neutrino masses $m_N \stackrel{<}{{}_\sim} 1$~TeV.  We
have checked  that for  $m_N \stackrel{<}{{}_\sim} 1$~TeV,  box graphs
due  to heavy  neutrinos  also dominate  the  process of  $\mu \to  e$
conversion  in Titanium  (not explicitly  shown  in Fig.~\ref{Fig11}).
Finally, the  two lower pannels show the  individual contributions due
to      heavy      neutrinos~$N_{1,2,3}$     [blue~(solid)      line],
sneutrinos~$\widetilde{N}_{1,2,\dots,12}$~[red~(dashed) line] and soft
SUSY-breaking LFV terms~[violet~(dotted)  line].  From these two lower
pannels,   it  is  obvious   that  for   heavy  neutrino   masses  $m_N
\stackrel{>}{{}_\sim} 1$~TeV, the  soft SUSY-breaking effects dominate
the CLFV  formfactors, which  are tagged with  the superscripts  SB in
Appendix~\ref{sec:olff}.   Instead,   for  $m_N  \stackrel{<}{{}_\sim}
1$~TeV, heavy-neutrino effects start becoming the leading contribution
to the CLFV observables associated with the muon.  Note that the green
(horizontal) lines in the middle  and lower pannels serve as reference
values obtained by assuming that only the $G^{L,R}_\gamma$ formfactors
contribute to the amplitudes.

An important consistency check for  our numerical analysis has been to
{\em  analytically} show that  all soft  SUSY-breaking effects  on the
formfactors~(\ref{gammaSB}), (\ref{ZSB}) and~(\ref{ffboxSB}) vanish in
the limit of degenerate charged  slepton masses. On the other hand, RG
effects from $M_{\rm GUT}$ to  $M_Z$ induce sizeable deviations to the
charged slepton  mass matrix from  the unit matrix. As  a consequence,
unitarity cancellations due to the so-called GIM mechanism become less
effective in this  case and so render the SB  part of the formfactors,
such  as $F_{l'lZ}^{L,{\rm SB}}$  and $F_{l'lZ}^{R,{\rm  SB}}$, rather
large.

Another essential check was to show that under the assumptions adopted
in~\cite{Hirsch:2012ax}, our formfactor $F_{l'lZ}^{L,\tilde{N}}$ given
in~(\ref{ZNt}) reduces to $\frac{2c_W}{g} F_L^c$, where $F_L^c$ is one
of  the formfactors defined  in~(6) of~\cite{Hirsch:2012ax},  which in
turn can be shown  to vanish.  The assumptions in~\cite{Hirsch:2012ax}
are:  (i)~the   standard  seesaw  mechanism   with  ultra-heavy  right
neutrinos,  (ii)~no charged  wino  or higgsino  mixing, and  (iii)~the
dominance of  the wino  contribution.  Under these  three assumptions,
the    interaction    vertices    occurring    in    the    formfactor
$F_{l'lZ}^{L,\tilde{N}}$ simplify as follows:
\begin{equation}
  \label{eq:Hirsch}
\tilde{B}_{lmA}^{R,1},\, \tilde{B}_{lmA}^{R,2}\ \to\ - U_{lk}\, , 
\qquad 
\tilde{C}_{AB}^1,\,  \tilde{C}_{AB}^2,\, \tilde{C}_{AB}^3,\, \tilde{C}_{AB}^4\
\to\ -\frac{1}{2}\, \delta_{kk'}\,,\qquad 
V_{mk}^{\tilde{\chi}^- R}\ \to\ c^2_w\; ,
\end{equation}
where $A,B$ assume now the restricted range of values $k,k'=1,2,3$ and
$U$  is  a $3\times  3$  unitary  matrix.   Given the  simplifications
in~(\ref{eq:Hirsch}),       we       recover      the       expression
of~\cite{Hirsch:2012ax},     resulting     in     the     replacement:
$F_{l'lZ}^{L,\tilde{N}}\to    \frac{2c_W}{g}   F_L^c$.     The   above
non-trivial  checks provide firm  support for  the correctness  of our
analytical  and  numerical results  that  we  have  presented in  this
section.

%\newpage

\setcounter{equation}{0}
\section{Conclusions}\label{concl}

We have analyzed Charged Lepton Flavour Violation in the MSSM extended
by low-scale singlet heavy  neutrinos, paying special attention to the
individual loop contributions  due to the heavy neutrinos~$N_{1,2,3}$,
sneutrinos~$\widetilde{N}_{1,2,\dots,12}$   and   soft   SUSY-breaking
terms.   In our  analysis, we  have included  for the  first  time the
complete  set of  box  diagrams, in  addition  to the  photon and  the
$Z$-boson mediated  interactions.  We  have also derived  the complete
set of  chiral amplitudes and  their associate formfactors  related to
the neutrinoless three-body  CLFV decays of the muon  and tau, such as
$\mu \to eee$, $\tau \to  \mu\mu\mu$, $\tau \to e\mu\mu$ and $\tau \to
ee\mu$,  and to the  coherent $\mu  \to e$  conversion in  nuclei. Our
analytical results are  general and can be applied to  most of the New
Physics  models with  CLFV.  In  this context,  we emphasize  that our
systematic  analysis  has  revealed  the  existence  of  two  new  box
formfactors,  which have not  been considered  before in  the existing
literature of New Physics theories with CLFV.

Our detailed  study has shown  that the soft SUSY-breaking  effects in
the  $Z$-boson-mediated  graphs  dominate  the CLFV  observables,  for
appreciable  regions of  the  $\nu_R$MSSM parameter  space in  mSUGRA.
Nevertheless, there  is a significant  portion of parameter  space for
heavy  neutrino  masses  $m_N\stackrel{<}{{}_\sim} 1$~TeV,  where  box
diagrams involving heavy  neutrinos in the loop can  be comparable to,
or even  larger than the corresponding  $Z$-boson-exchange diagrams in
$\mu   \to  eee$   and   in   $\mu  \to   e$   conversion  in   nuclei
(cf.~Fig.~\ref{Fig11}).   In   the  same  kinematic   regime,  due  to
accidental cancellations,  we have also observed a  suppression of the
branching ratios for  the photonic CLFV decays $\mu  \to e \gamma$, as
well as for  the decays $\tau \to e \gamma$  and $\tau \to \mu\gamma$.
As  mentioned in  the Introduction,  such a  suppression  in low-scale
seesaw models is a consequence  of a cancellation between particle and
sparticle contributions due to the approximate realization of the SUSY
no-go  theorem due  to  Ferrara and  Remiddi~\cite{FR74}. Instead,  in
high-scale  seesaw  models  such  cancellations  can  only  occur  for
particular   choices   of   the  neutrino-Yukawa   and   Majorana-mass
textures~\cite{AH05,EHLR}.    Hence,  the   results   obtained  within
supersymmetric   low-scale    seesaw   type-I   models,    with   $m_N
\stackrel{<}{{}_\sim}  10$~TeV,   corroborate  the  original  findings
in~\cite{IPPRD},   where   the   usual   paradigm  with   the   photon
dipole-moment operators dominating  the CLFV observables in high-scale
seesaw models~\cite{HMTY,softLFV}  gets radically modified,  such that
$\mu \to eee$ and $\mu  \to e$ conversion may also represent sensitive
probes of CLFV.

We have  found that  unlike heavy neutrinos,  CLFV effects  induced by
sneutrinos  remain subdominant  for the  entire region  of  the mSUGRA
parameter  space. In  addition, the  perturbativity constraint  on the
neutrino Yukawa  couplings~${\bf h}_\nu$ up  to the GUT  scale renders
the quartic  coupling contributions of order  $({\bf h}_\nu)^4$ small.
The present  study has focused on providing  numerical predictions for
relatively   small    and   intermediate   values    of   $\tan\beta$,
i.e.~$\tan\beta     \stackrel{<}{{}_\sim}    20$,     where    neutral
Higgs-mediated interactions constrained by the recent LHCb observation
of  the decay  $B_s  \to \mu\mu$  are  not expected  to give  sizeable
contributions.   A~global  analysis  that includes  large  $\tan\beta$
effects on  CLFV observables  and LHC constraints  will be given  in a
forthcoming communication.

\subsection*{Acknowledgements}
\vspace{-3mm}
\noindent
We   thank  George   Lafferty  for   valuable   discussions  regarding
experimental sensitivities at the intensity frontier of~CLFV. 
%and Asmaa Abada for clarifying comments.  
The    work    of    AP     is    supported    in    part    by    the
Lancaster--Manchester--Sheffield  Consortium  for Fundamental  Physics
under STFC  grant ST/J000418/1.  AP also  acknowledges partial support
by an IPPP  associateship from Durham University.  The  work of AI and
LP was  supported by  the Ministry of  Science, Sports  and Technology
under contract 119-0982930-1016.

\vfill\eject

\begin{appendix}
\renewcommand{\theequation}{\Alph{section}.\arabic{equation}}

\section{INTERACTION VERTICES}\label{vertices}

In this  appendix, we list the Lagrangians  describing the interaction
vertices required to calculate  the transition amplitudes for the CLFV
processes under study.  The corresponding interaction vertices for the
SM and the MSSM are obtained by adopting the conventions of the public
code {\tt  FeynArts-3.3.~FVMSSM.mod}.  The Lagrangians  of interest to
us include:
\medskip

\noindent
1.  Vertices from 2HDM sector of the MSSM involving SM particles only,
\begin{eqnarray}
{\cal L}_{\overline{d}uH^-} + {\rm H.c.}
 &=&
 \frac{g_w}{\sqrt{2} M_W} V^*_{ij}\: 
 \overline{d}_j
   \Big(t_\beta m_{d_j} P_L + t_\beta^{-1} m_{u_i} P_R\Big)
 u_i H^- + {\rm H.c.}
\end{eqnarray}
Here  $H^-$  is  the  negatively  charged Higgs  scalar,  $V$  is  the
Cabibbo--Kobayashi--Maskawa matrix, $m_{d_i}$  and $m_{u_i}$ are the quark
masses and $c_w=\cos\theta_w$.
\medskip

\noindent
2. Vertices of singlet neutrinos in the $\nu_R$SM sector of the MSSM,
\begin{eqnarray}
{\cal L}_{\overline{e}nG^-} + {\rm H.c.} 
 &=&
 \frac{g_w}{\sqrt{2} M_W} B_{ia}\: 
 \overline{e}_i
   \Big( - m_{e_i} P_L + m_{n_a} P_R \Big)
 n_a G^- + {\rm H.c.}\ ,
\\
{\cal L}_{\overline{e}nW^-} + {\rm H.c.}
 &=&
 -\frac{g_w}{\sqrt{2}} B_{ia}\: 
 \overline{e}_i
  \gamma^\mu P_L
 n_a W^-_\mu + {\rm H.c.}\ ,
\\
{\cal L}_{\overline{n}nZ} 
 &=&
 -\frac{g_w}{2c_W} C_{ab}\: 
 \overline{n}_a
   \gamma^\mu P_L
 n_b\, Z_\mu\ .
\end{eqnarray}
Here  $n_a$ and  $m_{n_a}$  denote the  neutrino mass-eigenstates  and
their  respective masses  and $B$  and $C$  are lepton  flavour mixing
matrices  defined in~\cite{AZPC}  and~\cite{IPNPB}.  The  matrices $B$
and $C$ satisfy the following set of relations:
\begin{eqnarray}
B_{la}B_{l'a}^* &=& \delta_{ll'}, \qquad C_{ac}C_{bc} \ =\ C_{ab},
\qquad B_{lb}C_{ba} \ =\ B_{la},\qquad B_{la}^* B_{lb} \ =\ C_{ab},
\nonumber\\
m_a C_{ac} C_{bc} &=& 0, \qquad m_a B_{lb} C^*_{ba} \ =\ 0,\qquad m_a
B_{la}B_{l'a}\ =\ 0\ . 
\end{eqnarray} 
\medskip

\noindent
3.   Vertices from  the 2HDM  sector  of the  MSSM involving  Majorana
neutrinos,
\begin{eqnarray}
{\cal L}_{\overline{e} n H^-} + {\rm H.c.}
 &=&
 \frac{g_w}{\sqrt{2} M_W} B_{ia} \:
 \overline{e}_i
  \Big(t_\beta m_{e_i} P_L + t_\beta^{-1} m_{n_a} P_R\Big)
 n_a H^- + {\rm H.c.}
\end{eqnarray}
4. MSSM vertices with sparticles,
\begin{eqnarray}
{\cal L}_{\overline{d} \tilde{\chi}^- \tilde{u}} + {\rm H.c.}
 &=&
 g_w\, \overline{d}_j \Big(\tilde{V}_{jm a}^{-dL} P_L + \tilde{V}_{jm
   a}^{-dR} P_R\Big) \tilde{\chi}_m ^- \tilde{u}_a + {\rm H.c.}\ , 
\\
{\cal L}_{\overline{u} \tilde{\chi}^+ \tilde{d}} + {\rm H.c.}
 &=&
 g_w\, \overline{u}_j \Big(\tilde{V}_{jm a}^{+uL} P_L + \tilde{V}_{jm
   a}^{+uL} P_R\Big) \tilde{\chi}_m ^+ \tilde{d}_a + {\rm H.c.}\ ,  
\\
{\cal L}_{\overline{\tilde{\chi}^-}\tilde{\chi}^-A} + {\cal
  L}_{\overline{\tilde{\chi}^+}\tilde{\chi}^+A} 
 &=& 
   e\:\overline{\tilde{\chi}^-_m} \gamma^\mu \tilde{\chi}^-_m A_\mu  
 - e\: \overline{\tilde{\chi}^+_m} \gamma^\mu \tilde{\chi}^+_m A_\mu, 
\\ 
{\cal L}_{\overline{\tilde{\chi}^-}\tilde{\chi}^-Z} + {\cal
  L}_{\overline{\tilde{\chi}^+}\tilde{\chi}^+Z} 
 &=&
\frac{g_w}{c_W}\,
\overline{\tilde{\chi}_m^-}
 \gamma^\mu \Big( V_{mk}^{\tilde{\chi}^-L} P_L +
 V_{mk}^{\tilde{\chi}^-R} P_R \Big)\tilde{\chi}_k^- Z_\mu 
\nonumber\\
&-&
\frac{g_w}{c_W}\,
\overline{\tilde{\chi}_m^+} 
 \gamma^\mu \Big( V_{mk}^{\tilde{\chi}^-L*} P_R +
 V_{mk}^{\tilde{\chi}^-R*} P_L \Big) \tilde{\chi}_k^+ Z_\mu\ , 
\\
% change 24112012
{\cal L}_{\bar{\tilde{\chi}}^0 \tilde{\chi}^0 Z}
 &=&
 \frac{g}{c_w} \bar{\tilde{\chi}}^0_m
 ( \gamma^\mu P_L V_{mk}^{\tilde{\chi}^0 L}
 + \gamma^\mu P_R V_{mk}^{\tilde{\chi}^0 R})
 \tilde{\chi}^0_k Z_\mu
\\
{\cal L}_{\tilde{e}^*\tilde{e}Z}
 &=&
 g_w \tilde{V}^{\tilde{e}}_{ab}\; \tilde{e}_a^*\, 
 i \hspace{-3pt}
 \stackrel{\leftrightarrow}{\partial}\hspace{-.25em}\vspace{.6em}^\mu 
\hspace{-2pt}\,  
 \tilde{e}_b\: Z_\mu\ , 
\\
{\cal L}_{\bar{e}\tilde{\chi}^0\tilde{e}} + {\cal
  L}_{\overline{\tilde{\chi}^0} e \tilde{e}^*} 
 &=&
g_w \,
  \overline{e}_j (P_L \tilde{V}_{jma}^{0 e L} + P_R \tilde{V}_{jma}^{0
    e R})\tilde{\chi}^0 \tilde{e}_a + {\rm H.c.}\ ,  
\\
{\cal L}_{\bar{u}\tilde{\chi}^0\tilde{u}} + {\cal
  L}_{\overline{\tilde{\chi}^0} u \tilde{u}^*} 
 &=&
g_w \,
  \overline{u}_j (P_L \tilde{V}_{jma}^{0 u L} + P_R \tilde{V}_{jma}^{0
    u R})\tilde{\chi}^0 \tilde{u}_a + {\rm H.c.}\ , 
\\
{\cal L}_{\bar{d}\tilde{\chi}^0\tilde{d}} + {\cal
  L}_{\overline{\tilde{\chi}^0} d \tilde{d}^*} 
 &=&
g_w \,
  \overline{d}_j (P_L \tilde{V}_{jma}^{0 d L} + P_R \tilde{V}_{jma}^{0
    d R})\tilde{\chi}^0 \tilde{d}_a + {\rm H.c.}\ , 
\end{eqnarray}
where 
\begin{eqnarray}
\tilde{V}_{jm a}^{-dL}
 &=&
 \frac{m_{d_j}}{\sqrt{2} c_\beta M_W} {\cal U}_{m 2}^* 
  V^*_{ij} (R_L^{\tilde{u}})^*_{ai}\ ,
\nonumber\\
\tilde{V}_{jm a}^{-dR} 
 &=&
 - {\cal V}_{m 1} V^*_{ij} (R_L^{\tilde{u}})^*_{ai}
 + \frac{m_{u_i}}{\sqrt{2} s_\beta M_W} {\cal V}_{m 2} 
   V^*_{ij} (R_R^{\tilde{u}})^*_{ai}\ ,
\\
\tilde{V}_{jm a}^{+uL}
 &=&
 \frac{m_{u_j}}{\sqrt{2} s_\beta M_W} {\cal V}_{m 2}^* V_{ji}
 (R_L^{\tilde{d}})^*_{ai}\ , 
\nonumber\\
\tilde{V}_{jm a}^{+uR}
 &=&
 - {\cal U}_{m 1} V_{ji} (R_L^{\tilde{d}})^*_{ai}
 + \frac{m_{d_i}}{\sqrt{2} c_\beta M_W} {\cal U}_{m 2} V_{ji}
 (R_R^{\tilde{d}})^*_{ai}\ , 
\\
V_{mk}^{\tilde{\chi}^-L} 
 &=&
   {\cal U}_{m 1} {\cal U}^*_{k 1}
 + \frac{1}{2} {\cal U}_{m 2} {\cal U}^*_{k 2}
 - \delta_{m k} s_w^2,
\nonumber\\
V_{mk}^{\tilde{\chi}^-R}
 &=&
   {\cal V}^*_{m 1} {\cal V}_{k 1}
 + \frac{1}{2} {\cal V}^*_{m 2} {\cal V}_{k 2}
 - \delta_{m k} s_w^2\ ,
\\
% change 24112012
V_{mk}^{\tilde{\chi}^0 L}
 &=&
 -\frac{1}{4} (Z_{m3} Z^*_{k3}-Z_{m4} Z^*_{k4})\ ,
\\
V_{mk}^{\tilde{\chi}^0 R}
 &=&
 \frac{1}{4} (Z^*_{m3} Z_{k3}-Z^*_{m4} Z_{k4})\ ,
\\
\tilde{V}^{\tilde{e}}_{ab}
 &=&
 \frac{c_{2w}}{c_w} 
 ( R^{\tilde{e}}_L)_{ai} (R^{\tilde{e}}_L)_{bi}^* 
 - \frac{s_w^2}{c_w} (R^{\tilde{e}}_R)_{ai} (R^{\tilde{e}}_R)_{bi}^* \ ,
\\
\tilde{V}^{0 \ell L}_{j m a}
 &=&
 -\sqrt{2} t_w Z^*_{m1} (R_R^{\tilde{e}})^*_{aj}
% change 24112012 Z_{m3} -> Z^*_{m3}
 - \frac{(m_e)_j}{\sqrt{2} c_\beta M_W} Z^*_{m3}
 (R_L^{\tilde{e}})^*_{aj}\ ,
\\
\tilde{V}^{0 \ell R}_{j m a}
 &=&
\frac{1}{\sqrt{2} c_W}(c_W Z_{m2} +s_W Z_{m1}) (R^{\tilde{e}}_L)^*_{aj}
- \frac{(m_e)_j}{\sqrt{2}c_\beta M_W} Z_{m3}
(R_R^{\tilde{e}})^*_{aj}\ ,
\\
\tilde{V}^{0 u L}_{j m a}
 &=&
 \frac{2\sqrt{2}}{3} t_w Z^*_{m1} (R_R^{\tilde{u}})^*_{aj}
 - \frac{(m_u)_j}{\sqrt{2} s_\beta M_W} Z^*_{m4}
 (R_L^{\tilde{u}})^*_{aj}\ ,
\\
\tilde{V}^{0 u R}_{j m a}
 &=&
- \frac{1}{2 c_W}(c_W Z_{m2} +\frac{1}{3} s_W Z_{m1}) (R^{\tilde{u}}_L)^*_{aj}
- \frac{(m_u)_j}{\sqrt{2}s_\beta M_W} Z_{m4}
(R_R^{\tilde{u}})^*_{aj}\ ,
\\
\tilde{V}^{0 d L}_{j m a}
 &=&
 -\frac{\sqrt{2}}{3} t_w Z^*_{m1} (R_R^{\tilde{d}})^*_{aj}
 - \frac{(m_d)_j}{\sqrt{2} c_\beta M_W} Z_{m3}
 (R_L^{\tilde{d}})^*_{aj}\ ,
\\
\tilde{V}^{0 d R}_{j m a}
 &=&
\frac{1}{2 c_W}(c_W Z_{m2} -\frac{1}{3} s_W Z_{m1}) (R^{\tilde{d}}_L)^*_{aj}
- \frac{(m_d)_j}{\sqrt{2}c_\beta M_W} Z_{m3}
(R_R^{\tilde{d}})^*_{aj}\ ,
\end{eqnarray}
and  $c_{2w} = \cos{2\theta_w}$.   The unitary  matrices diagonalizing
the chargino  mass matrix  ${\cal U}$ and  ${\cal V}$ and  the unitary
matrix  diagonalizing  the  neutralino   mass  matrix  $Z$  are  taken
from~\cite{DGR_Dress}. The matrices
\begin{equation}
R^{\tilde{f}L}_{ak}\ \equiv\ U^{\tilde{f}}_{ia} U^{f_L*}_{ik}\;,\qquad  
R^{\tilde{f}R}_{ak}\ \equiv\ U^{\tilde{f}}_{i+3\,a} U^{f_R*}_{ik}\;,
\end{equation}
with $f = d,\, u,\,  e$, $a=1,2,\dots,6$ and $i,k=1,2,3$, quantify the
disalignment   between  fermions   and  sfermions.    Here  $U^{f_L}$,
$U^{f_R}$  and $U^{\tilde{f}}$ are  unitary matrices  that diagonalize
the fermion and sfermion mass matrices, respectively.
\medskip

\noindent
5. Sneutrino vertices in the $\nu_R$MSSM,
\begin{eqnarray} 
{\cal L}_{\overline{e}\tilde{\chi}^- \tilde{N}} + {\rm H.c.}
 &=&
 g_w \tilde{N}_A \overline{\ell}_l \Big(P_L  \frac{m_l}{\sqrt{2}
   c_\beta M_W} \tilde{B}_{lm A}^{L,1} 
  + P_R \tilde{B}_{lm A}^{R,1}\Big) \tilde{\chi}_m^- + {\rm H.c.}
\nonumber\\ 
 &=&
 g_w \tilde{N}_A^* \overline{\ell}_l \Big(P_L  \frac{m_l}{\sqrt{2}
   c_\beta M_W} \tilde{B}_{lm A}^{L,2} 
  + P_R \tilde{B}_{lm A}^{R,2}\Big) \tilde{\chi}_m^- + {\rm H.c.}\ ,
\\
{\cal L}_{\tilde{N}\tilde{N}Z}
 &=&
 \frac{g_w}{c_W} \tilde{C}^1_{AB}\:
 \tilde{N}_A^* i  
 \hspace{-2pt}
 \stackrel{\leftrightarrow}{\partial}\hspace{-.25em}\vspace{.6em}^\mu 
\hspace{-2pt}\, 
 \tilde{N}_B Z_\mu 
\nonumber\\
 &=&
 \frac{g_w}{c_W} \tilde{C}^2_{AB}\:
 \tilde{N}_A^* i 
 \hspace{-2pt}
 \stackrel{\leftrightarrow}{\partial}\hspace{-.25em}\vspace{.6em}^\mu 
\hspace{-2pt}\, 
 \tilde{N}_B^* Z_\mu 
\nonumber\\
 &=&
 \frac{g_w}{c_W} \tilde{C}^3_{AB}\:
 \tilde{N}_A i 
 \hspace{-2pt}
 \stackrel{\leftrightarrow}{\partial}\hspace{-.25em}\vspace{.6em}^\mu 
\hspace{-2pt}\, 
 \tilde{N}_B Z_\mu 
\nonumber\\
 &=&
 \frac{g_w}{c_W} \tilde{C}^4_{AB}\: 
 \tilde{N}_A i 
 \hspace{-2pt}
 \stackrel{\leftrightarrow}{\partial}\hspace{-.25em}\vspace{.6em}^\mu 
\hspace{-2pt}\, 
 \tilde{N}_B^* Z_\mu, 
\end{eqnarray}
where
\begin{eqnarray}
%BL1
\tilde{B}_{lm A}^{L,1}
 &=&
 {\cal U}_{m 2} U^{\ell_R *}_{il} 
  {\cal U}_{i A}^{\tilde{\nu}},
\nonumber\\
%BL2
\tilde{B}_{lm A}^{L,2}
 &=&
 {\cal U}_{m 2} U^{\ell_R *}_{il} 
  {\cal U}_{i+6 A}^{\tilde{\nu}*},
\nonumber\\
%BR1
\tilde{B}_{lm A}^{R,1}
 &=&
 - {\cal U}_{i A}^{\tilde{\nu}} U^{\ell_L*}_{il}
   {\cal V}_{m 1}
 + \frac{m_{n_a}}{\sqrt{2} s_\beta M_W} 
   {\cal V}_{m 2}
   {\cal U}_{i+9 A}^{\tilde{\nu}} U^{\nu *}_{i+3 a} B_{la},
\nonumber\\
%BR2
\tilde{B}_{lm A}^{R,2}
 &=&
 - {\cal U}_{i+6 A}^{\tilde{\nu}*} U^{\ell_L*}_{il} 
   {\cal V}_{m 1}
 + \frac{m_{n_a}}{\sqrt{2} s_\beta M_W} 
   {\cal V}_{m 2}
   {\cal U}_{i+3 A}^{\tilde{\nu} *} U^{\nu *}_{i+3 a} B_{la},
\nonumber\\
%C1
\tilde{C}^1_{AB} 
 &=&
 - \frac{1}{2}
 {\cal U}_{i A}^{\tilde{\nu}*}
 {\cal U}_{i B}^{\tilde{\nu}},
\nonumber\\
%C2
\tilde{C}^2_{AB} 
 &=&
 - \frac{1}{2}
 {\cal U}_{i A}^{\tilde{\nu}*}
 {\cal U}_{i+6 B}^{\tilde{\nu}*},
\nonumber\\
%C3
\tilde{C}^3_{AB} 
 &=&
 - \frac{1}{2}
 {\cal U}_{i+6 A}^{\tilde{\nu}}
 {\cal U}_{i B}^{\tilde{\nu}},
\nonumber\\
%C4
\tilde{C}^6_{AB} 
 &=&
 - \frac{1}{2}
 {\cal U}_{i+6 A}^{\tilde{\nu}}
 {\cal U}_{i+6 B}^{\tilde{\nu}*}\ .
\end{eqnarray}
In  the   above,  ${\cal  U}^{\tilde{\nu}}$  is   the  unitary  matrix
diagonalizing the sneutrino mass matrix.

Notice that we have factored out the weak coupling constant $g_w$ from
all   interaction   vertices  defined   above.   To~better   identify
chirality-flip  mass effects  in  the CLFV  amplitudes,  we have  also
pulled out a factor  $m_l/(\sqrt{2} c_\beta M_W)$ from the interaction
vertex $\tilde{B}_{lm A}^{L}$.

\section{LOOP FUNCTIONS}\label{sec:lf}

The CLFV  amplitudes are expressed in terms  of leading-order one-loop
functions.  We expand  the loop functions with respect  to the momenta
and masses  of the  external charged leptons,  while keeping  only the
leading non-zero  terms. The leading  terms may then be  expressed, in
terms of the dimensionless loop integrals
\begin{eqnarray}
\overline{J}^m_{n_1n_2\dots n_k} (\lambda_1,\lambda_2,\dots,\lambda_k) 
&=&
\frac{(\mu^2)^{2-D/2}}{(M^2_W)^{-D/2-m+\sum_i n_i}} 
 \int\frac{d^D\ell}{(2\pi)^D} \frac{(\ell^2)^m}{\prod_{i=1}^k (\ell^2
   - m_i^2)^{n_i}} 
\nonumber\\
&=&
 \frac{i(-1)^{m-\sum_i n_i}}{(4\pi)^{D/2}\Gamma(\frac{D}{2})}
 \Big(\frac{\mu^2}{M_W^2}\Big)^{2-D/2} 
 \int_0^\infty \frac{dx x^{D/2-1+m}}{\prod_{i=1}^k (x+\lambda_i)^{n_i}},
\end{eqnarray}
where $m_i$ are  loop particle masses, $n_i$ are  the exponents of the
propagator  denominators,  $\lambda_i=m_i^2/M_W^2$  are  dimensionless
mass parameters  and $\mu$ is  't Hooft's renormalization  mass scale.
We choose $\mu$ to be $M_W$, even though any other scale can be chosen
equally well as  a reference scale for any of  the integrals.  For the
amplitudes  we  have  been   calculating,  the  integrals  are  either
divergent and satisfy $m+2-\sum_in_i = 0$, or they are convergent with
$m+2-\sum_in_i<0$.  For convergent integrals, we may set $D=4$, whilst
for  divergent integrals we  take $D=4-2\varepsilon$.   We pull  out a
factor  $i/(4\pi)^2$ from  all  integrals. Thus,  we  have for  finite
integrals,
\begin{equation}
\overline{J}^m_{n_1n_2\dots n_k} (\lambda_1,\lambda_2,\dots,\lambda_k) 
\ \equiv\
\frac{i}{(4\pi)^2}\ J^m_{n_1n_2\dots n_k} (\lambda_1,\lambda_2,\dots,\lambda_k)\; .
\end{equation}
Instead,  the divergent  integrals  are written  down  as a  sum of  a
divergent+constant term and a finite mass-dependent term:
\begin{equation}
\overline{J}^m_{n_1n_2\dots n_k}
(\lambda_1,\lambda_2,\dots\lambda_k)\ \equiv\ 
 \frac{i}{(4\pi)^2}\ \Big(\frac{1}{\varepsilon} + \mbox{const} +
 J^m_{n_1n_2\dots n_k} (\lambda_1,\lambda_2,\dots,\lambda_k)\Big)\; .
\end{equation}
In  the CLFV amplitudes,  the divergent+constant  terms vanish  in the
total sum,  or as  a result of  a GIM-like mechanism.   Therefore, all
CLFV amplitudes  can be  expressed in terms  of finite  mass dependent
functions  $J^m_{n_1n_2\dots}  (\lambda_1,\lambda_2,\dots)$, which  we
call  them the  {\it basic  integrals}.  The  CLFV amplitudes  and the
corresponding  formfactors considered  here are  described by  9 basic
integrals,  four for the  photonic amplitude:  $J^0_{31}$, $J^1_{31}$,
$J^1_{41}$  and   $J^2_{41}$,  three  for   the  $Z$-boson  amplitude:
$J^0_{11}$, $J^0_{111}$  and $J^1_{111}$, and two  for box amplitudes:
$J^0_{1111}$ and $J^1_{1111}$.

\vfill\eject

\section{ONE--LOOP FORMFACTORS}\label{sec:olff}

Here we present  the complete analytical form of  the CLFV formfactors
$F_\gamma$, $F_Z$ and $F_{\rm  box}$ defined in Section~\ref{CLFV}, in
the Feynman--'t  Hooft gauge.  In  the following, the  usual summation
convention over repeated indices is implied.  The interaction vertices
and loop  functions used  here are given  in Appendices~\ref{vertices}
and \ref{sec:lf}, respectively.

\subsection{Photon Formfactors}

The   formfactors   $F_\gamma^L$,   $F_\gamma^R$,   $G_\gamma^L$   and
$G_\gamma^R$ may be explicitly written as follows:
\begin{eqnarray}
(F_\gamma^L)_{l'l}
 &=&
 F^N_{l'l\gamma} + F^{L,\tilde{N}}_{l'l\gamma}
 + F^{L,{\rm SB}}_{l'l\gamma}\ ,
\nonumber\\
(F_\gamma^R)_{l'l}
 &=&
 F^N_{l'l\gamma} + F^{R,\tilde{N}}_{l'l\gamma}
 + F^{R,{\rm SB}}_{l'l\gamma}\ ,
\nonumber\\
(G_\gamma^L)_{l'l} 
 &=& 
 m_{l'} (G^N_{l'l\gamma} + G^{L,\tilde{N}}_{l'l\gamma})
 + G^{L,{\rm SB}}_{l'l\gamma}\ ,
\nonumber\\
(G_\gamma^R)_{l'l}
 &=&
 m_{l} (G^N_{l'l\gamma} + G^{R,\tilde{N}}_{l'l\gamma})
 + G^{R,{\rm SB}}_{l'l\gamma}\ ,
\end{eqnarray}
where
\begin{eqnarray}
F_{l'l\gamma}^{N}
 &=&
 B_{l' a} B^*_{l a} 
 \bigg[ 2 \bigg(J^1_{31}(1,\lambda_{n_a}) -
   \frac{1}{6}J^2_{41}(1,\lambda_{n_a})\bigg) 
    - \frac{1}{6} \lambda_{n_a} J^2_{41}(1,\lambda_{n_a})
    - \frac{1}{6 t^2_\beta} \lambda_{n_a} J^2_{41}(\lambda_{H^+},\lambda_{n_a})
 \bigg], 
\nonumber\\
G_{l'l\gamma}^{N}
 &=&
 B_{l' a} B^*_{l a}
 \bigg[ J^1_{31}(1,\lambda_{n_a}) + J^2_{41}(1,\lambda_{n_a})
    + \lambda_{n_a} \bigg(\frac{1}{2} J^1_{41}(1,\lambda_{n_a}) -
    J^0_{31}(1,\lambda_{n_a})\bigg) 
\nonumber\\
   &+& \lambda_{n_a} \lambda_{H^+} 
      \bigg( \frac{1}{2t^2_\beta}
      J^1_{41}(\lambda_{H^+},\lambda_{n_a}) +
      J^0_{31}(\lambda_{H^+},\lambda_{n_a}) \bigg) 
 \bigg],
\\
F_{l'l\gamma}^{L,\tilde{N}}
 &=&
 \frac{1}{2}(\tilde{B}^{R,1}_{l' k A} \tilde{B}^{R*}_{l k A} +
 \tilde{B}^{R,2}_{l' k A} \tilde{B}^{R,2*}_{l k A}) 
 \bigg[
   - \frac{2}{3} J^2_{41}(\lambda_{\tilde{\chi}_k},\lambda_{\tilde{N}_A})
   + \lambda_{\tilde{\chi}_k} J^1_{41}(\lambda_{\tilde{\chi}_k},\lambda_{\tilde{N}_A})
 \bigg], 
\nonumber\\
F_{l'l\gamma}^{R,\tilde{N}}
 &\equiv&
 \frac{m_l m_{l'}}{4 c_\beta^2 M_W^2}
 (\tilde{B}^{L,1}_{l' k A} \tilde{B}^{L,1*}_{l k A} +
 \tilde{B}^{L,2}_{l' k A} \tilde{B}^{L,2*}_{l k A}) 
 \bigg[
   - \frac{2}{3} J^2_{41}(\lambda_{\tilde{\chi}_k},\lambda_{\tilde{N}_A})
   + \lambda_{\tilde{\chi}_k} J^1_{41}(\lambda_{\tilde{\chi}_k},\lambda_{\tilde{N}_A})
 \bigg], 
\nonumber\\
G_{l'l\gamma}^{L,\tilde{N}}
 &=&
 \frac{1}{2} (\tilde{B}^{L,1}_{l' k A} \tilde{B}^{L,1*}_{l k A} +
 \tilde{B}^{L,2}_{l' k A} \tilde{B}^{L,2*}_{l k A}) 
 \bigg[
   - \frac{m_l^2}{2c^2_\beta M_W^2}
   \ \lambda_{\tilde{\chi_k}} J^1_{41}(\lambda_{\tilde{\chi}_k},\lambda_{\tilde{N}_A})
 \bigg] 
\nonumber\\
 &+&
 \frac{1}{2}(\tilde{B}^{R,1}_{l' k A} \tilde{B}^{R,1*}_{l k A} +
 \tilde{B}^{R,2}_{l' k A} \tilde{B}^{R,2*}_{l k A}) 
 \bigg[
   - \lambda_{\tilde{\chi}_k}
   J^1_{41}(\lambda_{\tilde{\chi}_k},\lambda_{\tilde{N}_A})
 \bigg]
\nonumber\\
 &+&
 \frac{1}{2}(\tilde{B}^{29L,1}_{l' k A} \tilde{V}^{R,1*}_{l k A} +
 \tilde{B}^{L,2}_{l' k A} \tilde{V}^{R,2*}_{l k A}) 
 \bigg[
   \frac{\sqrt{2}}{c_\beta} \sqrt{\lambda_{\tilde{\chi_k}}}
   J^1_{31}(\lambda_{\tilde{\chi}_k},\lambda_{\tilde{N}_A})
 \bigg],
\nonumber\\
G_{l'l\gamma}^{R,\tilde{N}}
 &=&
 \frac{1}{2}(\tilde{B}^{L,1}_{l' k A} \tilde{B}^{L,1*}_{l k A} +
 \tilde{B}^{L,2}_{l' k A} \tilde{B}^{L,2*}_{l k A}) 
 \bigg[
   - \frac{m_{l'}^2}{2c^2_\beta M_W^2}
   \ \lambda_{\tilde{\chi_k}} J^1_{41}(\lambda_{\tilde{\chi}_k},\lambda_{\tilde{N}_A})
 \bigg]
\nonumber\\
 &+&
 \frac{1}{2}(\tilde{B}^{R,1}_{l' k A} \tilde{B}^{R,1*}_{l k A} +
 \tilde{B}^{R,2}_{l' k A} \tilde{B}^{R,2*}_{l k A}) 
 \bigg[
   - \lambda_{\tilde{\chi}_k}
   J^1_{41}(\lambda_{\tilde{\chi}_k},\lambda_{\tilde{N}_A})
 \bigg]
\nonumber\\
 &+&
 \frac{1}{2}(\tilde{B}^{R,1}_{l' k A} B^{L,1*}_{l k A} +
 \tilde{B}^{R,2}_{l' k A} \tilde{B}^{L,2*}_{l k A}) 
 \bigg[
   \frac{\sqrt{2}}{c_\beta} \sqrt{\lambda_{\tilde{\chi_k}}}
   J^1_{31}(\lambda_{\tilde{\chi}_k},\lambda_{\tilde{N}_A})
 \bigg],
\\
F_{l'l\gamma}^{L,{\rm SB}}
 &=&
 \tilde{V}^{0 \ell R}_{l' m a} \tilde{V}^{0 \ell R*}_{l m a}
 \bigg[
   -\frac{1}{3} J^2_{41}(\lambda_{\tilde{e}_a},\lambda_{\tilde{\chi}^0_m})
 \bigg], 
\nonumber\\
F_{l'l\gamma}^{R,{\rm SB}}
 &=&
 \tilde{V}^{0 \ell L}_{l' m a} \tilde{V}^{0 \ell L*}_{l m a}
 \bigg[
   -\frac{1}{3} J^2_{41}(\lambda_{\tilde{e}_a},\lambda_{\tilde{\chi}^0_m})
 \bigg],
\nonumber\\
G_{l'l\gamma}^{L,{\rm SB}}
 &=&
 \tilde{V}^{0 \ell R}_{l' m a} \tilde{V}^{0 \ell R*}_{l m a}
 \bigg[
   m_{l'} \lambda_{\tilde{e}_a}
   J^1_{41}(\lambda_{\tilde{e}_a},\lambda_{\tilde{\chi}^0_m}) 
 \bigg]
 +
 \tilde{V}^{0 \ell L}_{l' m a} \tilde{V}^{0 \ell L*}_{l m a}
 \bigg[
   m_{l} \lambda_{\tilde{e}_a} J^1_{41}(\lambda_{\tilde{e}_a},\lambda_{\tilde{\chi}^0_m})
 \bigg]
\nonumber\\
 &+&
 \tilde{V}^{0 \ell L}_{l' m a} \tilde{V}^{0 \ell R*}_{l m a}
 \bigg[
   + 2 m_{\tilde{\chi}^0_m} \lambda_{\tilde{e}_a}
   J^0_{31}(\lambda_{\tilde{e}_a},\lambda_{\tilde{\chi}^0_m}) 
 \bigg],
\nonumber\\
G_{l'l\gamma}^{R,{\rm SB}}
 &=&
 \tilde{V}^{0 \ell L}_{l' m a} \tilde{V}^{0 \ell L*}_{l m a}
 \bigg[
   m_{l'} \lambda_{\tilde{e}_a}
   J^1_{41}(\lambda_{\tilde{e}_a},\lambda_{\tilde{\chi}^0_m}) 
 \bigg]
 +
 \tilde{V}^{0 \ell R}_{l' m a} \tilde{V}^{0 \ell R*}_{l m a}
 \bigg[
   m_{l} \lambda_{\tilde{e}_a} J^1_{41}(\lambda_{\tilde{e}_a},\lambda_{\tilde{\chi}^0_m})
 \bigg] 
\nonumber\\
 &+&
 \tilde{V}^{0 \ell R}_{l' m a} \tilde{V}^{0 \ell L*}_{l m a}
 \bigg[
   + 2 m_{\tilde{\chi}^0_m} \lambda_{\tilde{e}_a}
   J^0_{31}(\lambda_{\tilde{e}_a},\lambda_{\tilde{\chi}^0_m}) 
 \bigg] .
% change 24112012
\label{gammaSB}
\end{eqnarray}

\subsection{{\boldmath $Z$}-Boson Formfactors}

The formfactors $F_Z^L$ and $F_Z^R$ may be decomposed as follows:
\begin{eqnarray}
(F_Z^L)_{l'l}
 &=&
 F^N_{l'lZ} + F^{L,\tilde{N}}_{l'lZ}
 + F^{L,{\rm SB}}_{l'lZ}\ ,
\nonumber\\
(F_\gamma^R)_{l'l}
 &=&
 F^N_{l'lZ} + F^{R,\tilde{N}}_{l'lZ}
 + F^{R,{\rm SB}}_{l'lZ}\ ,
\end{eqnarray}
where
\begin{eqnarray}
F_{l'l Z}^{L,N}
 &=&
 B_{l'a} B^*_{l a} 
 \bigg[
   \frac{5}{2} \lambda_{n_a} J^0_{21}(1,\lambda_{n_a}) 
 \bigg]
\nonumber\\
 &+&
 B_{l'b} C_{ba} B^*_{l a}
 \bigg[
   - \frac{1}{2} J^1_{111}(1,\lambda_{n_b},\lambda_{n_a})
   + \frac{1}{2} \lambda_{n_a}\lambda_{n_b} J^0_{111}(1,\lambda_{n_b},\lambda_{n_a})
   +\frac{1}{2 t^2_\beta} \lambda_{n_a}\lambda_{n_b}
   J^0_{111}(\lambda_{H^+},\lambda_{n_b},\lambda_{n_a}) 
 \bigg],
\nonumber\\
F_{l'l Z}^{R,N}
 &=&
 - \frac{m_l m_{l'} t^2_\beta}{4 M_W^2}\ 
 B_{l'b} C_{ba} B^*_{l a}
   J^1_{111}(\lambda_{H^+},\lambda_{n_b},\lambda_{n_a}),
\\
F_{l'l Z}^{L,\tilde{N}}
 &=&
 \frac{1}{2}
 ( \tilde{B}^{R,1}_{l' m A} \tilde{V}^{\tilde{\chi}^- R}_{mk} \tilde{B}^{R,1*}_{l k A} 
 + \tilde{B}^{R,2}_{l' m A} \tilde{V}^{\tilde{\chi}^- R}_{mk} \tilde{B}^{R,2*}_{l k A} )
  J^1_{111}(\lambda_{\tilde{\chi}_m},\lambda_{\tilde{\chi}_k},\lambda_{\tilde{N}_A}) 
\nonumber\\
 &-&
  ( \tilde{B}^{R,1}_{l' m A} \tilde{V}^{\tilde{\chi}^- L}_{mk} \tilde{B}^{R,1*}_{l m A} 
  + \tilde{B}^{R,2}_{l' m A} \tilde{V}^{\tilde{\chi}^- L}_{mk} \tilde{B}^{R,2*}_{l m A})
  \sqrt{\lambda_{\tilde{\chi}_m}\lambda_{\tilde{\chi}_k}}
  J^0_{111}(\lambda_{\tilde{\chi}_m},\lambda_{\tilde{\chi}_k},\lambda_{\tilde{N}_A})
\nonumber\\
 &+&
 \frac{1}{2}
 ( \tilde{B}^{R,1}_{l' m A}  \tilde{B}^{R,1*}_{l k A} 
 + \tilde{B}^{R,2}_{l' m A}  \tilde{B}^{R,2*}_{l k A} )
 \Big(\frac{1}{2} - s_w^2\Big) 
 ( J^1_{21}(\lambda_{\tilde{\chi}_k},\lambda_{\tilde{N}_A})
 - 2 J^0_{11}(\lambda_{\tilde{\chi}_k},\lambda_{\tilde{N}_A}) )
\nonumber\\
 &+&
 \frac{1}{4}
 ( \tilde{B}^{R,1}_{l' k A} \tilde{C}^{1}_{BA} \tilde{B}^{R,1*}_{l k A}
 + \tilde{B}^{R,1}_{l' k A} \tilde{C}^{2}_{BA} \tilde{B}^{R,2*}_{l k A}
 + \tilde{B}^{R,2}_{l' m A} \tilde{C}^{3}_{BA} \tilde{B}^{R,1*}_{l k A}
\nonumber\\
 &&\  
 + \tilde{B}^{R,2}_{l' m A} \tilde{C}^{4}_{BA} \tilde{B}^{R,2*}_{l k A} )
  J^1_{111}(\lambda_{\tilde{\chi}_k},\lambda_{\tilde{N}_B},\lambda_{\tilde{N}_A}),
\nonumber\\
F_{l'l Z}^{R,\tilde{N}}
 &=&
 \frac{m_l m_{l'}}{2 s_\beta^2 M_W^2}
 \bigg[
 \frac{1}{2}
 ( \tilde{B}^{L,1}_{l' m A} \tilde{V}^{\tilde{\chi}^- L}_{mk} \tilde{B}^{L,1*}_{l k A} 
 + \tilde{B}^{L,2}_{l' m A} \tilde{V}^{\tilde{\chi}^- L}_{mk} \tilde{B}^{L,2*}_{l k A} )
  J^1_{111}(\lambda_{\tilde{\chi}_m},\lambda_{\tilde{\chi}_k},\lambda_{\tilde{N}_A})
\nonumber\\
 &-&
  ( \tilde{B}^{L,1}_{l' m A} \tilde{V}^{\tilde{\chi}^- R}_{mk} \tilde{B}^{L,1*}_{l m A} 
  + \tilde{B}^{L,2}_{l' m A} \tilde{V}^{\tilde{\chi}^- R}_{mk} \tilde{B}^{L,2*}_{l m A})
  \sqrt{\lambda_{\tilde{\chi}_m}\lambda_{\tilde{\chi}_k}}
  J^0_{111}(\lambda_{\tilde{\chi}_m},\lambda_{\tilde{\chi}_k},\lambda_{\tilde{N}_A})
\nonumber\\
 &+&
 \frac{1}{2}
 ( \tilde{B}^{L,1}_{l' m A}  \tilde{B}^{L,1*}_{l k A} 
 + \tilde{B}^{L,2}_{l' m A}  \tilde{B}^{L,2*}_{l k A} ) ( - s_w^2)
 ( J^1_{21}(\lambda_{\tilde{\chi}_k},\lambda_{\tilde{N}_A})
 - 2 J^0_{11}(\lambda_{\tilde{\chi}_k},\lambda_{\tilde{N}_A}) )
\nonumber\\
 &+&
 \frac{1}{4}
 ( \tilde{B}^{L,1}_{l' k A} \tilde{C}^{1}_{BA} \tilde{B}^{L,1*}_{l k A}
 + \tilde{B}^{L,1}_{l' k A} \tilde{C}^{2}_{BA} \tilde{B}^{L,2*}_{l k A}
 + \tilde{B}^{L,2}_{l' m A} \tilde{C}^{3}_{BA} \tilde{B}^{L,1*}_{l k A}
 \nonumber\\
 &&+\  
 \tilde{B}^{L,2}_{l' m A} \tilde{C}^{4}_{BA} \tilde{B}^{L,2*}_{l k A} )
  J^1_{111}(\lambda_{\tilde{\chi}_k},\lambda_{\tilde{N}_B},\lambda_{\tilde{N}_A})
 \bigg],
% change 24112012
\label{ZNt}
\\
F_{l'l Z}^{L,{\rm SB}}
 &=&
 - 
  \tilde{V}^{0\ell R}_{l' m a} \tilde{V}^{\tilde{\chi}^0 R}_{m k}
  \tilde{V}^{0\ell R*}_{l k a} 
   J^1_{111}(\lambda_{\tilde{\chi}^0_m},\lambda_{\tilde{\chi}^0_k},\lambda_{\tilde{e}_a}) 
 +
  \tilde{V}^{0\ell R}_{l' m a} \tilde{V}^{\tilde{\chi}^0 L}_{m k}
  \tilde{V}^{0\ell R*}_{l k a} 
  \ 2\sqrt{\lambda_{\tilde{\chi}_m}\lambda_{\tilde{\chi}_k}}
   J^0_{111}(\lambda_{\tilde{\chi}^0_m},\lambda_{\tilde{\chi}^0_k},\lambda_{\tilde{e}_a})
\nonumber\\
 &+& 
  \tilde{V}^{0\ell R}_{l' k a} \tilde{V}^{0\ell R*}_{l k a} 
 \Big(\frac{1}{2} - s_w^2\Big)
 \Big( 
   -\frac{1}{2} J^1_{21}(\lambda_{\tilde{\chi}^0_k},\lambda_{\tilde{e}_a})
   + J^0_{11}(\lambda_{\tilde{\chi}^0_k},\lambda_{\tilde{e}_a})
 \Big)
 - \frac{1}{2}
 \tilde{V}^{0\ell L}_{l' k b} \tilde{V}^{\tilde{e}}_{ba} \tilde{V}^{0\ell L*}_{l k a}
   J^1_{111}(\lambda_{\tilde{\chi}^0_k},\lambda_{\tilde{e}_b},\lambda_{\tilde{e}_a})\ ,
\nonumber\\
F_{l'l Z}^{R,{\rm SB}}
 &=&
 -
  \tilde{V}^{0\ell L}_{l' m a} \tilde{V}^{\tilde{\chi}^0 L}_{m k}
  \tilde{V}^{0\ell L*}_{l k a} 
   J^1_{111}(\lambda_{\tilde{\chi}^0_m},\lambda_{\tilde{\chi}^0_k},\lambda_{\tilde{e}_a})
 +
  \tilde{V}^{0\ell L}_{l' m a} \tilde{V}^{\tilde{\chi}^0 R}_{m k}
  \tilde{V}^{0\ell L*}_{l k a} 
  \ 2\sqrt{\lambda_{\tilde{\chi}_m}\lambda_{\tilde{\chi}_k}}
   J^0_{111}(\lambda_{\tilde{\chi}^0_m},\lambda_{\tilde{\chi}^0_k},\lambda_{\tilde{e}_a})
\nonumber\\
 &+&
  \tilde{V}^{0\ell L}_{l' k a} \tilde{V}^{0\ell L*}_{l k a}
 (- s_w^2)
 \Big(
   -\frac{1}{2} J^1_{21}(\lambda_{\tilde{\chi}^0_k},\lambda_{\tilde{e}_a})
   + J^0_{11}(\lambda_{\tilde{\chi}^0_k},\lambda_{\tilde{e}_a})
 \Big)
 - \frac{1}{2}
 \tilde{V}^{0\ell R}_{l' k b} \tilde{V}^{\tilde{e}}_{ba} \tilde{V}^{0\ell R*}_{l k a}
   J^1_{111}(\lambda_{\tilde{\chi}^0_k},\lambda_{\tilde{e}_b},\lambda_{\tilde{e}_a})\ .
% change 24112012
\label{ZSB}
\end{eqnarray}

\vfill\eject

\subsection{Leptonic Box Formfactors}

The  leptonic box  amplitudes are  expressed  in terms  of the  chiral
structures:   $\bar{l}'\Gamma_A^X    l\   \bar{l}_1\Gamma_A^Y   l^C_2$
[cf.~(\ref{Tbl_2}].   There  are  two  distinct contributions  to  the
chiral amplitudes.   The first one  has {\it direct} relevance  to the
original  structure given  above and  we  denote it  with a  subscript
$D$. The second contribution comes from a chiral amplitude of the form
$\bar{l}_1\Gamma_A^X  l\ \bar{l}'\Gamma_A^Y l^C_2$,  which contributes
to the  original amplitude $\bar{l}'\Gamma_A^X  l\ \bar{l}_1\Gamma_A^Y
l^C_2$,    after   performing    a    Fierz   transformation.     This
Fierz-transformed contribution is indicated with a subscript $F$. More
explicitly, the leptonic box formfactors are given by
\begin{eqnarray}
 B_{\ell V}^{LL} &=& B_{\ell V,D}^{LL} + B_{\ell V,F}^{LL},
 \qquad\qquad\qquad\quad\ \,
 B_{\ell V}^{RR} \ =\ B_{\ell V,D}^{RR} + B_{\ell V,F}^{RR},
\nonumber\\
 B_{\ell V}^{LR} &=& B_{\ell V,D}^{LR} - \frac{1}{2} B_{\ell S,F}^{LR},
 \qquad\qquad\qquad\quad\! 
 B_{\ell V}^{RL} \ =\ B_{\ell V,D}^{RL} - \frac{1}{2} B_{\ell S,F}^{RL},
\nonumber\\
 B_{\ell S}^{LL} &=& B_{\ell S,D}^{LL} + \frac{1}{2} B_{\ell S,F}^{LL}
 + \frac{3}{2} B_{\ell T,F}^{LL}, 
 \qquad\
 B_{\ell S}^{RR} \ =\ B_{\ell S,D}^{RR} + \frac{1}{2} B_{\ell
   S,F}^{RR} + \frac{3}{2} B_{\ell T,F}^{RR}, 
\nonumber\\
 B_{\ell S}^{LR} &=& B_{\ell S,D}^{LR} - 2 B_{\ell V,F}^{LR},
 \qquad\qquad\qquad\quad\,
 B_{\ell S}^{RL} \ =\ B_{\ell S,D}^{RL} - 2 B_{\ell V,F}^{RL},
\nonumber\\
 B_{\ell T}^{LL} &=& B_{\ell T,D}^{LL} - \frac{1}{2} B_{\ell T,F}^{LL}
 + \frac{1}{2} B_{\ell S,F}^{LL}, 
 \qquad\
 B_{\ell T}^{RR} \ =\ B_{\ell T,D}^{RR} - \frac{1}{2} B_{\ell
   T,F}^{RR} + \frac{1}{2} B_{\ell S,F}^{RR}\ . 
\end{eqnarray}
The   {\it  direct}   and  Fierz-transformed   contributions   to  the
formfactors are related by the exchange of outgoing leptons
\begin{eqnarray}
 B_{\ell A,F}^{XY} &=& B_{\ell A,D}^{XY} (l'\leftrightarrow l_1)\ .
\end{eqnarray}
The {\it direct}  contributions have {\it direct} $N$,  ${\rm SB}$ and
Fierz-transformed $\tilde{N}$ contributions:
\begin{eqnarray}
 B_{\ell V,D}^{LL} &=& B_{\ell V,D}^{LL,N} + B_{\ell
   V,F}^{LL,\tilde{N}} + B_{\ell V,D}^{LL,{\rm SB}},  
 \qquad
 B_{\ell V,D}^{RR} \ =\ B_{\ell V,D}^{RR,{\rm SB}},
\nonumber\\
 B_{\ell V,D}^{LR} &=& B_{\ell V,D}^{LR,{\rm SB}},
\qquad\qquad\qquad\qquad\qquad\,
 B_{\ell V,D}^{RL} \ =\ -\frac{1}{2} B_{\ell S,F}^{RL,\tilde{N}} +
 B_{\ell V,D}^{RL,{\rm SB}}, 
\nonumber\\ 
 B_{\ell S,D}^{LL} &=& B_{\ell S,D}^{LL,{\rm SB}},
\qquad\qquad\qquad\qquad\qquad\,
 B_{\ell S,D}^{RR} \ =\ \frac{1}{2} B_{\ell S,F}^{RR,\tilde{N}} +
 B_{\ell S,D}^{RR,{\rm SB}}, 
\nonumber\\
 B_{\ell S,D}^{LR} &=& B_{\ell S,D}^{LR,{\rm SB}},
\qquad\qquad\qquad\qquad\qquad\,
 B_{\ell S,D}^{RL} \ =\ B_{\ell S,D}^{RL,N} - 2 B_{\ell
   V,F}^{RL,\tilde{N}} + B_{\ell S,D}^{RL,{\rm SB}}, 
\nonumber\\
 B_{\ell T,D}^{LL} &=& B_{\ell T,D}^{LL,{\rm SB}},
\qquad\qquad\qquad\qquad\qquad\,
 B_{\ell T,D}^{RR} \ =\ \frac{1}{2} B_{\ell S,F}^{RR,\tilde{N}} +
 B_{\ell T,D}^{RR,{\rm SB}}\ . 
\label{FF_NtNSB}
\end{eqnarray}
The formfactor contributions from~(\ref{FF_NtNSB}) read:
\begin{eqnarray}
%A1BN BlVDLLN
B_{\ell V,D}^{LL,N}
 &=&
 B^*_{l a} B^*_{l_2 b} B_{l' a} B_{l_1 b}
 \bigg[
  - \bigg( 1 + \frac{\lambda_{n_a} \lambda_{n_b}}{4} \bigg) 
    J^1_{211}(1,\lambda_{n_a},\lambda_{n_b})
  + 2 \lambda_{n_a} \lambda_{n_b} J^0_{211}(1,\lambda_{n_a},\lambda_{n_b})
\nonumber\\
 &-&
    2 \lambda_{n_a} \lambda_{n_b} t_\beta^{-2}
    J^0_{1111}(1,\lambda_{H^+},\lambda_{n_a},\lambda_{n_b}) 
  - \frac{1}{2} \lambda_a \lambda_b t_\beta^{-2}
  J^1_{1111}(1,\lambda_{H^+},\lambda_{n_a},\lambda_{n_b}) 
\nonumber\\
 &-& \frac{1}{4} \lambda_a \lambda_b t_\beta^{-4}
J^1_{211}(\lambda_{H^+},\lambda_{n_a},\lambda_{n_b}) 
 \bigg],
\nonumber\\
%A8BN BLSDRLN
B_{\ell V,D}^{RL,N}
 &=&
 -  B^*_{l a} B^*_{l_2 b} B_{l' a} B_{l_1 b}\:
  \frac{m_l m_{l_1} t^2_\beta}{M_W^2}
  \bigg(
     J^1_{1111}(1,\lambda_{H^+},\lambda_{n_a},\lambda_{n_b})
   + \lambda_a \lambda_b J^0_{1111}(\lambda_{H^+},\lambda_{n_a},\lambda_{n_b})
  \bigg), 
\\
\nonumber\\
%A1AN BlVFLLNt 
B_{\ell V,F}^{LL,\tilde{N}}
 &=&
  ( \tilde{B}^{R,1}_{l_1 k B} \tilde{B}^{R,1*}_{l_2 m B} +
\tilde{B}^{R,2}_{l_1 k B} \tilde{B}^{R,2*}_{l_2 m B} ) 
  ( \tilde{B}^{R,1}_{l' m A} \tilde{B}^{R,1*}_{l k A} +
\tilde{B}^{R,2}_{l' m A} \tilde{B}^{R,2*}_{l k A} ) 
J^1_{1111}(\lambda_{\tilde{\chi}_k},\lambda_{\tilde{\chi}_m},\lambda_{\tilde{N}_A},
\lambda_{\tilde{N}_B}), 
\nonumber\\
%A4AN BlVFRLNt
B_{\ell V,F}^{RL,\tilde{N}}
 &=& 
  ( \tilde{B}^{L,1}_{l_1 k B} \tilde{B}^{R,1*}_{l_2 m B} +
\tilde{B}^{L,2}_{l_1 k B} \tilde{B}^{R,2*}_{l_2 m B} ) 
  ( \tilde{B}^{R,1}_{l' m A} \tilde{B}^{L,1*}_{l k A} +
\tilde{B}^{R,2}_{l' m A} \tilde{B}^{L,2*}_{l k A} ) 
\nonumber\\
 &\times&
 \frac{m_{l'} m_l}{2 c_\beta^2 M_W^2}
J^1_{1111}(\lambda_{\tilde{\chi}_k},\lambda_{\tilde{\chi}_m},\lambda_{\tilde{N}_A},
\lambda_{\tilde{N}_B}),
\nonumber\\
%A6AN BlSFRRNt
B_{\ell S,F}^{RR,\tilde{N}}
 &=& 
  ( \tilde{B}^{R,1}_{l_1 k B} \tilde{B}^{L,1*}_{l_2 m B} +
\tilde{B}^{R,2}_{l_1 k B} \tilde{B}^{L,2*}_{l_2 m B} ) 
  ( \tilde{B}^{R,1}_{l' m A} \tilde{B}^{L,1*}_{l k A} +
\tilde{B}^{R,2}_{l' m A} \tilde{B}^{L,2*}_{l k A} ) 
\nonumber\\
 &\times&
 \frac{2 m_{l_2} m_l}{c_\beta^2 M_W^2}
 \sqrt{\lambda_{\tilde{\chi}_k}\lambda_{\tilde{\chi}_m}} 
 J^0_{1111}(\lambda_{\tilde{\chi}_k},\lambda_{\tilde{\chi}_m},\lambda_{\tilde{N}_A},
\lambda_{\tilde{N}_B}), 
\nonumber\\
%A8AN BlSFRLNt
B_{\ell S,F}^{RL,\tilde{N}}
 &=& 
  ( \tilde{B}^{R,1}_{l_1 k B} \tilde{B}^{R,1*}_{l_2 m B} +
\tilde{B}^{R,2}_{l_1 k B} \tilde{B}^{R,2*}_{l_2 m B} ) 
  ( \tilde{B}^{L,1}_{l' m A} \tilde{B}^{L,1*}_{l k A} +
\tilde{B}^{L,2}_{l' m A} \tilde{B}^{L,2*}_{l k A} ) 
\nonumber\\
 &\times&
 \frac{2 m_{l_1} m_l}{c_\beta^2 M_W^2}
 \sqrt{\lambda_{\tilde{\chi}_k}\lambda_{\tilde{\chi}_m}} 
 J^0_{1111}(\lambda_{\tilde{\chi}_k},\lambda_{\tilde{\chi}_m},\lambda_{\tilde{N}_A},
\lambda_{\tilde{N}_B}), \\
\nonumber\\
% Bl1 BlVDLLSB
B_{\ell V,D}^{LL,{\rm SB}}
 &=& 
 - \tilde{V}^{0\ell R}_{l_1 m b} \tilde{V}^{0\ell R*}_{l m a}
 \tilde{V}^{0\ell R}_{l' n a} \tilde{V}^{0\ell R*}_{l_2 n b}  %1 
  J^1_{1111}(\lambda_{\tilde{\chi}^0_m},\lambda_{\tilde{\chi}^0_n},\lambda_{\tilde{e}_a},
\lambda_{\tilde{e}_b})  
\nonumber\\
 &-& 
   2 \tilde{V}^{0\ell R}_{l_2 m b} \tilde{V}^{0\ell R*}_{l m a}
   \tilde{V}^{0\ell R}_{l' n a} \tilde{V}^{0\ell R*}_{l_1 n b} %1/2 
 \sqrt{\lambda_{\tilde{\chi}^0_m} \lambda_{\tilde{\chi}^0_k}}
 J^0_{1111}(\lambda_{\tilde{\chi}^0_m},\lambda_{\tilde{\chi}^0_n},\lambda_{\tilde{e}_a},
\lambda_{\tilde{e}_b}),
\nonumber\\
% Bl2 BlVDRRSB
B_{\ell V,D}^{RR,{\rm SB}}
 &=&
 - \tilde{V}^{0\ell L}_{l_1 m b} \tilde{V}^{0\ell L*}_{l m a}
 \tilde{V}^{0\ell L}_{l' n a} \tilde{V}^{0\ell L*}_{l_2 n b} %1 
  J^1_{1111}(\lambda_{\tilde{\chi}^0_m},\lambda_{\tilde{\chi}^0_n},\lambda_{\tilde{e}_a},\lambda_{\tilde{e}_b}) 
\nonumber\\
 &-&
   2 \tilde{V}^{0\ell L}_{l_2 m b} \tilde{V}^{0\ell L*}_{l m a}
   \tilde{V}^{0\ell L}_{l' n a} \tilde{V}^{0\ell L*}_{l_1 n b} %1/2 
 \sqrt{\lambda_{\tilde{\chi}^0_m} \lambda_{\tilde{\chi}^0_k}}
 J^0_{1111}(\lambda_{\tilde{\chi}^0_m},\lambda_{\tilde{\chi}^0_n},\lambda_{\tilde{e}_a},
\lambda_{\tilde{e}_b}),
\nonumber\\
% Bl3 BlVDLRSB
B_{\ell V,D}^{LR,{\rm SB}}
 &=&
 2 \tilde{V}^{0\ell L}_{l_1 m b} \tilde{V}^{0\ell R*}_{l m a}
 \tilde{V}^{0\ell R}_{l' n a} \tilde{V}^{0\ell L*}_{l_2 n b} %-1/2 
 \sqrt{\lambda_{\tilde{\chi}^0_m} \lambda_{\tilde{\chi}^0_k}}
 J^0_{1111}(\lambda_{\tilde{\chi}^0_m},\lambda_{\tilde{\chi}^0_n},\lambda_{\tilde{e}_a},
\lambda_{\tilde{e}_b}) 
\nonumber\\
 &+&
 \tilde{V}^{0\ell L}_{l_2 m b} \tilde{V}^{0\ell R*}_{l m a}
 \tilde{V}^{0\ell R}_{l' n a} \tilde{V}^{0\ell L*}_{l_1 n b} %-1 
 J^1_{1111}(\lambda_{\tilde{\chi}^0_m},\lambda_{\tilde{\chi}^0_n},\lambda_{\tilde{e}_a},
\lambda_{\tilde{e}_b}), 
\nonumber\\
% Bl4 BlVDRLSB
B_{\ell V,D}^{RL,{\rm SB}}
 &=&
 2 \tilde{V}^{0\ell R}_{l_1 m b} \tilde{V}^{0\ell L*}_{l m a}
 \tilde{V}^{0\ell L}_{l' n a} \tilde{V}^{0\ell R*}_{l_2 n b} %-1/2 
 \sqrt{\lambda_{\tilde{\chi}^0_m} \lambda_{\tilde{\chi}^0_k}}
 J^0_{1111}(\lambda_{\tilde{\chi}^0_m},\lambda_{\tilde{\chi}^0_n},\lambda_{\tilde{e}_a},
\lambda_{\tilde{e}_b})
\nonumber\\
 &+&
 \tilde{V}^{0\ell R}_{l_2 m b} \tilde{V}^{0\ell L*}_{l m a}
 \tilde{V}^{0\ell L}_{l' n a} \tilde{V}^{0\ell R*}_{l_1 n b} %-1 
 J^1_{1111}(\lambda_{\tilde{\chi}^0_m},\lambda_{\tilde{\chi}^0_n},\lambda_{\tilde{e}_a},
\lambda_{\tilde{e}_b}), 
\nonumber\\
%Bl5 BlSDLLSB
B_{\ell S,D}^{LL,{\rm SB}}
 &=&
 -2 \tilde{V}^{0\ell L}_{l_1 m b} \tilde{V}^{0\ell R*}_{l m a}
 \tilde{V}^{0\ell L}_{l' n a} \tilde{V}^{0\ell R*}_{l_2 n b} %1/2 
 \sqrt{\lambda_{\tilde{\chi}^0_m} \lambda_{\tilde{\chi}^0_k}} 
 J^0_{1111}(\lambda_{\tilde{\chi}^0_m},\lambda_{\tilde{\chi}^0_n},\lambda_{\tilde{e}_a},
\lambda_{\tilde{e}_b})
\nonumber\\ 
 &-&
 2 \tilde{V}^{0\ell R}_{l_2 m b} \tilde{V}^{0\ell R*}_{l m a}
 \tilde{V}^{0\ell L}_{l' n a} \tilde{V}^{0\ell L*}_{l_1 n b} %1/2 
 \sqrt{\lambda_{\tilde{\chi}^0_m} \lambda_{\tilde{\chi}^0_k}}
 J^0_{1111}(\lambda_{\tilde{\chi}^0_m},\lambda_{\tilde{\chi}^0_n},\lambda_{\tilde{e}_a},
\lambda_{\tilde{e}_b}), 
\nonumber\\
%Bl6 BlSDRRSB
B_{\ell S,D}^{RR,{\rm SB}}
 &=&
 -2 \tilde{V}^{0\ell R}_{l_1 m b} \tilde{V}^{0\ell L*}_{l m a}
 \tilde{V}^{0\ell R}_{l' n a} \tilde{V}^{0\ell L*}_{l_2 n b} %1/2 
 \sqrt{\lambda_{\tilde{\chi}^0_m} \lambda_{\tilde{\chi}^0_k}}
 J^0_{1111}(\lambda_{\tilde{\chi}^0_m},\lambda_{\tilde{\chi}^0_n},\lambda_{\tilde{e}_a},
\lambda_{\tilde{e}_b}) 
\nonumber\\
 &-& 
  2 \tilde{V}^{0\ell L}_{l_2 m b} \tilde{V}^{0\ell L*}_{l m a}
  \tilde{V}^{0\ell R}_{l' n a} \tilde{V}^{0\ell R*}_{l_1 n b} %1/2 
 \sqrt{\lambda_{\tilde{\chi}^0_m} \lambda_{\tilde{\chi}^0_k}}
 J^0_{1111}(\lambda_{\tilde{\chi}^0_m},\lambda_{\tilde{\chi}^0_n},\lambda_{\tilde{e}_a},
\lambda_{\tilde{e}_b}),
\nonumber\\
%Bl7 BlSDLRSB
B_{\ell S,D}^{LR,{\rm SB}}
 &=&
 2 \tilde{V}^{0\ell R}_{l_1 m b} \tilde{V}^{0\ell R*}_{l m a}
 \tilde{V}^{0\ell L}_{l' n a} \tilde{V}^{0\ell L*}_{l_2 n b} %-2 
  J^1_{1111}(\lambda_{\tilde{\chi}^0_m},\lambda_{\tilde{\chi}^0_n},\lambda_{\tilde{e}_a},
\lambda_{\tilde{e}_b})
\nonumber\\
 &+&
  2 \tilde{V}^{0\ell L}_{l_2 m b} \tilde{V}^{0\ell R*}_{l m a}
  \tilde{V}^{0\ell L}_{l' n a} \tilde{V}^{0\ell R*}_{l_1 n b} %-2 
 J^1_{1111}(\lambda_{\tilde{\chi}^0_m},\lambda_{\tilde{\chi}^0_n},\lambda_{\tilde{e}_a},
\lambda_{\tilde{e}_b}),
\nonumber\\
%Bl8 BlSDRLSB
B_{\ell S,D}^{RL,{\rm SB}}
 &=&
 2 \tilde{V}^{0\ell L}_{l_1 m b} \tilde{V}^{0\ell L*}_{l m a}
 \tilde{V}^{0\ell R}_{l' n a} \tilde{V}^{0\ell R*}_{l_2 n b} %-2 
  J^1_{1111}(\lambda_{\tilde{\chi}^0_m},\lambda_{\tilde{\chi}^0_n},\lambda_{\tilde{e}_a},
\lambda_{\tilde{e}_b})
\nonumber\\
 &+&
 2 \tilde{V}^{0\ell R}_{l_2 m b} \tilde{V}^{0\ell L*}_{l m a}
 \tilde{V}^{0\ell R}_{l' n a} \tilde{V}^{0\ell L*}_{l_1 n b} %-2 
 J^1_{1111}(\lambda_{\tilde{\chi}^0_m},\lambda_{\tilde{\chi}^0_n},\lambda_{\tilde{e}_a},
\lambda_{\tilde{e}_b}),
\nonumber\\
%Bl9 BlTDLLSB
B_{\ell T,D}^{LL,{\rm SB}}
 &=&
 -2 \tilde{V}^{0\ell L}_{l_1 m b} \tilde{V}^{0\ell R*}_{l m a}
 \tilde{V}^{0\ell L}_{l' n a} \tilde{V}^{0\ell R*}_{l_2 n b} %1/2 
 \sqrt{\lambda_{\tilde{\chi}^0_m} \lambda_{\tilde{\chi}^0_k}} 
 J^0_{1111}(\lambda_{\tilde{\chi}^0_m},\lambda_{\tilde{\chi}^0_n},\lambda_{\tilde{e}_a},
\lambda_{\tilde{e}_b})
\nonumber\\ 
 &+&
  2 \tilde{V}^{0\ell R}_{l_2 m b} \tilde{V}^{0\ell R*}_{l m a}
  \tilde{V}^{0\ell L}_{l' n a} \tilde{V}^{0\ell L*}_{l_1 n b} %1/2 
 \sqrt{\lambda_{\tilde{\chi}^0_m} \lambda_{\tilde{\chi}^0_k}}
 J^0_{1111}(\lambda_{\tilde{\chi}^0_m},\lambda_{\tilde{\chi}^0_n},\lambda_{\tilde{e}_a},
\lambda_{\tilde{e}_b}),
\nonumber\\
%Bl10 BlTDRRSB
B_{\ell T,D}^{RR,{\rm SB}}
 &=&
 - 2 \tilde{V}^{0\ell R}_{l_1 m b} \tilde{V}^{0\ell L*}_{l m a}
 \tilde{V}^{0\ell R}_{l' n a} \tilde{V}^{0\ell L*}_{l_2 n b} %1/2 
 \sqrt{\lambda_{\tilde{\chi}^0_m} \lambda_{\tilde{\chi}^0_k}} 
 J^0_{1111}(\lambda_{\tilde{\chi}^0_m},\lambda_{\tilde{\chi}^0_n},\lambda_{\tilde{e}_a},
\lambda_{\tilde{e}_b})
\nonumber\\ 
 &+&
  2 \tilde{V}^{0\ell L}_{l_2 m b} \tilde{V}^{0\ell L*}_{l m a}
  \tilde{V}^{0\ell R}_{l' n a} \tilde{V}^{0\ell R*}_{l_1 n b} %1/2 
 \sqrt{\lambda_{\tilde{\chi}^0_m} \lambda_{\tilde{\chi}^0_k}}
 J^0_{1111}(\lambda_{\tilde{\chi}^0_m},\lambda_{\tilde{\chi}^0_n},\lambda_{\tilde{e}_a},
\lambda_{\tilde{e}_b})\ .
\label{ffboxSB}
\end{eqnarray}

\subsection{Semileptonic Box Formfactors}

Semileptonic form  factors have only {\it  direct} contributions, with
the following $N$, $\tilde{N}$ and ${\rm SB}$ content:
\begin{eqnarray}
B_{dV}^{LL} &=& B_{dV}^{LL,N} + B_{dV}^{LL,\tilde{N}} +B_{dV}^{LL,{\rm SB}}\ ,
                                                             \nonumber\\    
B_{uV}^{LL} &=& B_{uV}^{LL,N} + B_{uV}^{LL,\tilde{N}} +  B_{uV}^{LL,{\rm SB}}\ ,
\end{eqnarray}
and
\begin{equation}
\quad\, B_{dA}^{XY}\ =\ B_{dA}^{XY,{\rm SB}}\;,\qquad 
B_{uA}^{XY}\ =\ B_{uA}^{XY,{\rm SB}}\ , 
\end{equation}
for $(X,Y,A)\neq (L,L,V)$.  The  $N$ and $\tilde{N}$ contributions are
given by
\begin{eqnarray}
% A1dN BdVLLN
B_{dV}^{LL,N}  &=&  B_{l' a} B^*_{l a} (V^*)_{bd_1}(V)_{bd_2}
 \bigg[
    - \bigg( 1 + \frac{\lambda_{n_a} \lambda_{u_b}}{4} \bigg)
    J^1_{211}(1,\lambda_{n_a},\lambda_{u_b}) 
 + 2 \lambda_{n_a} \lambda_{u_b}
 J^0_{211}(1,\lambda_{n_a},\lambda_{u_b})
\nonumber\\
 &+&
    \frac{1}{2 t_\beta^2} \lambda_{n_a}\lambda_{u_b}
    J^0_{1111}(1,\lambda_{H^+},\lambda_{n_a},\lambda_{u_b}) 
  - \frac{1}{2 t_\beta^2} \lambda_{n_a}\lambda_{u_b}
  J^1_{1111}(1,\lambda_{H^+},\lambda_{n_a},\lambda_{u_b}) 
\nonumber\\
 &-& \frac{1}{4 t_\beta^4} \lambda_{n_a}\lambda_{u_b}
J^1_{211}(\lambda_{H^+},\lambda_{n_a},\lambda_{u_b}) 
 \bigg],
\nonumber\\
% A1uN BuVLLN
B_{uV}^{LL,N}
 &=&
 B_{l' a} B^*_{l a} (V^*)_{d_2b}(V)_{d_1b}
 \bigg[
    \bigg( 4 + \frac{\lambda_{n_a} \lambda_{d_b}}{4} \bigg)
    J^1_{211}(1,\lambda_{n_a},\lambda_{d_b}) 
  - 2 \lambda_{n_a} \lambda_{d_b} J^0_{211}(1,\lambda_{n_a},\lambda_{d_b})
\nonumber\\
 &+&
    \frac{1}{2}\lambda_{n_a}\lambda_{d_b}
    J^0_{1111}(1,\lambda_{H^+},\lambda_{n_a},\lambda_{d_b}) 
  - \frac{1}{2} \lambda_{n_a}\lambda_{d_b}
  J^1_{1111}(1,\lambda_{H^+},\lambda_{n_a},\lambda_{d_b}) 
\nonumber\\
 &+& \frac{1}{4} \lambda_{n_a}\lambda_{d_b}
J^1_{211}(\lambda_{H^+},\lambda_{n_a},
\lambda_{d_b})
 \bigg],
\\
% A1dNt BdVLLNt
B_{dV}^{LL,\tilde{N}}    
 &=&
 - \frac{1}{2} \tilde{V}_{d_1ka}^{-dR}\tilde{V}_{d_2ma}^{-dR*} 
   ( \tilde{V}_{lkA}^{-\ell R,1*}\tilde{V}_{l'mA}^{-\ell R,1} 
   + \tilde{V}_{lkA}^{-\ell R,2*}\tilde{V}_{l'mA}^{-\ell R,2})\:
   J^1_{1111}(\lambda_{\lambda_{\tilde{\chi}_k}},\lambda_{\tilde{\chi}_m},
\lambda_{\tilde{N}_A},\lambda_{\tilde{u}_a}),
\nonumber\\
% A1uNt BuVLLNt
B_{uV}^{LL,\tilde{N}}
 &=&
 - \tilde{V}_{u_1ka}^{-uR}\tilde{V}_{u_2ma}^{-uR*} 
   ( \tilde{V}_{lkA}^{-\ell R,1*}\tilde{V}_{l'mA}^{-\ell R,1}
   + \tilde{V}_{lkA}^{-\ell R,2*}\tilde{V}_{l'mA}^{-\ell R,2})\:
   \sqrt{\lambda_{\tilde{\chi}_k} \lambda_{\tilde{\chi}_m}} 
   J^0_{1111}(\lambda_{\lambda_{\tilde{\chi}_k}},\lambda_{\tilde{\chi}_m},
\lambda_{\tilde{N}_A},\lambda_{\tilde{d}_a})\ .
\end{eqnarray}
The   ${\rm    SB}$   form   factors    $B_{dA}^{XY,{\rm   SB}}$   and
$B_{uA}^{XY,{\rm  SB}}$,  with  $X=L,R$,  $Y=L,R$ and  $A=V,S,T$,  are
obtained   from  the  {\it   direct}  leptonic   formfactors  $B_{\ell
  A}^{XY,{\rm SB}}$, by making  the replacements: $\ell\to d$, $l_1\to
d$, $l_2\to  d$, $\tilde{e}\to\tilde{d}$ and $\ell\to  u$, $l_1\to u$,
$l_2\to u$, $\tilde{e}\to\tilde{u}$,  in both the interaction vertices
and the arguments  of the $J$-loop functions that  carry the index $b$
in~(\ref{ffboxSB}).

\end{appendix}


\vspace{3mm}

\begin{thebibliography}{99}

\bibitem{sol-nuosc}
  B. T.~Cleveland {\it et al.} [Homeostake], 
  Astrophys.\ J. 498 (1998) 505;\\
  J. N.~Abdurashitov {\it et al.} [SAGE Collaboration], 
  Phys.\ Rev.\ C~{\bf 80} (2009) 015807
  [arXiv:0901.2200 [nucl-ex]];\\
  J.~Hosaka {\it et al.} [Super-Kamiokande Collaboration] 
  Phys.\ Rev.\ D {\bf 73} (2006) 112001
  [hep-ex/0508053];\\
  J. P.~Cravens {\it et al.} [Super-Kamiokande Collaboration], 
  Phys.\ Rev.\ D {\bf 78} (2008) 032002   [arXiv:0801.0776 [hep-ex]];\\
  B.~Aharmim {\it et al.} [SNO Collaboration], 
  Phys.\ Rev.\ C {\bf 72} (2005) 055502;\\
  B.~Aharmim {\it et al.} [SNO Collaboration], 
  Phys.\ Rev.\ C {\bf 81} (2010) 055504.

\bibitem{atm-nuosc}
  Y.~Fukuda {\it et al.} [Super-Kamiokande collaboration], 
  Phys.\ Rev.\ Lett. {\bf 81} (1998) 1562;\\
  J.~Hosaka {\it et al.} [Super-Kamiokande Collaboration], 
  Phys.\ Rev.\ D~{\bf 74} (2006) 032002
  [hep-ex/0604011];\\
  J.~Ashie {\it et al.} [Super-Kamiokande Collaboration], 
  Phys.\ Rev.\ D~{\bf 71} (2005) 112005
  [hep-ex/0501064].

\bibitem{acc-nuosc}
  M.H. Ahn {\it et al.} [K2K Collaboration], 
  Phys.\ Rev.\ D {\bf 74} (2006) 072003
  [hep-ex/0606032];\\
  P. Adamson {\it et al.} [MINOS Collaboration], 
  Phys.\ Rev.\ Lett.\ {\bf 101} (2008) 131802
  [arXiv:0806.2237 [hep-ex]];\\
  P. Adamson {\it et al.} [MINOS Collaboration], 
  Phys.\ Rev.\ Lett.\ {\bf 101} (2008) 221804
  [arXiv:0807.2424 [hep-ex]];\\
  A. Habig {\it et al.} [MINOS Collaboration],
  Nucl.\ Phys.\ Proc.\ Suppl.\ {\bf 218} (2011) 320;\\
  M.~Hartz [T2K Collaboration], arXiv:1201.1846 [hep-ex];\\
  P.~Adamson {\it et al.} [MINOS Collaboration],
  Phys.\ Rev.\ Lett.\ {\bf 108} (2012) 191801,
  [arXiv:1202.2772 [hep-ex]].

\bibitem{reac-nuosc}
  S.~Abe {\it et al.} [KamLAND Collaboration], 
  Phys. Rev. Lett. {\bf 100} (2008) 221803.

\bibitem{nu13} Y. Abe {\it et al.} [DOUBLE-CHOOZ Collaboration], 
  Phys.\ Rev.\ Lett.\ {\bf 108} (2012) 131801
  [arXiv:1112.6353 [hep-ex]];\\
  F. P.~ An {\it et al.} [DAYA-BAY Collaboration],
  Phys.\ Rev.\ Lett.\ {\bf 108} (2012) 171803
  [arXiv:1203.1669 [hep-ex]];\\
  J. K.~Ahn {\it et al.} [RENO Collaboration],
  Phys.\ Rev.\ Lett.\ {\bf 108} (2012) 191802
  [arXiv:1204.0626 [hep-ex]].

\bibitem{PMNS} B. Pontecorvo, Zh.\ Eksp.\ Teor.\ Fiz.\ {\bf 33} (1957)
  549 [JETP {\bf 6} (1957) 429];\\ Z. Maki, M. Nakagawa and S. Sakata,
  Prog.\ Theor.\ Phys.\ {\bf 28} (1962) 870.

\bibitem{MEG11} J. Adam {\it et al.} [MEG Collaboration],
  Phys.\ Rev.\ Lett. {\bf 107} (2011) 171801 
 [arXiv:1107.5547 [hep-ex]].


\bibitem{SINDRUM88} U. Bellgart {\it et al.} [SINDRUM Collaboration],
  Nucl.\ Phys.\ B~{\bf 299} 1988 1. 

\bibitem{Titanium}  C. Dohmen {\it et al.} [SINDRUM II collaboration],
  Phys.\ Lett.\ B~{\bf 317} (1993) 631.

\bibitem{Gold} W. Bertl {\it et al.}, Eur.\ Phys.\ J.\ C~{\bf 47}
  (2006) 337. 

\bibitem{LFV-tau}
  Y.~Miyazaki {\it et al.} [Belle Collaboration], 
  arXiv:1206.5595 [hep-ex];\\
  Y. Miyazaki {\it et al.} [Belle Collaboration], 
  Phys.\ Lett.\ B {\bf 699} (2011) 251
  [arXiv:1101.0755 [hep-ex]];\\
  B. Aubert {\it et al.} [BaBar Collaboration], 
  Phys.\ Rev.\ Lett.\ {\bf 103} (2009) 021801;\\
  K. Hayasaka {\it et al.} [Belle Collaboration], 
  Phys.\ Lett.\ B {\bf 687} (2010) 139  
  [arXiv:1001.3221 [hep-ex]];\\
  J. P. Lees {\it et al.} [BaBar Collaboration], 
  Phys.\ Rev.\ D {\bf 81} (2010) 111101
  [arXiv:1002.4550 [hep-ex]];\\
  Y. Miyazaki {\it et al.} [Belle Collaboration], 
  Phys.\ Lett.\ B {\bf 692} (2010) 4
  [arXiv:1003.1183 [hep-ex]];\\
  B. Aubert {\it et al.} [BaBar Collaboration], 
  Phys.\ Rev.\ Lett.\ {\bf 104} (2010) 151802
  [arXiv:0908.2381 [hep-ex]];\\
  J. Beringer {\it et al.} [Particle Data Group], 
  Phys.\ Rev.\ D~{\bf 86} (2012) 010001.

\bibitem{MEGPhD12}
  B.~A.~Golden,
  PhD Thesis, 2012, http://meg.web.psi.ch/docs/theses/BenThesis.pdf;\\
  J. Adam,
  PhD Thesis, 2012,
  http://meg.web.psi.ch/docs/theses/PhDThesis$_{}$Adam$_{}$2012.pdf. 

\bibitem{Hew12} J. L.~Hewett {\it et al.}, 1205.2671 [hep-ex].

\bibitem{mu3eBerger11}
  N. Berger [$\mu$3e Collaboration], arXiv:1110.1504 [hep-ex].

\bibitem{COMET}
  A. Kurup [COMET Collaboration],
  Nucl.~Phys.~Proc.~Suppl. {\bf 218}, 38 (2011);\\
  R. Akhmetshin et al. [COMET]
  http://j-parc.jp/researcher/Hadron/en/pac$_{}$1203/pdf/COMET-PhaseI-LoI.pdf.

\bibitem{Mu2e}
R.~J.~Abrams {\it et al.}  [Mu2e Collaboration],
  %``Mu2e Conceptual Design Report,''
  arXiv:1211.7019 [physics.ins-det];\\
  %%CITATION = ARXIV:1211.7019;%%
  R.~K.~Kurschke, arXiv:1112.0242 [hep-ex];\\
  E.~C.~Dukes, Nucl.~Phys.~Proc.~Suppl. {\bf 218}, 32 (2011).

\bibitem{PRISM} Y.~Kuno,   %``PRISM/PRIME,''
  Nucl.\ Phys.\ Proc.\ Suppl.\  {\bf 149} (2005) 376;\\
  R.~J.~Barow, 
  Nucl.\ Phys.\ Proc.\ Suppl.\  {\bf 218} (2011) 44. 

\bibitem{Hay09} K. Hayasaka, Journal of Physics: Conference Series
  {\bf 171} (2009) 012079.

\bibitem{Bon07}  M. Bona  {\it et  al.}  [SuperB  Collaboration] Pisa,
  Italy:    INFN   (2007)    453    p.    www.pi.infn.it/SuperB/?q=CDR
  [arXiv:0709.0451  [hep-ex]].  With  great dismay,  we  have recently
  been informed that the SuperB project will be discontinued.

\bibitem{Review} For reviews, see, H.P. Nilles, Phys.\ Rept.\ {\bf 110}
  (1984) 1;\\ H. Haber and G.  Kane, Phys.\ Rept.\ {\bf 117} (1985) 75.

\bibitem{HdisATL} G. Aad {\it et al.}, 1207.7214 [hep-ex]

\bibitem{HdisCMS} S. Chatrchyan {\it et al.}, 1207.7235 [hep-ex]

\bibitem{seesaw}  P.  Minkowski,   Phys.\  Lett.\  B~{\bf  67}  (1977) 421;\\   
M.   Gell-Mann, P.   Ramond and  R.  Slansky,  in  {\it Supergravity},
eds.\   D.Z.   Freedman   and  P.~van   Nieuwenhuizen  (North-Holland,
Amsterdam, 1979);\\  
T.  Yanagida, in Proc.\ of the {\it Workshop on the Unified Theory and
  the  Baryon   Number  in   the  Universe},  Tsukuba,   Japan,  1979,
eds.\ O.~Sawada  and A.~Sugamoto;\\ 
S.L.   Glashow,   in  {\it   Quarks  and  Leptons,   Carg\`ese  1979},
eds.   M.    L\'evy,   et   al.,   (Plenum   1980    New   York),   p.
707;\\ 
R.~N.~Mohapatra  and G.~Senjanovi\'c, Phys.\  Rev.\ Lett.\ {\bf 44}
(1980) 912;\\    
J.  Schechter and J.  W.  F.  Valle, Phys.\ Rev.\ D~{\bf 22} (1980)
2227.

\bibitem{seesawII} 
  W. Konetschny and W. Kummer, Phys.\ Lett.\ B~{\bf 70} (1977) 433;\\
  J. Schechter and J. W. F. Valle, Phys.\ Rev.\ D~{\bf 22} (1980) 2227;\\
  G. Lazarides, Q. Shafi and C. Wetterrich, Nucl.\ Phys.\ B~{\bf 181}
  (1981) 287;\\ 
  R. N. Mohapatra and G. Senjanovic, Phys.\ Rev.\ D~{\bf 23} (1981) 165;\\
  T. P. Cheng and L.-F. Li, Phys.\ Rev.\ D~{\bf 22} (1980) 2860;\\
  J. Schechter, J. W. F. Valle, Phys.\ Rev.\ D~{\bf 25} (1982) 774.

\bibitem{seesawIII} R. Foot, H. Lew, F.G. He and G.C. Joshi,
  Z.\ Phys.\ C~{\bf 44} (1989) 441.

\bibitem{WW} D.~Wyler and L. Wolfenstein, Nucl.\ Phys.\ B~{\bf 218} (1983) 205.

\bibitem{MV} R.~N.~Mohapatra, Phys.\ Rev.\ Lett.\ {\bf 56} (1986) 561;\\
  R.~N. Mohapatra and J. W. F. Valle, Phys.\ Rev.\ D~{\bf 34} (1986)
  1642;\\ S. Nandi and U. Sarkar, Phys.\ Rev.\ Lett.\ {\bf 56} (1986)
  564.

\bibitem{BGL} G.~C.~Branco, W.~Grimus and L.~Lavoura, Nucl.\ Phys.\
  B~{\bf 312} (1989) 492.

\bibitem{AZPC} A. Pilaftsis, Z.\ Phys.\ C~{\bf 55} (1992) 275; \\
P.~S.~B. Dev and A.~Pilaftsis,
  %``Minimal Radiative Neutrino Mass Mechanism for Inverse Seesaw Models,''
  Phys. Rev. D {\bf 86} (2012) 113001 [arXiv:1209.4051 [hep-ph]].
%%CITATION = ARXIV:1209.4051;%%

\bibitem{APRLtau} 
A.~Pilaftsis,
%``Resonant tau-leptogenesis with observable lepton number violation,''
Phys.\ Rev.\ Lett.\  {\bf 95} (2005) 081602 [hep-ph/0408103];\\
A.~Pilaftsis and T.~E.~J.~Underwood,
%``Electroweak-scale resonant leptogenesis,''
Phys.\ Rev.\ D {\bf 72} (2005) 113001 [hep-ph/0506107];\\
F.~F.~Deppisch and A.~Pilaftsis,
%``Lepton Flavour Violation and theta(13) in Minimal Resonant Leptogenesis,''
Phys.\ Rev.\ D {\bf 83} (2011) 076007 [arXiv:1012.1834 [hep-ph]].

\bibitem{LFVp}
  E. Nardi, E. Roulet and D. Tommasini
  Phys.\ Lett.\ B~{\bf 327} (1994) 319
  [hep-ph/9402224];\\
  S. Antusch, J. P. Baumann and E. Fern\'andez-Martinez,
  Nucl.\ Phys.\ B~{\bf 810} (2008) 369
  [arXiv:0807.1003 [hep-ph]].

\bibitem{BK} S.~Bergmann and A.~Kagan, Nucl.\ Phys.\ B~{\bf 538}
  (1999) 368;\\ F.~del Aguila, J.~de Blas and M.~Perez-Victoria,
  Phys.\ Rev.\ D~{\bf 78} (2008) 013010.

\bibitem{IPNPB} A.~Ilakovac and A.~Pilaftsis,
  %``Flavour violating charged lepton decays in seesaw-type models,''
  Nucl.\ Phys.\ B {\bf 437} (1995) 491
  [hep-ph/9403398].

\bibitem{IPPRD} A.~Ilakovac and A.~Pilaftsis,
  %``Supersymmetric Lepton Flavour Violation in Low-Scale Seesaw Models,''
  Phys.\ Rev.\  D~{\bf 80} (2009) 091902
  [arXiv:0904.2381 [hep-ph]].
  %CITATION = PHRVA,D80,091902;%%

\bibitem{IPNPBP} A.~Ilakovac and A.~Pilaftsis,
  Nucl.\ Phys.\ Proc.\ Suppl.\ {\bf 218} (2011) 26
  [arXiv:1012.2823 [hep-ph]].

\bibitem{BM} F. Borzumati and A. Masiero, Phys.\ Rev.\ Lett.\ {\bf 57}
  (1986) 961.

\bibitem{HMTY} J. Hisano, T. Moroi, K. Tobe and M. Yamaguchi, Phys.\
  Rev.\ D~{\bf 53} (1996) 2442 [hep-ph/9510309].

\bibitem{softLFV} J. Hisano, T. Moroi, K. Tobe, M. Yamaguchi and
  T. Yanagida, Phys. Lett. B~{\bf 357} (1995) 579 [hep-ph/9501407];\\
J. Hisano and D. Nomura, Phys.\ Rev.\ D~{\bf 59} (1999) 116005
[hep-ph/9810479];\\  
D.~F.~Carvalho, J.~R.~Ellis, M.~E.~Gomez and S.~Lola,
%``Charged lepton flavor violation in the CMSSM in view of the muon
%anomalous magnetic moment,'' 
  Phys.\ Lett.\ B {\bf 515} (2001) 323
  [hep-ph/0103256];\\
J. Hisano, M. Nagai, P. Paradisi and Y. Shimizu, 
JHEP~{\bf 0912} (2009) 030 [arXiv:0904.2080 [hep-ph]].

\bibitem{EHLR}
J.~R.~Ellis, J.~Hisano, S.~Lola and M.~Raidal,
%``CP  violation  in  the  minimal  supersymmetric  seesaw  model,''
Nucl.\ Phys.\ B~{\bf 621} (2002) 208~[hep-ph/0109125].

\bibitem{CDM} 
C. Arina, F. Bazzocchi, N. Fornengo, J. C. Romao and J. W. F. Valle, 
Phys.\ Rev.\ Lett.\ {\bf 101} (2008) 161802 [arXiv:0806.3225 [hep-ph]];\\  
F. Deppisch and A. Pilaftsis,  
JHEP {\bf 0810} (2008) 080 [arXiv:0808.0490 [hep-ph]];\\
F.~-X.~Josse-Michaux and E.~Molinaro,
  Phys.\ Rev.\ D~{\bf 84} (2011) 125021 [arXiv:1108.0482 [hep-ph]];\\ 
H. An, P. S. B. Dev, Y. Cai and R. N. Mohapatra, Phys.\ Rev.\ Lett.\ 
{\bf 108} (2012) 081806 [arXiv:1110.1366 [hep-ph]];\\
B.~Dumont, G.~Belanger, S.~Fichet, S.~Kraml and T.~Schwetz,
%``Mixed sneutrino dark matter in light of the 2011 XENON and LHC results,''
JCAP {\bf 1209} (2012) 013 [arXiv:1206.1521 [hep-ph]].
%%CITATION = ARXIV:1206.1521;%%

\bibitem{ResLG} A.~Pilaftsis,
  %``CP violation and baryogenesis due to heavy Majorana neutrinos,''
  Phys.\ Rev.\ D {\bf 56} (1997) 5431
  [hep-ph/9707235];\\ 
A. Pilaftsis  and T. E. J. Underwood,  Nucl.\ Phys.\ B
  {\bf 694} (2004) 303 [hep-ph/0309342].

\bibitem{EWBAU}
  T.~Cohen, D.~E.~Morrissey and A.~Pierce,
  %``Electroweak Baryogenesis and Higgs Signatures,''
  Phys.\ Rev.\ D~{\bf 86} (2012) 013009  [arXiv:1203.2924 [hep-ph]];\\
  %%CITATION = ARXIV:1203.2924;%%
  M.~Carena, G.~Nardini, M.~Quiros and C.~E.~M.~Wagner,
  %``MSSM Electroweak Baryogenesis and LHC Data,''
  arXiv:1207.6330 [hep-ph].
  %%CITATION = ARXIV:1207.6330;%%

\bibitem{FR74} S.~Ferrara and E.~Remiddi,
%``Absence Of The Anomalous Magnetic Moment
%In A Supersymmetric Abelian Gauge Theory,''
  Phys.\ Lett.\ B {\bf 53} (1974) 347.

%\cite{Hirsch:2012ax}
\bibitem{Hirsch:2012ax}
  M.~Hirsch, F.~Staub and A.~Vicente,
  %``Enhancing $l_i \to 3 l_j$ with the $Z^0$-penguin,''
  Phys.\ Rev.\ D {\bf 85} (2012) 113013
  [arXiv:1202.1825 [hep-ph]].
  %%CITATION = ARXIV:1202.1825;%%

\bibitem{Abada:2012cq}
A.~Abada, D.~Das, A.~Vicente and C.~Weiland,
%``Enhancing lepton flavour violation in the supersymmetric inverse
%seesaw beyond the dipole contribution,'' 
  JHEP {\bf 1209} (2012) 015  [arXiv:1206.6497 [hep-ph]].
  %%CITATION = ARXIV:1206.6497;%%

\bibitem{m-gtqt}
  G. Aad {\it et al.} [ATLAS Collaboration], arXiv:1206.1760;\\
  S. Chatrchyan {\it et al} [CMS Collaboration], arXiv:1207.1898.

\bibitem{KPS}
J.~Bernab\'eu, A.~Santamaria, J.~Vidal, A.~Mendez and J.~W.~F.~Valle,
  %``Lepton Flavour Nonconservation at High-Energies in a Superstring Inspired
  %Standard Model,''
  Phys.\ Lett.\  B {\bf 187} (1987) 303;\\
J.~G. K\"orner, A. Pilaftsis and K. Schilcher, Phys.\
  Lett.\ B {\bf 300} (1993) 381;\\
J.~Bernab\'eu, J.~G.~K\"orner, A.~Pilaftsis and K.~Schilcher,
  %``Universality breaking effects in leptonic Z decays,''
  Phys.\ Rev.\ Lett.\  {\bf 71} (1993) 2695.
%  [hep-ph/9307295].

\bibitem{DV} F.~Deppisch and J.~W.~F.~Valle,
 %``Enhanced lepton flavour violation in the supersymmetric inverse seesaw
 %model,''
  Phys.\ Rev.\  D~{\bf 72} (2005) 036001.

\bibitem{KS} 
J.~Kersten and A.~Y.~Smirnov,
%``Right-Handed Neutrinos at CERN LHC and the Mechanism of Neutrino
%Mass Generation,'' 
Phys.\ Rev.\ D {\bf 76} (2007) 073005 [arXiv:0705.3221 [hep-ph]].
%%CITATION = ARXIV:0705.3221;%%

\bibitem{P-PRD08} A.~Pilaftsis, Phys.\ Rev.\ D~{\bf 78} (2008) 013008
  [arXiv:0805.1677 [hep-ph]].

\bibitem{CL} T. P.~Cheng and L. F.~Li, Phys.\ Rev.\ Lett.\ {\bf 45}
  (1980) 1908.

\bibitem{AH05} E. Arganda and M. Herrero, Phys.\ Rev.\ D~{\bf 73} (2006) 055003.

\bibitem{COKFV} H.~C. Chiang, E. Oset, T. S. Kosmas, A. Faessler
   and J.~D.~Vergados, Nucl.~Phys.\ A~{\bf 559} (1993) 526.

\bibitem{KKO} R. Kitano, M. Koike and Y. Okada, Phys.\ Rev.\ D~{\bf
  66} (2002) 096002.

%\cite{Ilakovac:1995km}
\bibitem{Ilakovac:1995km}
A.~Ilakovac, B.~A.~Kniehl and A.~Pilaftsis,
%``Semileptonic lepton number / flavor violating tau decays in
%Majorana neutrino models,'' 
  Phys.\ Rev.\ D {\bf 52} (1995) 3993
  [hep-ph/9503456].
  %%CITATION = HEP-PH/9503456;%%


%\cite{Alonso:2012ji}
\bibitem{Alonso:2012ji}
  R.~Alonso, M.~Dhen, M.~B.~Gavela and T.~Hambye,
  %``Muon conversion to electron in nuclei in type-I seesaw models,''
  arXiv:1209.2679 [hep-ph].
  %%CITATION = ARXIV:1209.2679;%%

\bibitem{ElOl12} J. Ellis and K. A. Olive, Eur.\ Phys.\ J. C~{\bf 72}
  (2012) 2005 [arXiv:1202.3262]. 

\bibitem{CHHHWW12} M. Carena, H. E. Haber, S. Heinemeyer,
  W. Hollik, C. E. M. Wagner and G. Weiglein, Nucl.\ Phys. B {\bf
    580} (2000) 29 [hep-ph-0001002].

\bibitem{CPsuperH} 
J.~S.~Lee, M.~Carena, J.~Ellis, A.~Pilaftsis and C.~E.~M.~Wagner,
%``CPsuperH2.3: an Updated Tool for Phenomenology in the MSSM with
%Explicit CP Violation,'' 
  arXiv:1208.2212 [hep-ph];\\
  %%CITATION = ARXIV:1208.2212;%%
  J.~S.~Lee, M.~Carena, J.~Ellis, A.~Pilaftsis and C.~E.~M.~Wagner,
%``CPsuperH2.0: an Improved Computational Tool for Higgs Phenomenology
%in the MSSM with Explicit CP Violation,'' 
  Comput.\ Phys.\ Commun.\  {\bf 180} (2009) 312;\\
%%CITATION = ARXIV:0712.2360;%%
J.~S.~Lee, A.~Pilaftsis, M.~S.~Carena, S.~Y.~Choi, M.~Drees,
J.~R.~Ellis and C.~E.~M.~Wagner, 
%``CPsuperH: A Computational tool for Higgs phenomenology in the
%minimal supersymmetric standard model with explicit CP violation,'' 
  Comput.\ Phys.\ Commun.\  {\bf 156} (2004) 283.
%%CITATION = HEP-PH/0307377;%%
%\cite{Heinemeyer:2010eg}

\bibitem{Heinemeyer:2010eg}
S.~Heinemeyer, M.~J.~Herrero, S.~Penaranda and A.~M.~Rodriguez-Sanchez,
%``Higgs Boson Masses in the MSSM with Heavy Majorana Neutrinos,''
JHEP {\bf 1105} (2011) 063.
%  [arXiv:1007.5512 [hep-ph]].
%%CITATION = ARXIV:1007.5512;%%

\bibitem{Cha02} P. Chankowski and S. Pokorski,
  Int.\ J.\ Mod.\ Phys. A~{\bf 17} (2002) 575 
 [hep-ph/0110249].

\bibitem{Pet04} S. Petcov, S. Profumo, Y. Takanishi and C. Yaguna,
  Nucl.\ Phys.\ B~{\bf 676} (2004) 453-480 [hep-ph/0306195].

\bibitem{tau-8a} K.\ Hayasaka {\it et al.} [Belle Collaboration],
  Phys.\ Lett.\ B {\bf 687} (2010) 139.

\bibitem{tau-8b} B.\ Aubert {\it et al.} [Babar Collaboration], 
Phys.\ Rev.\ Lett.\ {\bf 104} (2010) 021802.

\bibitem{mueAl-16} The {\tt Mu2e} Project at Fermilab: http://mu2e.fnal.gov/.

\bibitem{mueTi-18a} Y.\ Mori {\it et al.} [The PRIME Working Group],
  "An experimental search for $\mu^-\to e^-$ conversion process at the
  ultimate sensitivity of the order of $10^{-18}$ with PRISM", LOI-25,
  http://www-ps.kek.jp/jhf-np/LOIlist/LOIlist.html.

\bibitem{mueTi-18b} The {\tt X}-Project: http://projectx.fnal.gov.

\bibitem{MEG} S.~Ritt {\it et al.} [MEG Collaboration],
  %``Status Of The Meg Expriment Mu $\to$ E Gamma,''
  Nucl.\ Phys.\ Proc.\ Suppl.\  {\bf 162} (2006) 279.

\bibitem{DGR_Dress} M. Drees, M. Goldbole and P. Roy, {\it Theory and
  Phenomenology of Sparticles}, World Slientific, 2004.


\end{thebibliography}
\end{document}